\documentclass[fleqn,usenatbib]{mnras}

\usepackage{newtxtext,newtxmath}

\usepackage[T1]{fontenc}

\DeclareRobustCommand{\VAN}[3]{#2}
\let\VANthebibliography\thebibliography
\def\thebibliography{\DeclareRobustCommand{\VAN}[3]{##3}\VANthebibliography}

\usepackage{graphicx}	
\usepackage{amsmath}	
\usepackage[normalem]{ulem}

\title[Three new hot Jupiters approaching the TAMS]{TOI-3664\,b, TOI-4034\,b \& TOI-6564\,b: Three new hot Jupiters around stars approaching the terminal age main sequence}

\author[M.~P.~Battley et al.]{
Matthew P. Battley,$^{1,2}$\thanks{E-mail: m.battley@qmul.ac.uk}
Marina Lafarga,$^{3,4}$
Edward Gillen,$^{1}$
Monika Lendl,$^{2}$
Solène Ulmer-Moll,$^{5}$
\newauthor
Cynthia S.\ K.\ Ho,$^{1}$
Emilio Marfil,$^{6}$
Sergio Sousa,$^{7,8}$
Yolanda Frensch,$^{2}$
Dimitri Veras,$^{3,4,9}$
François Bouchy,$^{2}$
\newauthor
Yann Carteret,$^{2}$
Ian J.\ M.\ Crossfield,$^{10}$ 
Tyler Fairnington,$^{11,12}$
Mathilde Houelle,$^{2}$
Dan Huber,$^{13,14}$ 
\newauthor
Marziye Jafariyazani,$^{15,16}$
Léna Parc,$^{2}$
Don Radford,$^{17}$
TG Tan,$^{18}$
Sara Tavella,$^{2}$
Rob Wittenmyer,$^{10}$
\newauthor
Duncan Wright$^{10}$
$\And$ 
George Zhou$^{10}$
\\
$^{1}$Astronomy Unit, Queen Mary University of London, Mile End Road, London E1 4NS, UK\\
$^{2}$Observatoire Astronomique de l’Université de Genève, Chemin Pegasi 51, CH-1290 Versoix, Switzerland\\
$^{3}$Department of Physics, University of Warwick, Gibbet Hill Road, Coventry CV4 7AL, UK\\
$^{4}$Centre for Exoplanets and Habitability, University of Warwick, Gibbet Hill Road, Coventry CV4 7AL, UK\\
$^{5}$Leiden Observatory, Leiden University, P.O. Box 9513, 2300 RA
Leiden, The Netherlands\\
$^{6}$Departamento de Ingenier\'{i}a Topogr\'{a}fica y Cartograf\'{i}a, E.T.S.I. en Topograf\'{i}a, Geodesia y Cartograf\'{i}a, Universidad Polit\'{e}cnica de Madrid, 28031, Madrid, Spain\\
$^{7}$Instituto de Astrofisica e Ciencias do Espaco, Universidade do Porto, CAUP, Rua das Estrelas, 4150-762 Porto, Portugal\\
$^{8}$Departamento de Fisica e Astronomia, Faculdade de Ciencias, Universidade do Porto, Rua do Campo Alegre, 4169-007 Porto, Portugal\\
$^{9}$Centre for Space Domain Awareness, University of Warwick, Gibbet Hill Road, Coventry CV4 7AL, UK\\
$^{10}$Department of Physics and Astronomy, University of Kansas, Lawrence, KS, USA\\
$^{11}$Centre for Astrophysics, University of Southern Queensland, Toowoomba QLD, 4350, Australia\\
$^{12}$Department of Astronomy \& Astrophysics, University of Chicago, Chicago, IL, USA\\
$^{13}$Institute for Astronomy, University of Hawai‘i, Honolulu, HI 96822, USA\\
$^{14}$Sydney Institute for Astronomy, School of Physics, University of Sydney NSW 2006, Australia\\
$^{15}$NASA Ames Research Center, Moffett Field, CA 94035, USA\\
$^{16}$SETI Institute, Mountain View, CA 94043, USA \\
$^{17}$Brierfield observatory, Bowral, NSW, Australia\\
$^{18}$Perth Exoplanet Survey Telescope, Perth, Western Australia, Australia\\
}

\date{Accepted XXX. Received YYY; in original form ZZZ}

\pubyear{\the\year{}}

\begin{document}
\label{firstpage}
\pagerange{\pageref{firstpage}--\pageref{lastpage}}
\maketitle

\begin{abstract}
Studying the evolution of hot Jupiters requires a sample of well-characterised systems across all evolutionary states. We present three new gas giant exoplanets around stars approaching the end of the main sequence, a comparatively unexplored epoch of hot Jupiter evolution. These planets were discovered by \textit{TESS} before being vetted and confirmed through dedicated spectroscopic follow-up programmes by CARMENES, CORALIE and MINERVA-Australis. TOI-3664\,b has a period of 3.30 days, a radius of 1.22 $\pm$ 0.03 R$_\mathrm{Jup}$ and a mass of 0.36 $\pm$ 0.12 M$_\mathrm{Jup}$. TOI-4034\,b is a short-period hot Jupiter with a period of 1.80 days, a radius of 1.58 $\pm$ 0.02 R$_\mathrm{Jup}$ and a mass of 0.87 $\pm$ 0.16 M$_\mathrm{Jup}$. Meanwhile TOI-6564\,b has a period of 3.99 days, radius of 1.46 $\pm$ 0.02 R$_\mathrm{Jup}$ and mass of 0.70 $\pm$ 0.07 M$_\mathrm{Jup}$. All three planets have radii larger than Jupiter  but sub-Jupiter masses, in line with slight inflation as their hosts increase in luminosity towards the end of the main sequence. These exoplanets' low densities and hosts' advanced evolutionary states make them interesting planets with which to study the later stages of hot Jupiter evolution. Careful analysis was undertaken to determine the ages of each system, considering astrometry, gyrochronology, stellar isochrones and lithium abundance, yielding ages of 9.0$^{+2.4}_{-2.1}$ Gyr, 5.7$\pm 0.5$ Gyr and 4.0$\pm 1.0$ Gyr for TOI-3664, TOI-4034 and TOI-6564 respectively, yet each system has a similar evolutionary state because of their differing stellar masses (0.98 $\pm$ 0.03, 1.19$^{+0.13}_{-0.03}$ and 1.18$^{+0.16}_{-0.03}$ $M_*$). These three planets add more steps to the "age-ladder" of exoplanetary evolution, building towards the community's goal of understanding how planets evolve over time.
\end{abstract}







\section{Introduction}
\label{sec:intro}

The discovery of 51 Pegasi b by \citet{Mayor1995} fundamentally changed the community’s understanding of how planets form and evolve, revealing a gas-giant exoplanet orbiting its host with a period of only four days, unlike anything seen in the Solar System. This seminal exoplanet, which orbits seven times closer to its star than Mercury, became the first in an enigmatic group of exoplanets called ‘hot Jupiters’, a rare but crucial group of exoplanets with masses of at least 0.25 Jupiter masses and orbital periods of up to ten days \citep[following][]{Dawson2018}. Since this first discovery, over 650 hot Jupiters have been discovered,\footnote{n.b. the number of hot Jupiters was drawn from the exoplanet archive, https://exoplanetarchive.ipac.caltech.edu/index.html, accessed on 24 November 2025} yet their formation and evolution are still not well understood (see \citealt{Dawson2018} for a full review). 

Three main formation pathways have been proposed to form hot Jupiters: in-situ formation, disk migration and high eccentricity migration. In-situ formation theories posit that hot Jupiters form in-place, already possessing their present-day orbital periods, either via gravitational instability in the disk \citep[e.g.][]{Boss1997}, or via core accretion, where a proto-planet accretes significant amounts of gas from its environment \citep[e.g.][]{Perri1974,Pollack1996}. While the high temperatures required for gravitational instability for such close-in exoplanets makes that theory less plausible  \citep[e.g. see ][]{Rafikov2005}, core accretion may still be able to form these planets if sufficient solids can be built up very early in the planet formation \citep{Dawson2018}. 

Meanwhile, disk migration and high-eccentricity migration theories circumvent the challenges of in-situ formation by suggesting that hot Jupiters form further out in the disk and then migrate inwards. Under disk migration, torques acting upon the planet from the disk can slow the planet down, resulting in a decreased orbital separation, a.k.a inward migration \citep[e.g. see ][]{Goldreich1980,Lin1996,Ida2004}. Such migration would typically lead to planets with smaller eccentricities due to damping from the disk \citep{Papaloizou2000}. On the other hand, in high-eccentricity migration, the planet loses angular momentum through eccentricity excitation by another body (typically via planet-planet scattering \citep[e.g.][]{Rasio1996,Weidenschilling1996,Chatterjee2008} or secular interactions such as the Kozai-Lidov effect \citep{Kozai1962SecularEccentricity,Lidov1962TheBodies}, before losing orbital energy via tidal interactions with its host star \citep[e.g.][]{Dawson2018}. This high-eccentricity migration pathway should initially lead towards higher eccentricities until these are damped away by tides at close orbital separations \citep[e.g.][]{Jackson2008}.

In order to understand how these theories interact and determine the most common evolutionary pathways, it is imperative that the community builds a population of hot Jupiters across the entire extent of their evolution. While exoplanets spend the vast majority of their time orbiting stars in their main-sequence phase of evolution, this is book-ended by two dynamic periods of stellar (and indeed planetary) evolution: the pre-main sequence and the red giant branch. Young exoplanets around pre-main sequence stars (and those up to several hundred Myr) are particularly valuable as they probe eras of ongoing planet formation and the majority of the proposed disk/high-eccentricity migration, however true hot-Jupiter progenitors remain elusive despite several dedicated searches \citep[e.g.][]{Newton2019,Nardiello2019,Battley2020_YSD,BattleyYOUNGSTER,Mantovan2024,Vach2024}. On the other hand, exoplanets around stars at and beyond the terminal age main sequence offer an opportunity to explore the latter stages of exoplanet evolution, at a time where their host stars are dramatically increasing in luminosity and beginning to expand. Such changes can lead to reinflation of exoplanets \citep[e.g.][]{Grunblatt2016,Grunblatt2017,Wittenmyer2022,Pereira2024}, reduction in orbital separations and even planetary engulfment \citep[e.g.][]{Villaver2009,Schlaufman2013}. Although the discovery of `hot' ($P$< 10 days) planets around stars from the terminal age main sequence to red giant phase is challenged by the shorter time periods spent in these evolutionary regions and potential engulfment of shorter-period planets, a small number of such planets have been found, typically around sub-giant stars e.g. HD 185269 b \citep{Johnson2006}, WASP-136 b \citep{Lam2017}, HD 202772A b \citep{Wang2019}, NGTS-12 b \citep{Bryant2020}, TOI-954 b \citep{Sha2021}, TOI-1842 b \citep{Wittenmyer2022}, TOI-4377 b \& TOI-4551 b \citep{Pereira2024}. These exoplanets often appear to have inflated radii, low densities and may be imminently at risk of being engulfed by their host star \citep{Bryant2025}; indeed \citet{Bryant2020} calculated that NGTS-12 b may be engulfed by its host star within only 500 Myr. 

Two particularly crucial parameters for anchoring hot Jupiter evolution are stellar age and the equivalent evolutionary phase (EEP) of their hosts. In their construction of the MESA isochrones, \citet{Dotter2016} and \citet{Choi2016} define EEP as a series of evolutionary points which correspond to physically-motivated stages of stellar evolution shared by most, if not all, stellar types. For example the terminal age main sequence, TAMS, represents the point at which hydrogen is exhausted in the core of the star, and the tip of the Red Giant branch, RGBTip, represents the point where the stellar luminosity is at its maximum \citep{Dotter2016}. Hence, while age is a valuable parameter for considering the timespans available for processes such as migration and atmospheric evolution, EEP also considers the stellar mass, accounting for the fact that more massive, hotter stars evolve more quickly than their cooler counterparts. Using this proxy for stellar evolution thus allows for planetary systems to be compared more easily across different host star types.

This paper presents three new hot Jupiter exoplanets orbiting host stars approaching the very end of the main sequence, representing a sparsely populated but important region of hot Jupiter evolution. All three systems were discovered originally by TESS, and then were followed up with high-precision spectrographs (CARMENES, CORALIE and MINERVA-Australis) to determine their masses. 

The remainder of this paper is organised as follows: section \ref{sec:obs} discusses the photometric, spectroscopic and community observations of the three candidate systems and their stellar hosts, section \ref{sec:analysis} discusses the data analysis conducted to determine the stellar and planetary parameters, including a joint fit of each planetary system, and section \ref{sec:results} presents the results of the joint fits for these three new hot Jupiter exoplanets, before comparing the new exoplanets to the known population. The paper concludes in section \ref{sec:conclusion}.





\section{Observations}
\label{sec:obs}

\subsection{TESS photometry}
\label{sec:TESS_photom}

\begin{figure*}
\centering
\includegraphics[width=\hsize]{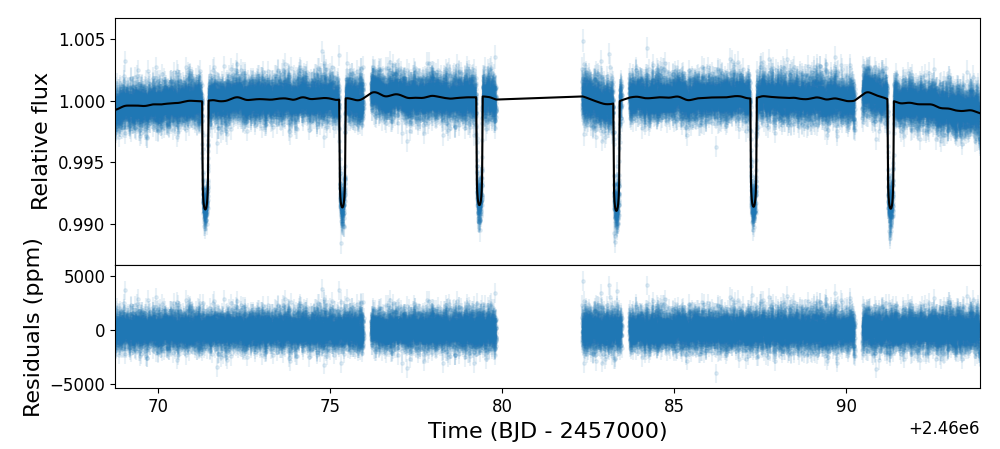}
  \caption{Full \textit{TESS} light-curve for TOI-6564, illustrating discovery of a repeated transit signal. \textit{Top}: original \textit{TESS} light-curve from Sector 65 (in blue), with joint model from section \ref{sec:joint_fit} over-plotted in black. \textit{Bottom}: Residuals to the joint fit. Note that the light-curves for TOI-3664 and TOI-4034 are similar in vein, with clear transits and minimal stellar variability, but are not instructive to reproduce here due to the large numbers of sectors of available data.}
     \label{fig:toi6564_original_lc}
\end{figure*}

All three planets considered in this work were originally detected by the \textit{Transiting Exoplanet Survey Satellite} \citep[\textit{TESS,}][]{Rickeretal.2014TheSatellite}. \textit{TESS} is a space telescope conducting a near all-sky transit survey by observing a different 24$^{\circ}$ x 96$^{\circ}$ `sector' of sky at approximately monthly intervals. The excellent photometric performance of the satellite and all-sky nature of TESS's search has resulted in  thousands of transiting exoplanet candidates,\footnote{7771 at time of writing; https://exoplanetarchive.ipac.caltech.edu/ - accessed 24 November 2025} or `TESS objects of interest' (TOIs), including TOI-3664\,b, TOI-4034\,b and TOI-6564\,b. 

Because of the evolving survey strategy of \textit{TESS}, each star on the sky receives a slightly different selection of observations. TOI-3664 (TIC 348437470) was observed by \textit{TESS} only once in its primary mission, with long-cadence (30min cadence) observations occurring in Sector 18. However, following its elevation to a TOI (see below), it was included in the \textit{TESS} short-cadence sample and observed in Sector 58 (2\,min cadence only) and Sectors 85/86 (20s and 2\,min cadence data). On the other hand, because of its northerly position close to the \textit{TESS} continuous viewing zone, TOI-4034 (TIC 375654303) was observed in four Sectors in \textit{TESS}'s Primary Mission (Sectors 17, 18, 24 and 25) with long cadence (30min data), and in nine sectors with short cadence (2\,min data) in the extended mission (Sectors 52, 57, 58, 59, 77, 78, 79, 84 \& 86). Finally, although TOI-6564 (TIC 453668803) fell on the \textit{TESS} detectors in three sectors, in the first two sectors (Sectors 11 and 38), the star fell right on the edge of the \textit{TESS} detectors, a region which is known to result in low-quality data dominated by systematics, where lightcurves are not routinely extracted. Hence light-curves for this target were generated for the first time in Sector 65, in year 3 of the \textit{TESS} mission (200s and 2\,min cadence data), meaning that only this sector is considered in this analysis. The full light-curve for TOI-6564 is shown in Figure \ref{fig:toi6564_original_lc} as an illustration of the transit discovery for all three targets.

Because of the brightness of its host star ($T_{\mathrm{mag}}$ = 9.57 mag), the candidate planet signal around TOI-6564 was first identified as part of the main search pipeline implemented by the \textit{TESS} Science Processing Operations Centre pipeline \citep[SPOC;][located at NASA Ames Research Center]{Jenkins2016}, which includes data calibration and extraction from the raw \textit{TESS} data, corrections for typical systematics through the Presearch Data Conditioning step, a dedicated transiting planet search and a suite of data validation steps  \citep[see][for full details]{Jenkins2016}. TOI-6564\,b passed all of the internal SPOC vetting tests and was alerted as a TOI on 20 July 2023.

In contrast, planetary candidate signals around the dimmer stars TOI-3664 and TOI-4034 were first identified as part of the \textit{TESS} Faint-star search \citep{Kunimoto2022}, an expanded search of the \textit{TESS} Primary Mission data  conducted using the Quick Look Pipeline \citep[QLP,][]{Huang2019, Huang2020a, Huang2020b,Kunimoto2021}. This search focused on fainter targets with TESS magnitudes between 10.5 $\leq T_{\mathrm{mag}} \leq$ 13.5 and featured light-curve extraction using an independent pipeline based on difference imaging \citep{Huang2019}, a search for transits using the Box-Least-Squares algorithm \citep{Kovacs2002} and a series of basic criteria to select the signals which passed the detection threshold (at least 5 points per transit, a signal to pink noise ratio of at least 9, and general signal to noise ratio of above 5 or 9 depending whether the \textit{TESS} magnitude was above or below 12). Signals that passed these tests were deemed `threshold crossing events' and passed through a series of additional vetting steps, including sine-wave identification, planet model template matching, depth and V-shape checks, centroid offset checks and human visual inspection of remaining candidates \citep{Kunimoto2021}. The candidate signals around TOI-3664 and TOI-4034 passed each of these checks and were both elevated to TOIs on 23 June 2021.


\subsection{CARMENES spectroscopy}\label{sec:CARMENES_obs}

The two northern hemisphere candidates, TOI-3664 \& TOI-4034, were followed up spectroscopically by the Calar Alto high-Resolution search for M dwarfs with Exoearths with Near-infrared and optical Echelle Spectrographs, \citep[CARMENES,][]{Quirrenbach2020}, which was chosen due to its high spectral resolution (R = 94,600 in VIS; 80,400 in NIR), northern latitude (+37:13:25) and multi-colour wavelength coverage (visible and infra-red wavelengths). CARMENES is installed on the 3.5m telescope at the Calar Alto Observatory in Almería (Spain) and consists of two separate spectrographs (deemed the VIS and NIR channels respectively) covering a wavelength range of 0.55 - 1.7 $\mu m$ and fed by fibers from the Cassegrain focus of the telescope. Observation time on CARMENES was secured as part of the Opticon Radionet Pilot (ORP) Trans-National Access programme (Grant Reference Number 101004719) in semester 2024A; PI M. Battley, CAHA programme number 051.
The observations were taken with one of the fibers on target and the other one fed by a Fabry-Pérot etalon for simultaneous calibration. The observations were reduced with the default CARMENES data reduction pipeline \texttt{caracal} \citep{Caballero2016}.

Because the majority of candidates alerted by \textit{TESS} typically have minimal spectroscopic observations, the radial velocity (RV) observations with CARMENES were split into two stages: an initial vetting stage to check for obvious false positives such as spectroscopic or eclipsing binaries, and an extended monitoring stage once it was determined that TOI-3664 and TOI-4034 were reliable targets.

\subsubsection{Vetting}

The initial vetting step was carried out by observing one stellar spectrum at each of the expected maximum and minimum radial velocity epochs, or as close to phases 0.25 and 0.75 in the orbit as possible, according to the photometric ephemeris from the TOI alert. Vetting observations carried out between expected maximum (phase 0.25) and minumum (phase 0.75) revealed offsets of only $\sim$ 100-150 m/s for both stars, and occured in phase with the planetary ephemeris, consistent with true planetary signals caused by orbiting bodies with sub-Jupiter masses. This meant that both TOI-3664 and TOI-4034 passed the vetting stage of the observations and continued to the longer monitoring stage described below.

Note that six additional targets were vetted as part of this CARMENES programme, but none were found to be suitable for further monitoring due to a range of false-positive scenarios and difficult observing constraints. Briefly, TOI-5459 and TOI-6382 were highlighted as spectroscopic binaries, TOI-6248 was found to be a likely eclipsing binary, TOI-2538 was found to have untenably wide spectral lines and the observations of TOI-3571 were found to be in anti-phase with the transit ephemeris, suggestive either of an eclipsing binary or complex stellar activity. Meanwhile due to its longer period, only one RV measurement was possible for TOI-4688, so conclusions on this object are limited. Further discussion of this wider vetting is presented in Appendix \ref{sec:CARM_vetting_append}.

\subsubsection{Monitoring}
\label{sec:CARM_monitoring}

Extended monitoring of TOI-3664 was carried out between 12 February 2024 and 5 March 2024 for a total of 20 observations, while observations of TOI-4034 occurred between 12 May 2024 and 29 June 2024 for a total of 22 observations. Observations were carried out in pairs, consisting of two 22.5 minute exposures taken in series with the VIS and NIR channels simultaneously to allow for either two independent measurements at similar phases or an averaged 45min exposure for higher signal to noise measurements. The raw data was analysed using both the \texttt{raccoon} \citep{Lafarga2020} and \texttt{serval} \citep{SERVAL2018} pipelines for comparison, however the final chosen radial velocities were those extracted by the \texttt{serval} pipeline, aligning with the bulk of other CARMENES discoveries. \texttt{serval} is a template matching pipeline that initially derives an approximate RV time series by performing a least squares fit between the spectra with the highest signal-to-noise ratio (S/N) and the rest of observations. These initial RVs are used to create a high S/N template by co-adding the RV-corrected observed spectra, which is subsequently used to derive new RVs in an iterative process (one iteration is usually sufficient). The template co-adding is performed by cubic B-spline regression. \texttt{serval} also accounts for the barycentric movement of the Earth and any instrumental drift corrections. The RVs are extracted order by order, and a final RV is computed as the weighted mean of the order RVs.

The full radial velocity results are plotted in Figures \ref{fig:rv_obs_toi-3664} and \ref{fig:rv_obs_toi-4034} for TOI-3664 and TOI-4034 respectively, and tabulated in Appendix \ref{sec:appendix_rvs}.

\begin{figure}
\centering
\includegraphics[width=\hsize]{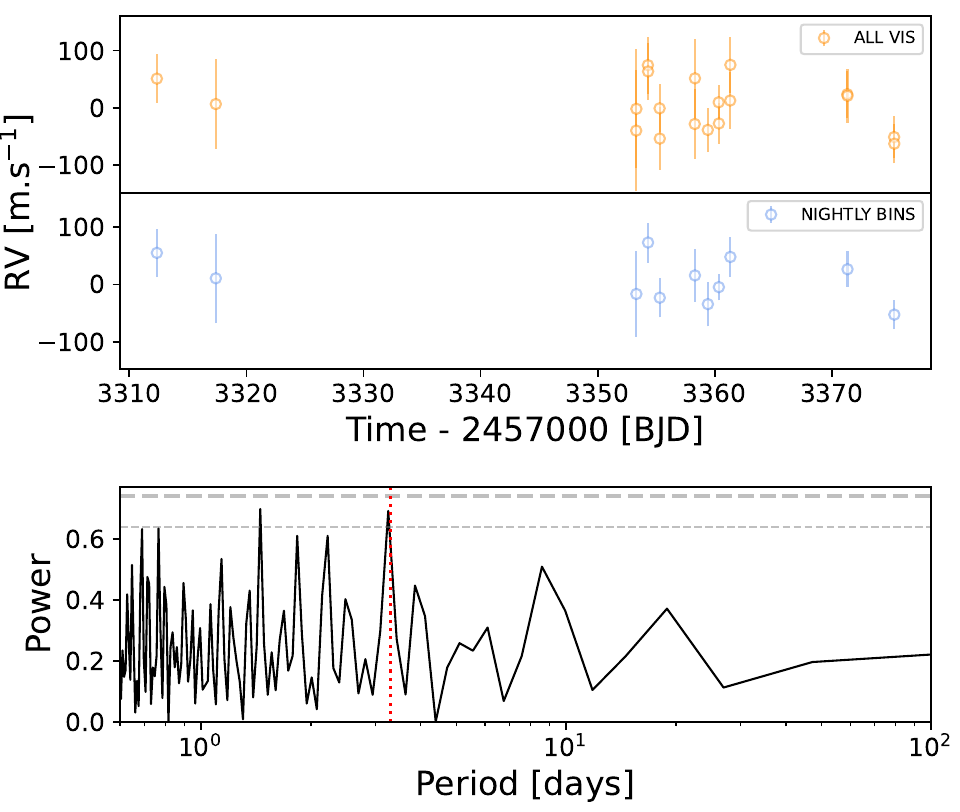}
  \caption{Overview of collected RV data for TOI-3664. \textit{Top:} All VIS RV measurements are shown in orange, while the same observations binned nightly are shown in blue. All points are shown as empty circles, with their 1$\sigma$ errors shown as straight lines. The mean stellar RV offset has been removed. \textit{Bottom:} Lomb-Scargle periodogram for all collected RV observations of TOI-3664. The 1\% and 10\% false alarm probability (FAP) values are shown by the dotted grey lines. The highest peak at 3.3 days corresponds to the planetary period seen from the photometric data (illustrated by the red dotted line), while the additional peak which passes the 10\% FAP value is an alias which disappears after fitting the main RV signal.}
     \label{fig:rv_obs_toi-3664}
\end{figure}

\begin{figure}
\centering
\includegraphics[width=\hsize]{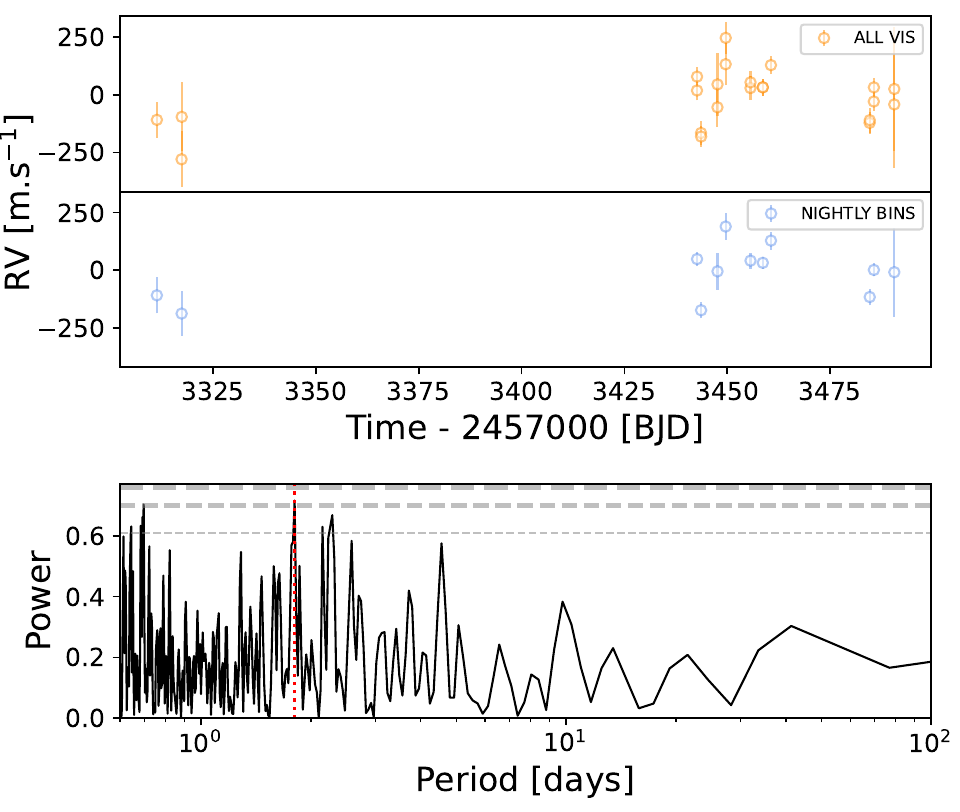}
  \caption{Overview of collected RV data for TOI-4034. Format is the same as Figure \ref{fig:rv_obs_toi-3664} except that the 0.1\% FAP value is also shown on the periodogram due to the slightly stronger periodic signals present. The strongest peak at 1.8 days aligns with the photometric period, shown as the red dotted line}.
     \label{fig:rv_obs_toi-4034}
\end{figure}

\subsection{CORALIE spectroscopy}
\label{sec:COR_spectra}

TOI-6564, the single southern-hemisphere target, was added to an existing programme on the CORALIE spectrograph \citep{Queloz2000} to follow up planet candidates from \textit{TESS}. CORALIE is a echelle spectrograph installed on the Swiss 1.2 metre Leonhard Euler Telescope at ESO's La Silla Observatory and is fed by two fibers to allow for simultaneous calibration with a Fabry-Pérot etalon. It covers a wavelength range of 0.39 - 0.68 $\mu m$ and has a spectral resolution of $\sim$60,000. Spectra were extracted from the detector using the standard CORALIE calibration reduction pipeline, before radial velocity measurements were determined through cross-correlation with a binary G2 mask \citep{Pepe2002TheVII}. Similar to the CARMENES observations, several vetting spectra were taken at opposite phases initially to search for obvious false positive signals, before an extended monitoring campaign was undertaken to determine the mass of the candidate. In total, TOI-6564 was observed 22 times between 25 April 2024 and 3 March 2025. Note that a slight $\sim$10 m/s offset was introduced to the data between the 2024 and 2025 observing seasons due to a new calibration lamp being installed, so this is accounted for in the analysis by treating the pre- and post-2024 data as two separate datasets.

Full radial velocity data for TOI-6564 is plotted in Figure \ref{fig:rv_obs_toi-6564} and tabulated in Appendix \ref{sec:appendix_rvs}.

\subsection{MINERVA-Australis spectroscopy}

Independently to the CORALIE efforts, TOI-6564 was also followed up by MINERVA-Australis Observatory. MINERVA-Australis is an array of four identical 0.7 m telescopes located at Mount Kent Observatory in Queensland, Australia. The four telescopes are linked via fibre feeds to a single KiwiSpec echelle spectrograph at a spectral resolving power of $R\sim$80,000 over the wavelength region of 5000-6300\AA\,\citep{Addison2019_Minerva}. The array is wholly dedicated to radial-velocity follow-up of TESS planet candidates \citep[e.g.][]{Nielsen2019,Addison2021,Wittenmyer2022}. Two simultaneous fibres provide wavelength calibration and correct for instrumental variations. The calibration fibres are illuminated by a quartz lamp through an iodine cell, eliminating contamination by Ar lines. Each epoch consists of 30 to 60 minute exposures from up to four individual telescopes. The radial velocities from each telescope are treated as coming from separate instruments to account for small velocity offsets between the fibres.

MINERVA-Australis observed TOI-6564 between 11-25 August 2023 for a total of seven observations. Two different radial velocity values were extracted from each observation; low precision least-squares RVS and higher precision CCFRVs. Low precision RVs were derived from a least-squares deconvolution of the observed spectrum against a synthetic non-rotating template (similar to the CHIRON pipeline; e.g. see \citet{Jones2019,Wang2019}; $\sim$20 m/s noise floor per telescope). CCFRVs are higher precision radial velocities derived from a cross correlation against an averaged spectrum of the target ($\sim$5 m/s noise floor per telescope), but are only produced for low $v\sin{i}$ stars (<10km/s) that have received more than five  observations. In this case the $v\sin{i}$ of TOI-6564 was found to be <1.0 km/s from the CORALIE spectra, so the higher-precision CCFRVs were extracted and used in this analysis.

The full MINERVA-Australis data are plotted alongside CORALIE data in Figure \ref{fig:rv_obs_toi-6564} and tabulated in Appendix B.

\begin{figure}
\centering
\includegraphics[width=\hsize]{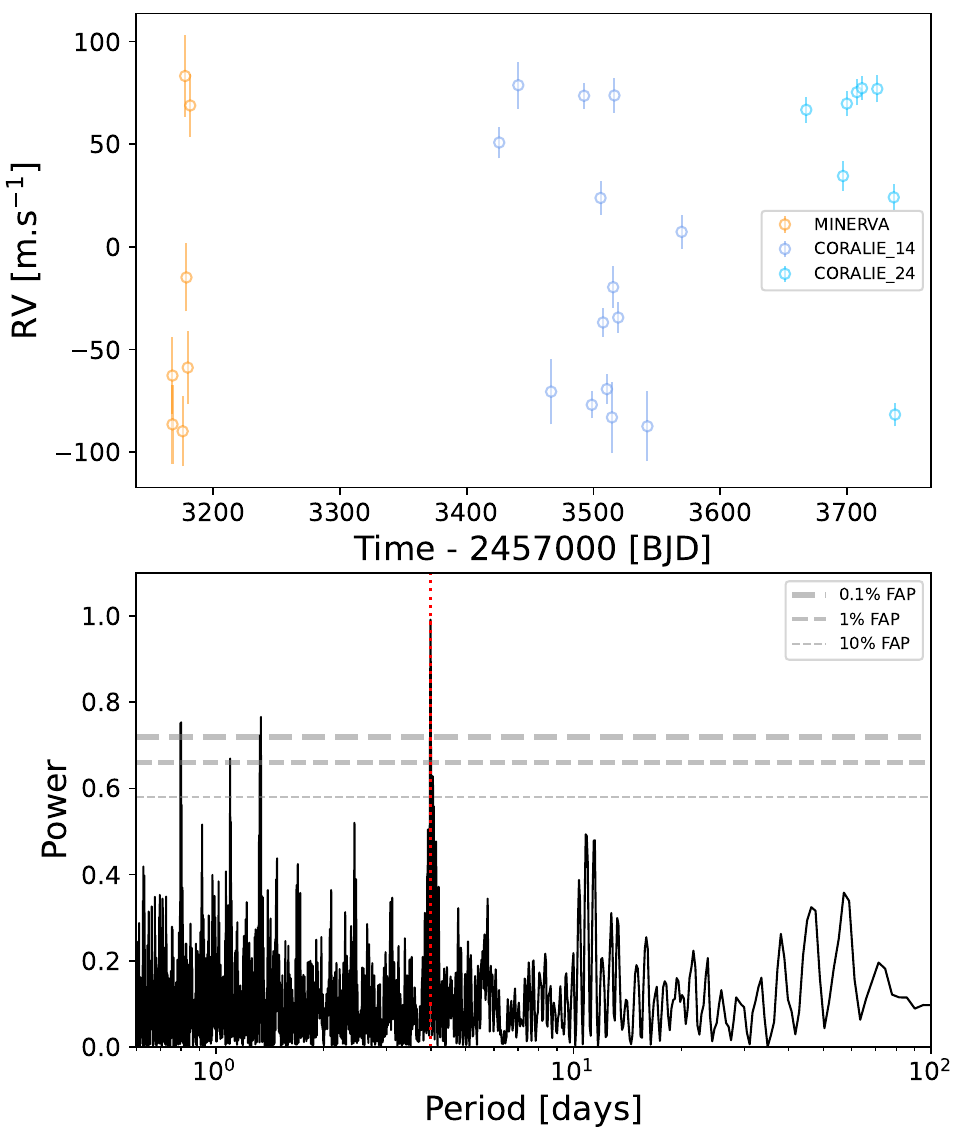}
  \caption{Overview of collected RV data for TOI-6564. \textit{Top:} All collected RV observations for TOI-6564. MINERVA-Australis data is plotted in orange, before the two different seasons of CORALIE data are plotted in purple and blue respectively. All points are shown as empty circles, with their 1$\sigma$ errors shown as straight lines. Inter-instrument offsets have been removed. \textit{Bottom:} Lomb-Scargle periodogram for the collected RV observations. The 10\%, 1\% and 0.1\% false alarm probabilities (FAP) are shown by the dotted grey lines. The highest peak corresponds to a period of 3.99 days, agreeing with the transit period shown as the red dotted line.}
     \label{fig:rv_obs_toi-6564}
\end{figure}

\subsection{Community observations}

Following elevation to TOIs, all planet candidates were uploaded to the Exoplanet Follow-up Observing Program website (ExoFOP),\footnote{https://exofop.ipac.caltech.edu/tess/} where community members are able to upload independent follow-up observations. A variety of additional follow-up observations were carried out for all three targets for vetting purposes, so are discussed further here, split up by observation type.

\subsubsection{Additional spectroscopy}

Both TOI-3664 and TOI-4034 received one spectroscopic vetting observation with the High Resolution Echelle Spectrometer \citep[HIRES][]{Vogt1994_HIRES}, installed on the Keck 10m telescope in Hawaii. HIRES is a grating cross-dispersed echelle spectrometer with an overall wavelength range of between 0.3 - 1.0 $\mu$m and a resolution of 25,000 to 85,000 depending on the slit plate used. The vetting observations were carried out with a spectral resolution of $R$ = 50,000 and a wavelength range of 0.36-0.90 $\mu$m. The obtained spectra from these observations revealed single-star systems with no obvious false positives, alike to the CARMENES spectra, verifying the stars' suitability for further monitoring. 

\subsubsection{Seeing-limited photometry}

Both TOI-4034 and TOI-6564 were additionally followed up by members of the TFOP SG1 Sub-Group for seeing-limited photometry in order to confirm that the transit signals were on target and to check for chromaticity suggestive of an eclipsing binary.

TOI-4034 was followed up in visible \& infra-red (V \& I-bands; Nikita Chazov on MASTER-Ural (0.4m)/ApogeeAltaU16m), red (R-band; Tomas Popajewski on a Newtonian reflector (0.25 m)/ ASI294MM PRO) and blue (B-band; Adam Popowicz on SUTO-UZPW50-0m5 (0.5 m)/ MORAVIAN G4-900) filters. These observations together showed that the signal was on time, on-target, uncontaminated by  other stars and without significant chromaticity, validating that the signal observed in \textit{TESS} was commensurate with an on-target transit signal from a body of planetary size. However, because of the relatively dim magnitude of TOI-4034 (Vmag = 13.3), these additional photometric observations were of much lower signal to noise than the \textit{TESS} observations (especially in the latter sectors of \textit{TESS} observations), so are not included in the joint analysis below (section \ref{sec:joint_fit}). Furthermore, because many sectors of \textit{TESS} data are available on either side of these follow-up observations, their inclusion would not appreciably improve the transit ephemerides.

TOI-6564 was followed up by Brierfield Observatory and the Perth Exoplanet Survey Telescope. Brierfield Observatory is located near Bowral, N.S.W. Australia. The 0.36m Planewave CDK14 telescope is equipped with a 4096 x 4096 Moravian 16 803 camera. The image scale is 0.735 arcsec per pixel, dramatically better than the 21 arcsec pixels of TESS, resulting in a 50 arcmin x 50 arcmin field of view. The collected photometric data included a single full transit of TOI-6564\,b observed using a Johnson Blue filter, split into 110 x 120-second exposures. Photometry was extracted on  UTC 2024 May 23 using the AstroImageJ (AIJ) software package \citep{Collins2017}, utilising a circular photometric aperture with a 5.1 arcsec radius. Note that no additional detrending parameters were used. The observed transit, plotted in Figure \ref{fig:toi6564_transit_fold}, was on time based on the \textit{TESS} ephemeris, and of comparable depth (9.5 ppt) to the \textit{TESS} transits, despite its bluer filter.

The Perth Exoplanet Survey Telescope (PEST) is located near Perth, Australia. The 0.3 m telescope is equipped with a $5544\times3694$ QHY183M camera.  Images are binned 2x2 in software giving an image scale of 0$\farcs$7 pixel$^{-1}$ resulting in a $32\arcmin\times21\arcmin$ field of view. A custom pipeline based on {\tt C-Munipack}\footnote{http://c-munipack.sourceforge.net} was used to calibrate the images and extract the differential photometry. This observation, also plotted in Figure \ref{fig:toi6564_transit_fold}, yielded an on-time transit in the red (rp) filter with a depth of 7.75 $\pm$ 0.32 ppt.

As well as proving that the transit signal was on target and achromatic, these two extra transits helped appreciably to improve the transit ephemeris because only one sector of \textit{TESS} data was available for TOI-6564, and hence they were included in the joint fit. 

\subsubsection{SAI Speckle Polarimeter imaging}

\begin{figure}
\centering
\includegraphics[width=\hsize]{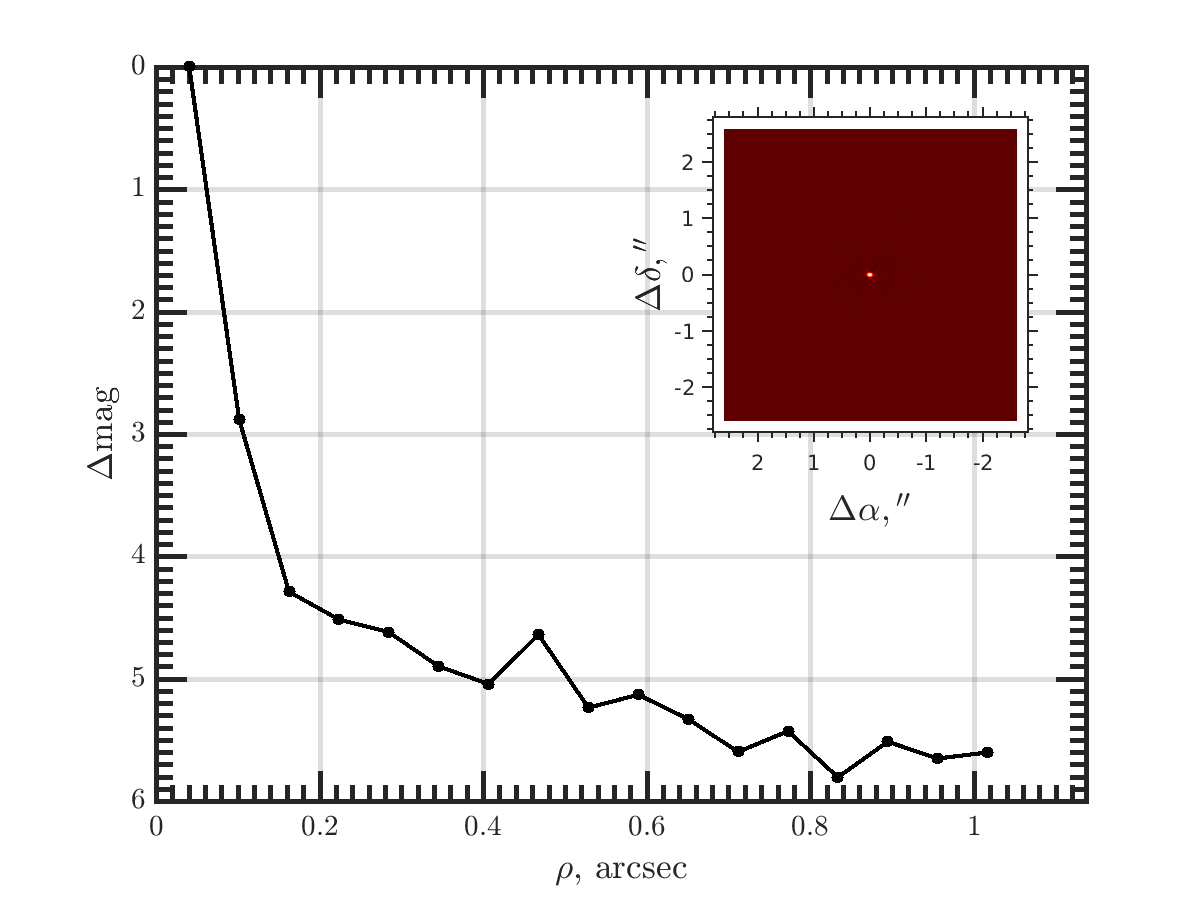}
  \caption{SAI speckle-imaging observations of TOI-4034. No additional sources were observed within 2$\arcsec$ of the target.}
     \label{fig:SAI_obs}
\end{figure}

High-angular resolution speckle imaging is a valuable tool to check for nearby companions to stars which may contaminate \textit{TESS} photometry, especially given the wide $\sim$21" pixel-scale of \textit{TESS}. Nearby stellar sources can create two troublesome signals in \textit{TESS}: background eclipsing binaries may mimic true planetary signals, while nearby non-eclipsing stars can result in excess flux in the \textit{TESS} aperture, leading to underestimated planetary radii. A search for stellar companions near TOI-4034 was conducted on the night of 27 October 2021 UT by using the Speckle Polarimeter installed on the 2.5m telescope at the Caucasian Observatory of Sternberg Astronomical Institute (SAI) \citep{Safonov2017}. The Cousins I-band, a bandpass very similar to that of the \textit{TESS} camera, was used for the observations resulting in the image shown in Figure \ref{fig:SAI_obs}. This observation was sensitive to signals of a 5.6-magnitude fainter star at 1", but revealed no such stellar companions nearby to TOI-4034.

\subsubsection{Gaia Photometry}

TOI-3664 and TOI-4034 were also included in a study by \citet{Panahi2022} who utilised the high angular resolution of the \textit{Gaia} satellite \citep{GaiaCollaboration2016GaiaProperties,GaiaCollaboration2018GaiaProperties} -- which allow it to resolve stars with at least $\sim$0.7 arcsecond separation -- and collected \textit{Gaia} epoch photometry to circumvent \textit{TESS}'s reasonably wide point spread function (PSF) to evaluate whether a large number of known TOI signals were on-target or caused by background eclipsing binaries. In order to do this, \citet{Panahi2022} compared photometric data taken by the \textit{Gaia} mission for all available stars within the \textit{TESS} PSF to the alerted TOI signal, and matched the transit depths, durations and ephemerides to the most likely \textit{Gaia} sources. In the ideal case where the TOI-hosting star is the brightest star in the PSF and the same transit signal is observable for the star in the \textit{Gaia} photometry, the background eclipsing binary hypothesis can be ruled out and the TOI signal can be confirmed as `on-target.' This was the case for both TOI-3664 and TOI-4034, which were confirmed to be on-target in Phase II of the ongoing analysis programme.

\section{Data analysis}
\label{sec:analysis}

\subsection{Stellar parameter determination}
\label{sec:stellar_params}


Up to date stellar parameters for TOI-3664, TOI-4034 and TOI-6564 were determined through a combination of spectroscopic analysis of the collected CARMENES/CORALIE/HIRES spectra and fitting the Spectral Energy Distribution of each star.

\begin{figure*}
\centering
\includegraphics[width=17cm]{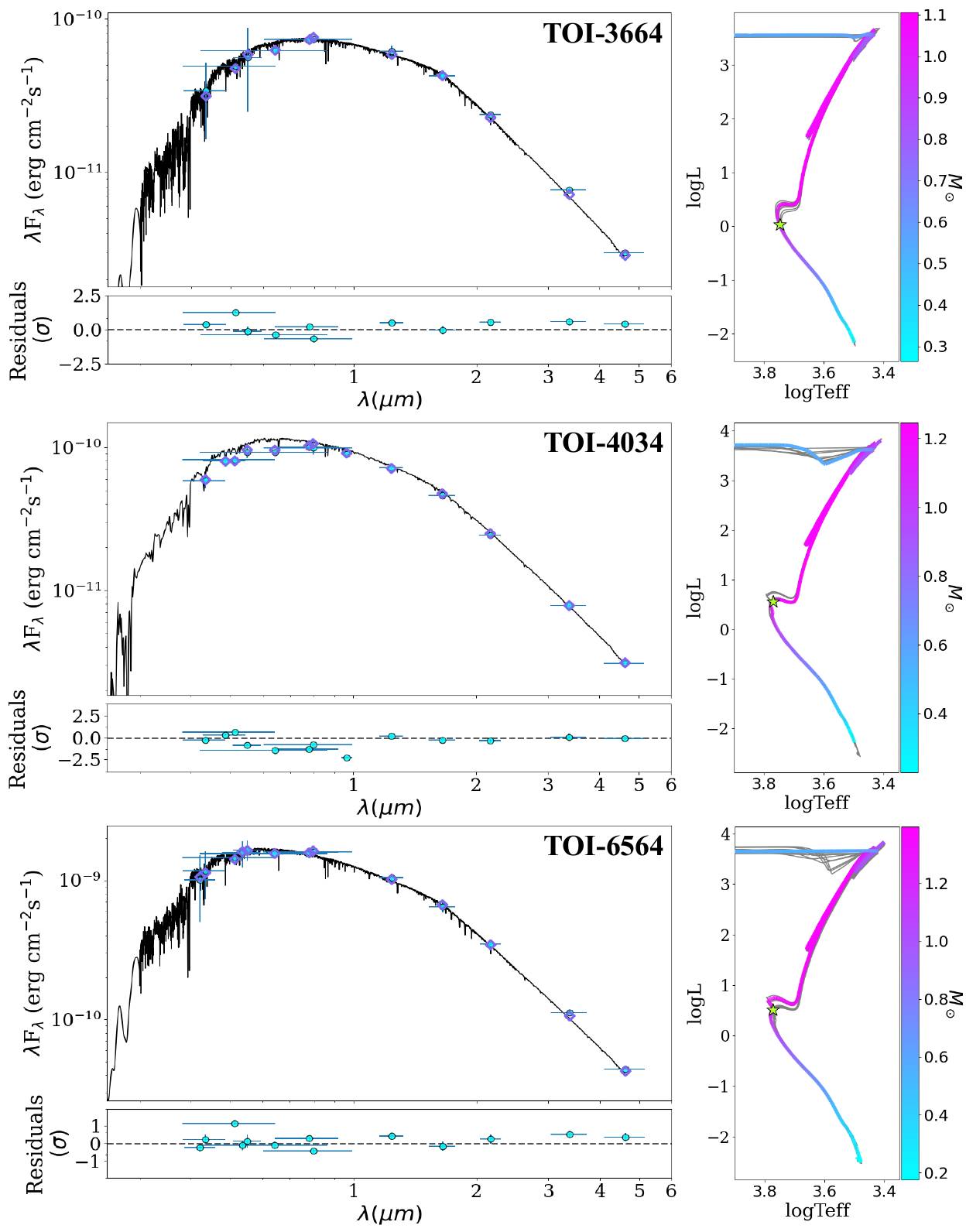}
  \caption{Characterisation of the host stars. Top to bottom: TOI-3664, TOI-4034 and TOI-6564. \textit{Left}: Spectral Energy Distribution fit results for each star. Photometric measurements are depicted as blue diamonds, with the best-fit stellar spectrum shown in black. \textit{Right}: Hertzsprung-Russell ($T_{\rm eff}$ vs luminosity) diagram for each star based on isochronal fitting. The best-fitting isochrone is shown in colour (with specific colours representing stellar mass), with ten additional isochronal samples shown in gray, based on best-fit stellar parameters drawn randomly from the posterior distributions of the ARIADNE stellar model. Each star is shown in green, appearing near the Terminal Main Sequence.}
     \label{fig:stellar_charact}
\end{figure*}

\begin{table*}
\centering                  
\caption{Stellar names, properties, and photometric magnitudes for all host stars} 
\label{stellar_table}      
\begin{tabular}{l c c c c}        
\hline
Parameter  &  & Value & & Source \\    
\hline                        
Names\\
TOI & \textbf{TOI-3664} & \textbf{TOI-4034} & \textbf{TOI-6564} & TESS \\
TIC ID & 348437470 & 375654303 & 453668803 & TESS \\
2MASS ID & J02211405+5942425 & J22203526+7146024 & J13431631-5609229 & 2MASS \\
\textit{Gaia} DR3 ID & 459319848632063360 & 2230550759343923456 & 6064308867810722560 & \textit{Gaia} DR3 \\ \\

Astrometric properties \\
RA  & 02:21:14.04   &  22:20:35.28 &  13:43:16.29 &  \textit{Gaia} DR3  \\
Dec & +59:42:42.68  & +71:46:02.45 & -56:09:23.22 &  \textit{Gaia} DR3 \\ 
pmRA [mas/yr] & -7.5543 ± 0.0125 & -0.1520 ± 0.0124 & 	-12.1358 ± 0.0084  &    \textit{Gaia} DR3\\
pmDec [mas/yr] & 8.6675 ± 0.0155 & 3.3590 ± 0.0115 & 	-15.0171 ± 0.0145 & \textit{Gaia} DR3\\
Parallax [mas] & 2.1483 ± 0.0153 & 	1.3056 ± 0.0102 &  	5.1705 ± 0.0142& \textit{Gaia} DR3\\
Distance [pc] & 465$^{+6}_{-4}$ & 760$^{+8}_{-6}$ & 193 ± 1 & Sect. \ref{sec:stellar_params}\\ \\

Magnitudes\\
TESS [mag] & 12.91 ± 0.0061 & 12.5798 ± 0.0159 & 9.5687 ± 0.006 & TESS\\
B [mag] & 14.754 ± 0.048 & 14.168 & 10.946 ± 0.072 & Tycho\\
V [mag] & 13.883 ± 0.114 & 13.262 ± 0.092 & 10.236 ± 0.005 & Tycho\\
G [mag] & 13.5461 ± 0.0003 & 13.1082 ± 0.000365 & 10.0315 ± 0.000245 & \textit{Gaia} DR3\\
J [mag] & 11.995 ± 0.022 & 11.825 ± 0.026 & 8.918 ± 0.019 & 2MASS\\
H [mag] & 11.618 ± 0.028 & 11.523 ± 0.031 & 8.656 ± 0.046 & 2MASS\\
Ks [mag] & 11.48 ± 0.021 & 11.459 ± 0.022 & 8.558 ± 0.021 & 2MASS\\
WISE1 [mag] & 11.401 ± 0.024 & 11.38 ± 0.023 & 8.483 ± 0.023 & WISE\\
WISE2 [mag] & 11.439 ± 0.02  & 11.408 ± 0.021 & 8.523 ± 0.02 & WISE\\
WISE3 [mag] & 11.366 ± 0.161 & 11.183 ± 0.086 & 8.515 ± 0.02 & WISE\\
WISE4 [mag] & 8.911 & 9.499 ± 0.444 & 8.809 ± 0.269 & WISE\\ \\

Bulk properties\\ 
$T_{\mathrm{eff}}$ [K]   & 5593 ± 32   & 5891 ± 40   & 5943 ± 64   & Sect. \ref{sec:stellar_params} \\
$[\mathrm{Fe/H}]$ [dex]               & 0.17 ± 0.02 & 0.12 ± 0.02 & 0.19 ± 0.04 & Sect. \ref{sec:stellar_params} \\
log g [cm/s$^{2}$]      & 4.37 ± 0.06 & 3.87 ± 0.08 & 4.04 ± 0.11 & Sect. \ref{sec:stellar_params} \\
$v \sin i$ [km/s] & 2.4 $\pm$ 1.0 & 7.9 $\pm$ 1.0  & $<$ 1.0   & Sect. \ref{sec:stellar_params} \\
M$_\star$ [M$_\odot$]    & 0.98 ± 0.03 & 1.19$^{+0.13}_{-0.03}$ & 1.18$^{+0.16}_{-0.03}$ & Sect. \ref{sec:stellar_params} \\
R$_\star$ [R$_\odot$]    & 1.11 ± 0.02 & 1.83 ± 0.02 & 1.70 ± 0.02 &  Sect. \ref{sec:stellar_params}\\
$\rho_*$ (g/cm$^3$) & 1.014 $\pm$ 0.046 & 0.277 $\pm$ 0.026 & 0.395 $\pm$ 0.023 & Sect. \ref{sec:joint_fit} \\
Age [Gyr]   & 9.0$^{+2.4}_{-2.1}$   & 5.7 $\pm$ 0.5 & 4.0 $\pm$ 1.0 & Sect. \ref{sec:age}\\
EEP [-]                  & 407$^{+13}_{-16}$  & 448$^{+6}_{-52}$  & 445$^{+8}_{-58}$  & Sect. \ref{sec:stellar_params}\\
\hline                                   
\\
\end{tabular}

\textbf{References:} TESS \citep{Stassun2019}; 2MASS \citep{Skrutskie2006_2MASS}; \textit{Gaia} DR3 \citep{GaiaDR3_2023}; Tycho \citep{Hog2000_Tycho2}; WISE \citep{Wright2010_WISE}.
\end{table*}

As a sanity check of the values provided in the TESS Input Catalogue \citep[TICv8.2,][]{Stassun2018,Stassun2019}, initial stellar parameters for each star were determined by fitting their Spectral Energy Distributionss (SEDs) using the \texttt{ARIADNE} code \citep{Vines2022_ARIADNE}. \texttt{ARIADNE} allows for fitting six different stellar models independently using the nested sampler \texttt{dynesty} \citep{Skilling2004_dynesty1,Skilling2006_dynesty2,Speagle2019}, which can then be combined via Bayesian Model Averaging over the entire sample of models to derive the final set of parameters for each star. The models fit by \texttt{ARIADNE} are Phoenix v2 \citep{Husser2013}, BT-Settl, BT-Cond \citep{Allard2012}, BT-NextGen \citep{Allard2012,Hauschildt1999}, \citet{Kurucz1993} and \citet{Castelli2004}. In order to find the magnitudes to fit for each star, \texttt{ARIADNE} searches through a variety of literature sources for broad-band photometry, including the Two Micron All-sky Survey catalogue \citep[2MASS,][]{Skrutskie2006_2MASS}, Gaia Data Release 2 \citep[Gaia DR2,][]{GaiaCollaboration2016GaiaProperties,GaiaCollaboration2018GaiaProperties}, Tycho-2 catalogue \citep{Hog2000_Tycho2}, ALL-WISE \citep{Wright2010_WISE}, the All-Sky Compiled Catalog (ASCC) of \citet{Kharchenko2001_ASCC}, Pan-STARRS1 survey \citep{Chambers2016}, Spitzer/GLIMPSE survey \citep{Churchwell2009}, GALEX catalogue \citep{Bianchi2011_GALEX}, APASS photometric survey \citep{Henden2014_APASS}, Sloan Digital Sky Survey \citep[SDSS DR12,][]{2015ApJS..219...12A}, Strömgren-Crawford catalogue \citep{Paunzen2015} and the \textit{TESS} Input Catalog \citep[TICv8,][]{Stassun2018,Stassun2019}.

Additional magnitudes known from literature but missed by the default \texttt{ARIADNE} search were then added by hand: Johnson B and V bands from TICv8 for TOI-3664 and TOI-4034, 2MASS\_J, 2MASS\_H and 2MASS\_Ks bands from 2MASS \citep{Skrutskie2006_2MASS} for all stars, and additionally the WISE RSR\_W3 and RSR\_W4 bands from the Wide-Field Infrared Survey Explorer \citep{Wright2010_WISE} in the cases of TOI-3664 and TOI-6564. Note that in the case of TOI-4034 the Pan-STARRS r, i and z bands were removed due to very large errors. 

For the purposes of the initial stellar parameter check, the default priors in \texttt{ARIADNE} were retained, before running the \texttt{dynesty} nested sampling using 500 live points, 4 threads and a dlogz of 0.5, implementing a random-walk sampling method. This initial SED fit revealed stellar parameters broadly consistent with those provided in the TESS Input Catalogue and Gaia DR3 \citep{GaiaDR3_2023}. 

The collected spectroscopic data gave the opportunity to refine these parameters further, by deriving precise stellar parameters for all stars from analysis of the spectra themselves.

In the case of TOI-3664 and TOI-4034, stellar parameters ($T_{\rm eff}$, $\log{g}$ and [Fe/H]) were obtained from the CARMENES VIS spectra by means of {\sc SteParSyn}, a Bayesian implementation of the spectral synthesis method \citep{Tabernero22}. Beginning with the raw CARMENES spectra, all CARMENES orders of the template FITS files were merged into a single spectrum for each star and corrected by the radial velocity shift of the targets with iSpec \citep{Blanco-Cuaresma14}. Since both stars are F- to G-type, {\sc SteParSyn} was run using the MARCS synthetic grid and the set of Fe~{\sc i} and Fe~{\sc ii} lines available in \citet{Tabernero22}. The final results are displayed in Table \ref{stellar_table} and agree well with those derived from SED fitting. 

In the case of TOI-6564 however, spectroscopic stellar parameters were determined from the CORALIE data using the ARES+MOOG spectroscopic analysis as described in \citet{Santos2013,Sousa2014,Sousa2021}. We used the latest version of ARES\footnote{The latest version, ARES v2, can be downloaded at \url{https://github.com/sousasag/ARES}} \citep{Sousa2007, Sousa2015} to consistently measure the equivalent widths (EW) for the iron line list presented in \citet{Sousa2008}. The best spectroscopic parameters are found using the ionization and excitation equilibrium. In this process, it is used for a grid of Kurucz model atmospheres \citep{Kurucz1993} and the radiative transfer code MOOG \citep{Sneden1973}. The results from this analysis are shown in Table \ref{stellar_table}, and once again agree well with the SED-based parameters.

As the initial $v \sin i$ values measured from the CARMENES and MINERVA-Australis data were found to be overestimated due to the somewhat low signal to noise spectra and wing-broadening of the stellar lines due to macroturbulence \citep[e.g. see][]{Doyle2014}, we recomputed $v \sin i$ values using higher signal to noise spectra. For TOI-3664 and TOI-4034, $v \sin i$ was extracted from the Keck/HIRES spectra using the SpecMatch-Synth code \citep{Petigura2015}, the standard pipeline for extracting stellar parameters from HIRES data, which characterises the star by fitting synthetic spectra to an observed spectrum. Meanwhile, a new $v \sin i$ measurement for TOI-6564 was determined from the Full-Width Half Maximum of the CORALIE Cross-Correlation Function, using the relation introduced by \citet{Santos2002}. This method is particularly valuable for $v \sin i$ determination because the width of the CCF at half its maximum is largely unaffected by macroturbulence. The extracted $v \sin i$ values are presented in Table \ref{stellar_table}.

In order to determine self-consistent stellar radii, masses, ages and Equivalent Evolutionary Phases (EEPs) for each star, a second ARIADNE run was completed for each star using the spectroscopically derived stellar parameters, which also included fitting the stars to their most likely isochrones using the \texttt{isochrones} code \citep{Dotter2016,Choi2016}. Priors were placed on $T_\mathrm{eff}$, logg and [Fe/H] based on the outputs derived above from the spectra, but priors on distance, radius and extinction, $\mathrm{A}_v$, were left as the default priors in \texttt{ARIADNE}, which sets the priors to the Gaia DR3 stellar radius, Gaia DR3 distance from the corrected \citet{Bailer-Jones2023} catalogue and the maximum line-of-sight extinction allowed by the SFD galactic dust map \citep{Schlegel1998,Schlafly2011}. As before, the \texttt{dynesty} nested sampling was run using 500 live points, 4 threads and a dlogz of 0.5, implementing a random-walk sampling method. The combination of models fitted by \texttt{ARIADNE} led to results shown in Table \ref{stellar_table} (and illustrated in Figure \ref{fig:stellar_charact}) and are consistent with, but more precise than, those within the TESS Input Catalogue.

Note however that isochrone fitting for both mass and age was complicated by the position of all three stars falling near the terminal main sequence, a region which is very sensitive to changes in stellar parameters. This was particularly challenging for TOI-4034 and TOI-6564, shown in Figure \ref{fig:stellar_charact}, which fell in the corner of the main-sequence/sub-giant branch and hence are highly susceptible to small changes in effective temperature and mass. Indeed, the initial ARIADNE results for TOI-4034 and TOI-6564 showed a dichotomy in age depending on the true stellar mass, largely due to the stars' isochronal positions being consistent with both younger and older ages depending whether they were evolving \textit{on} or \textit{off} the main sequence or before/after the main sequence turn-off. The true age of the system is thus crucially important to the system parameters and will be discussed in detail in the next section. 

\subsection{Age determination}
\label{sec:age}

Determining the age of these systems through isochrone fitting and/or gyrochronology alone is complicated by their position near the terminal main sequence and the lack of clear stellar activity periods in the collected \textit{TESS} data. Hence, additional stellar ageing methods are discussed here to determine the most reliable ages for each system.

\subsubsection{Kinematic ages}

One of the most precise methods of ageing stars is placing stars in stellar groups such as clusters and stellar associations. Both TOI-3664 and TOI-4034 were highlighted in \citet{Dias2014} as high likelihood members of young clusters (TOI-3664: 67\% member of ASCC 8, age $\sim$50Myr; TOI-4034: 85\% member of Collinder 471, age $\sim$7 Myr), which originally motivated their follow-up with CARMENES. Similarly, TOI-6564 was included in a \citet{Damiani2019} sample of high-likelihood members of the $\sim$5-20 Myr Scorpius Centaurus association (expected 10-30\% field-star contamination), based on a combination of spatial, kinematic and colour-magnitude cuts. However, these signs of young cluster membership contrasted with the lack of significant stellar activity signals seen in the collected photometric and spectroscopic data (see sections \ref{sec:phot_analysis} and \ref{sec:spec_analysis}), motivating a more in-depth look into the ages of these systems.

\begin{figure}
\centering
\includegraphics[width=\hsize]{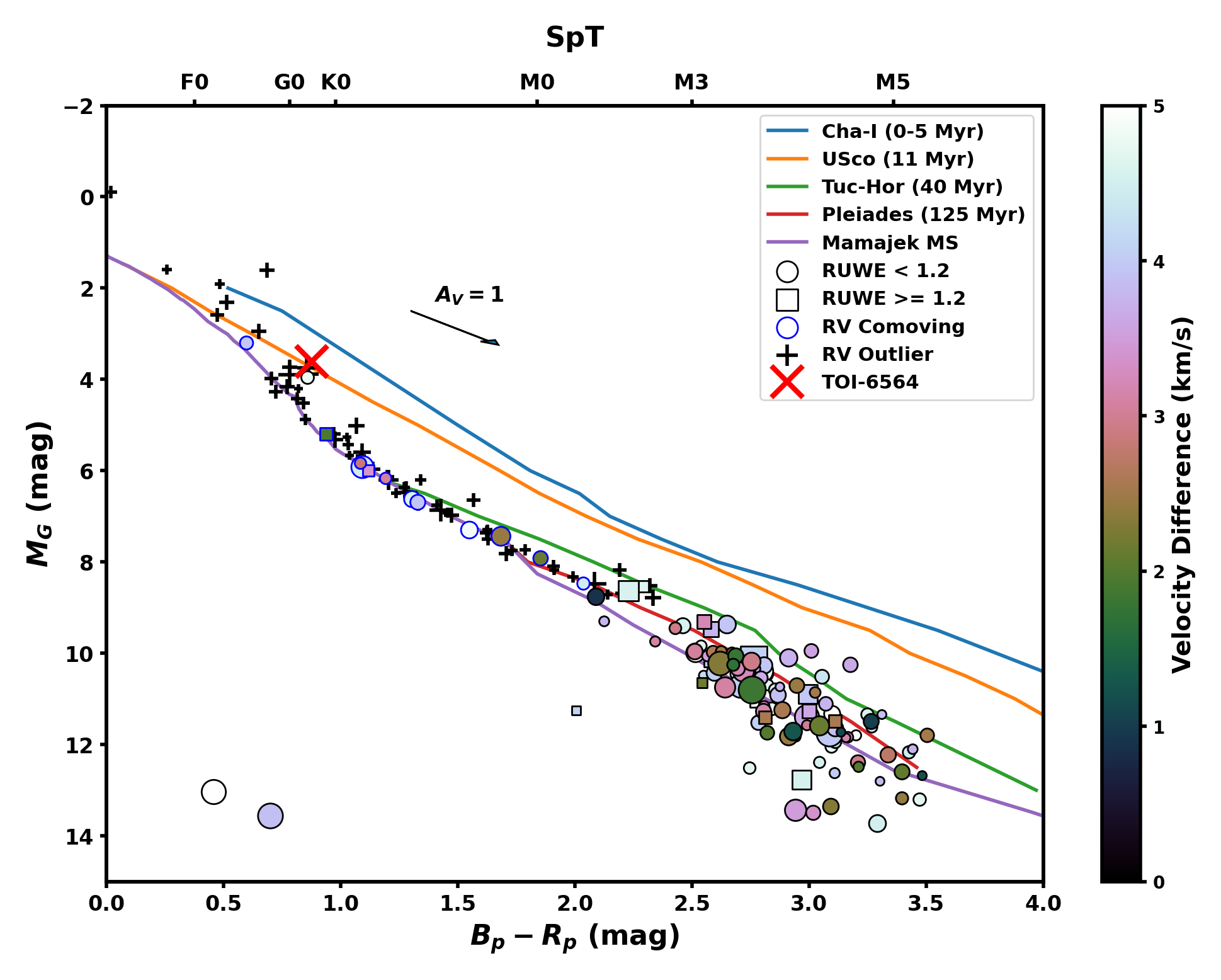}
  \caption{Colour-magnitude diagram for comoving stars to TOI-6564 generated by \texttt{comove}. TOI-6564 is shown as the red cross, with circles, squares and plus symbols representing the comoving stars. Note that the blue circles/crosses denote comoving stars with radial velocities consistent with TOI-6564, while the crosses show those with consistent proper motions but radial velocities outside the 5km/s limits. Isochrones for some well-known clusters/associations (Chamaeleon I, Upper-Scorpius, Tucana Horologium and the Pleiades) are shown as the blue, orange, green and red lines respectively, with the Main-Sequence from \citet{Pecaut2013} plotted in purple.}
     \label{fig:toi6564_CMD}
\end{figure}

The significantly improved astrometry delivered by the Gaia satellite \citep{GaiaCollaboration2016GaiaProperties,GaiaCollaboration2018GaiaProperties} offered the opportunity to revaluate cluster membership. Most notably, in a new study with Gaia DR2 data, \citet{Cantat-Gaudin_asterism} concluded that Collinder 471 was not a real cluster, and instead an asterism, debunking this claim of youth for TOI-4034. In order to examine whether the three host stars were likely associated with other co-moving groups of stars based on the more precise Gaia DR3 data, all three targets were analysed using the \texttt{comove} package,\footnote{https://github.com/adamkraus/Comove} a modified version of the FindFriends routine introduced in \citet{Tofflemire2021} which searches for comoving neighbours to a target star and returns useful information regarding their potential youth. In order to find the closest comoving stars to each target, \texttt{comove} was run using velocity limits of $<$5 km/s and distance limits of 30pc from the target. Although each star was in a crowded region of the sky, TOI-3664 and TOI-4034 were both found to have relatively few comoving companions within the 5km/s and 30pc limits (16 and 45 members from proper motion alone; 0 and 1 members including RV signals as well - although note that the number of targets with RV measurements in Gaia DR3 was limited by the magnitude of the stars), suggesting that they may be older background interlopers instead of true cluster members. This is supported by the relatively large distances found for these stars in section \ref{sec:stellar_params}: TOI-3664: 407$^{+13}_{-16}$ pc; TOI-4034: 760$^{+8}_{-6}$pc. On the other hand, TOI-6564 was found to have 208 comoving stars within the 5km/s and 30pc limits, of which at least 12 were also found to have consistent radial velocities (a number that is again limited by the available Gaia DR3 radial velocity measurements). A colour-magnitude diagram for these comoving stars is shown in Figure \ref{fig:toi6564_CMD}, compared to isochrones for some well-know young clusters. Although the placement of TOI-6564 on the plot gives some evidence as to why it might have been included in an Upper-Scorpius sample, as it was for \citet{Damiani2019}, the wider population of comoving stars on the colour-magnitude diagram appears consistent with the main sequence, suggesting that TOI-6564 is evolving \textit{off} the main sequence instead of on to it. It is also worth noting that the distance found for TOI-6564 of 193 $\pm$ 1 pc also places it behind the main Upper-Sco sample, which is know to be situated at $\sim$130 pc \citep[e.g.][]{He22_USco}.

\subsubsection{Gyrochronology}

Gyrochronology is a commonly used method to age field stars, based on the gradual spin down of stars as they evolve along the main sequence \citep[e.g. see][]{Kawaler1989,Barnes2003,Bouma2023}. However, gyrochronology is challenging for these three stars due to the difficulty in determining rotation periods for each target (as discussed in section \ref{sec:phot_analysis}), with a lack of activity in the \textit{TESS} photometry for these G-type stars suggestive of rotation periods $>$10-20 days. Similarly, no additional significant periods at <10 days are present in the radial velocity measurements discussed in section \ref{sec:spec_analysis}. The $v \sin i$ values obtained from the spectra for  TOI-3664 and TOI-6564 also suggest longer stellar rotation periods, with a measured $v \sin i$ value for TOI-3664 of 2.4 $\pm$ 1.0 km/s only slightly larger than the resolution-induced lower limit of 2.2 km/s for the HIRES spectrograph, and a $v \sin i$ measurement of <1.0 km/s from the CORALIE data of TOI-6564 also suggestive of a slowly-rotating star. This supports an older age hypothesis for both TOI-3664 and TOI-6564, however cannot provide very reliable age estimates due to the dependence on stellar inclination.

The situation for TOI-4034 is slightly more complex, with a measured $v \sin i$ value of 7.9 $\pm$ 1.0 km/s suggestive of a maximum rotation period of 11.7 days. Such a rotation period for a star with an effective temperature of 5891 $\pm$ 40 K would suggest an age on the order of 1 Gyr according to relations from \citet{Bouma2023}, contrasting with the older ages discussed in the next two sections. However, inspection of other similar hot Jupiter exoplanetary systems ($M_p$ > 0.25 $M_{\mathrm{Jup}}$) with very short periods (P $<$ 2 days) around similar G-type stars ($5300 < T_{\mathrm{eff}} < 6000$ K) reveals several other `old' systems with inflated $v \sin i$ values, for example  HAT-P-23 b \citep[][age = 4.0 $\pm$ 2.0 Gyr; $v \sin i$ = 8.1 km/s]{Bakos2011} and Kepler-41 b \citep[][age = 4.4 $\pm$ 1.3 Gyr; $v \sin i$ = 6 $\pm$ 2 km/s]{Santerne2011,Bonomo2015}. This trend is reinforced by the work of \citet{Tejada2021} who found that hot Jupiter host stars on average spin faster than similar G-type stars which host small planets and/or giant planets on wider orbits (e.g. see Figure 5 of their paper). They propose that this ‘spin-up’ is caused by tidal torques transferring angular momentum from the closely orbiting hot Jupiter to the host star, supporting earlier works showing tidal spin-up for hot Jupiter hosts \citep[e.g.][]{Brown2014,Poppenhaeger2014,Maxted2015}. This suggests that gyrochonology is unreliable for stars with very closely orbiting hot Jupiter exoplanets, and hence does not place any useful constraints on the age of the TOI-4034 system.

\subsubsection{Isochronal ages}

\begin{figure}
\centering
\includegraphics[width=\hsize]{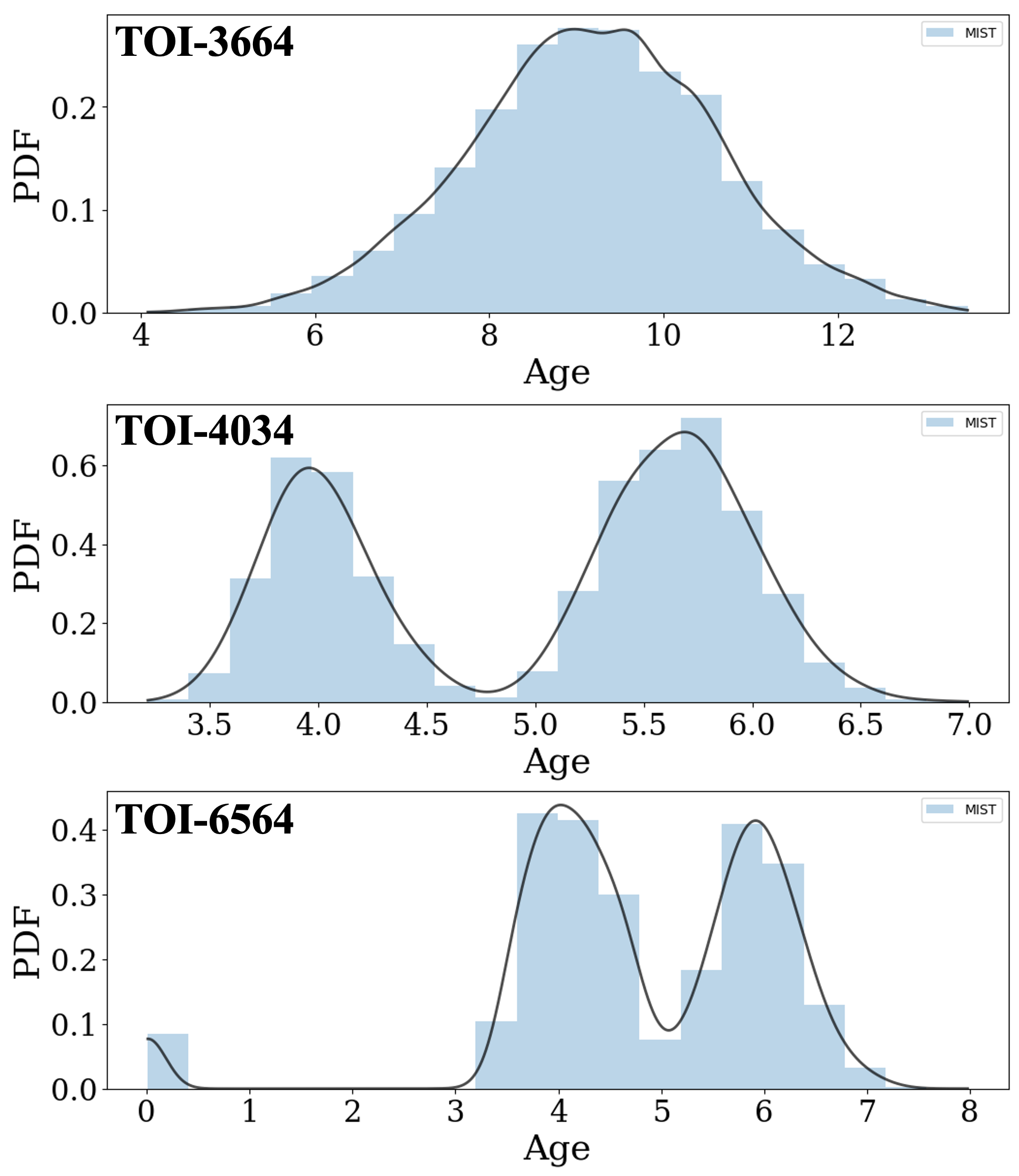}
  \caption{Age posteriors from the \texttt{ARIADNE/isochrones} stellar fits. Note that age is presented in Gyr. Both TOI-4034 and TOI-6564 show a clear dichotomy between two likely ages.}
     \label{fig:ariadne_ages}
\end{figure}

\begin{figure}
\centering
\includegraphics[width=\hsize]{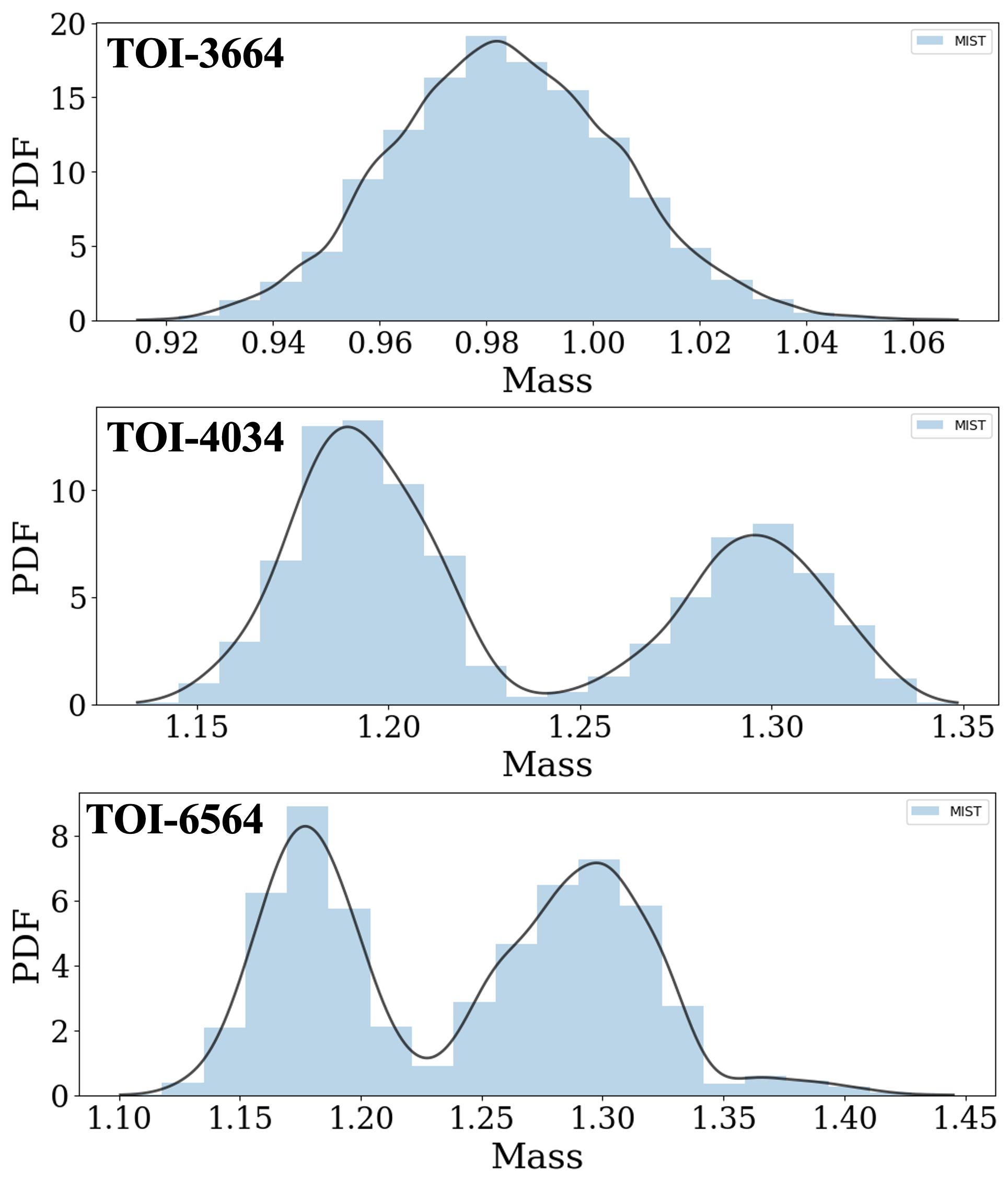}
  \caption{Stellar mass posteriors from the \texttt{ARIADNE/isochrones} stellar fits. Similar to the ages, both TOI-4034 and TOI-6564 show a clear dichotomy between two likely masses. Note that the mass and age are inversely related, so the older age peak is associated with the lower mass peak in each case.}
     \label{fig:stellar_mass_comparison}
\end{figure}

Age estimates were also obtained via isochrone fitting as part of the ARIADNE stellar fits discussed in section \ref{sec:stellar_params}. These fits, which considered the extracted stellar properties for each star, resulted in ages of 9.0$^{+2.4}_{-2.1}$, 5.7$^{+0.5}_{-2.1}$ and 4.0$^{+2.5}_{-0.6}$ Gyr for TOI-3664, TOI-4034 and TOI-6564 respectively, with the posterior distributions shown in Figure \ref{fig:ariadne_ages}. While the posterior for TOI-3664 exhibits a normal (if wide) distribution, it is informative to examine the posterior distributions for TOI-4034 and TOI-6564 more carefully, as each star fell in a very dynamic part of the Hertzsprung-Russel diagram, near the corner of the Terminal Age Main Sequence (TAMS) - see Figure \ref{fig:stellar_charact}. Because of this location and the relationship between age and stellar mass, two distinct peaks are seen in both the age and mass posteriors for TOI-4034 and TOI-6564 (Figures \ref{fig:ariadne_ages} and \ref{fig:stellar_mass_comparison} respectively), complicating a true age determination from isochrones alone. Nonetheless, the evolutionary state of each system is clearer for all three stars, with Equivalent Evolutionary Phase (EEP) values of 407$^{+13}_{-16}$,  448$^{+6}_{-52}$  \& 445$^{+8}_{-58}$ respectively confirming that all systems are approaching the Terminal Age Main-Sequence. Note that for the MIST isochrones with which this isochrone fitting was carried out, an EEP number of 202 corresponds with the Zero-Age Main-Sequence (ZAMS), an EEP value of 353 denotes the Intermediate Age Main-Sequence (IAMS) and a value of 454 represents the TAMS.

In an attempt to overcome the dichotomy in mass/age for TOI-4034 and TOI-6564, an independent determination of the stellar mass was estimated by combining the output stellar density from the joint fit carried out in section \ref{sec:joint_fit} (largely driven by the shape of the transit) with the radius determined from the ARIADNE fit to the spectral energy distribution in section \ref{sec:stellar_params}. This resulted in masses of 1.20 $\pm$ 0.12 $M_\odot$ for TOI-4034 and 1.38 $\pm$ 0.09 for TOI-6564, favouring the lower half of the mass posterior for TOI-4034 and the upper half for TOI-6564. These masses corresponded with the older age of 5.70 $\pm$ 0.5 Gyr  for TOI-4034 and the slightly younger age of 4.0 $\pm$ 1.0 Gyr for TOI-6564, giving further weight to the determined averages of the ARIADNE fits. 


\subsubsection{Lithium ages}

Another useful age indicator is photospheric lithium depletion, as photospheric lithium typically decreases with stellar age. While in less massive stars (T$_\mathrm{{eff}} < 4500 K$) lithium is often considered a youth indicator (indeed \citet{Jeffries2023} show that lithium is usually depleted within 1 Gyr in stars cooler than 4500 K), in hotter stars, modest amounts of lithium may be present until much older ages, with some stars showing lithium signatures with equivalent width of up to 100 m\AA\ beyond 5 Gyr \citep[see e.g.][]{Magrini2021,Sun2025}.

\citet{Jeffries2023} have developed the \texttt{EAGLES} code\footnote{https://github.com/robdjeff/eagles} which allows the user to derive age estimates and posterior age probability distributions based on measurements of the equivalent width of the lithium doublet around 6708\AA. Both TOI-3664 and TOI-6564 display clear lithium lines in their spectra, from which the lithium equivalent width (Li$_{EW}$) was determined by fitting a sum of three inverted Gaussian distributions to the spectra, centred on each of the lines in the lithium doublet and the nearby Fe I line which can become blended at lower spectrograph resolution or higher stellar activity levels. The results of this fitting revealed lithium equivalent width measurements of Li$_{EW}$ = 31.3 $\pm$ 5.6 m\AA\, for TOI-3664 and Li$_{EW}$ = 37.0 $\pm$ 1.5 m\AA\, for TOI-6564. Fitting these alongside the stellar parameters with the \texttt{EAGLES} code yielded the following age estimates: TOI-3664: 2.97$^{+4.15}_{-1.85}$ Gyr; TOI-6564: 3.04$^{+4.00}_{-2.23}$ Gyr, consistent with those from the isochronal fitting above. Note that the relatively large errors are driven by the fact that lithium abundances become much more disperse in older stars, leading to a wider range of plausible ages \citep[see e.g.][]{Jeffries2023}. 


On the other hand, TOI-4034 did not display any notable Lithium lines around 6708 m\AA, so it could not be included in the lithium analysis. However, the lack of lithium provides a lower limit on the age of the system, suggesting that the target was sufficiently old that the bulk of its original lithium had already been consumed (e.g. likely $\gg$ 1 Gyr). The decreased lithium abundance compared to TOI-6564, which has a similar stellar mass, also provides some evidence that TOI-4034 is older than TOI-6564, aligning with the isochronal ages.  

\subsubsection{Aggregate ages}

To summarise, although all three systems were initially suspected to be young due to inclusion in catalogues of young clusters/associations, the combination of better kinematic data, gyrochronology, isochronal fitting and lithium abundance reveal that these systems are actually much older, with the host stars approaching the terminal age main sequence. In the case of TOI-3664, a small $v \sin i$ value of 2.4 km/s and lithium age of  2.97$^{+4.15}_{-1.85}$ Gyr supports the old isochronal age of 9.0$^{+2.4}_{-2.1}$ Gyr, and so we adopt this value for the system. Meanwhile, although the gyrochronological age of such a close-in system as TOI-4034 appears unreliable due to tidal effects, the lack of observable lithium in the spectrum and isochronal age of 5.7$^{+0.5}_{-2.1}$ Gyr supports an early G-type star approaching the very end of the main sequence. Finally, the isochronal age of 4.0$^{+2.5}_{-0.6}$ Gyr for TOI-6564 is supported by a lithium age of 3.04$^{+4.00}_{-2.23}$ Gyr, and a low $v \sin i$ value of 1.0km/s commensurate with an old, inactive star. One final complication is that both TOI-4034 and TOI-6564 display a mass/age degeneracy due to their position near the corner of the Terminal Age Main Sequence, but this degeneracy was solved by independently determining the stellar mass from the joint fit stellar density and the SED-driven radius, setting a narrower age of 5.7 $\pm$ 0.5 Gyr for TOI-4034 and 4.0 $\pm$ 1.0 Gyr for TOI-6564.

The most likely ages for these three systems are thus: TOI-3664: 9.0$^{+2.4}_{-2.1}$ Gyr, TOI-4034: 5.7 $\pm$ 0.5 Gyr and TOI-6564: 4.0 $\pm$ 1.0 Gyr, representing a wide range of ages despite their similar evolutionary phases, driven by their different host star masses.

The discrepancy between the young literature ages and old ages found for all of these targets highlights the importance of careful age vetting of clusters and individual systems during the follow-up process of exoplanet candidates. Given the key importance of stellar/planetary ages for anchoring exoplanetary evolution theories, is recommended that the community uses multiple ageing methods wherever possible to determine the true age of exoplanetary systems.

\subsection{Photometric analysis}
\label{sec:phot_analysis}

Wherever possible, the light-curves analysed in this work were drawn from the standard \textit{TESS} SPOC extraction pipeline \citep{Jenkins2016}, using the PDC\_SAP flux \citep{Stumpe2012,Stumpe2014,Smith2012}, however where these were not available (e.g. during the Primary Mission) they were drawn from the QLP pipeline instead \citep{Huang2019}. After downloading from the Mikulski Archive for Space Telescopes (MAST),\footnote{https://mast.stsci.edu/portal/Mashup/Clients/Mast/Portal.html} individual sectors of data were cleaned using a custom-built cleaning pipeline to ensure that similar cleaning steps were applied to all sectors of data regardless of which data extraction pipeline had been used (e.g. SPOC 2\,min or QLP 30min). Briefly, this pipeline removes poor-quality data based on the quality flags automatically generated by the SPOC and QLP pipeline, data-points close to spacecraft momentum dumps, epochs of excess scattered light and data at times where spacecraft pointing was compromised. Three additional epochs of particularly noisy data caused by scattered light from the Earth and Moon were removed in sectors 18 and 86, between 1791.37 to 1815.03 TBJD, 3637.90 to 3641.60 TBJD and 3651.46 to 3655.20 TBJD respectively. Note that in the case of TOI-3664, the excessive photometric scatter in Sector 86 was found to wash out all other signals and hence this sector was removed from this analysis. Following this cleaning step, the individual sectors of data were combined into a single light-curve for each target using the LightCurve object included within the \texttt{lightkurve} package\footnote{https://lightkurve.github.io/lightkurve/index.html} \citep{Lightkurve2018}. 

To make use of the increased data quantity available since the initial detection of the planet candidates, an updated planet-search through the light-curves for each system was undertaken, both with and without masking the known planet. This had the dual benefit of providing updated transit ephemerides for inputting into the joint fit (see Section \ref{sec:joint_fit}), and allowing for signals due to potential additional planets in the system to be searched for. Planet searches were conducted using all sectors of available data for all three targets by first detrending the data using the custom Young Star Detrending (YSD) code described in \citet{Battley2020_YSD}, before searching for planets using the Box Least Squares  algorithm \citep{Kovacs2002} implemented as the \texttt{BoxLeastSquares} function in \texttt{astropy} \citep{astropy:2013,astropy:2018,astropy:2022}. In all cases, no additional candidate signals were found in this search.


In addition to the transit search, a Lomb-Scargle \citep{Lomb1976Least-squaresData, Scargle1982StudiesData} search for sinusoidal periodicities was undertaken in the light-curves for all three targets to search for the most likely rotation period of each star. For the rotation-specific analysis, the QLP data was analysed because of its better performance at separating stellar signals in crowded regions \citep{Huang2020a,Huang2020b}. Note that unlike in the main transit search and joint analysis, for the rotation period search the QLP Simple Aperture Photometry (SAP) flux was chosen to best preserve any stellar activity signals, as these are largely removed by the detrending applied to the Kepler-Spline SAP (KSPSAP) data. Two alternative period searches were carried out; one using all available light-curves and a second using only the longest continuous section of data (typically 1-2 sectors). Unfortunately, in all cases the \textit{TESS} data was found to be dominated by systematic signals rather than reliable rotation signals (indeed the primary rotation period was found to vary on a sector by sector basis), suggesting that these stars are reasonably inactive and have stellar rotation periods longer than half a \textit{TESS} sector.

As \textit{TESS} is known to produce significantly less reliable stellar activity periods beyond periods of $\sim$12 days \citep{Boyle2025_TESS_rotation}, a similar lomb-scargle search was carried out on data from the ASAS-SN survey \citep{ASAS-SN1,ASAS-SN2}, however no significant rotation period peaks (i.e. those over the 10\% false alarm probability level) were seen in the ASAS-SN periodograms.

\begin{table*}
\caption{Planet parameters for all systems from juliet: median and 68\% confidence interval}             
\label{planetary_table}      
\centering                          
\resizebox{\textwidth}{!}{%
\begin{tabular}{l l l c c c}        
\hline               
Parameter  & & Prior distribution$^*$ & TOI-3664\,b & TOI-4034\,b & TOI-6564\,b \\    
\hline                        
$P$ \dotfill & Period (days) \dotfill & $N$($P_{photom}$,0.1) \dotfill  & 3.2974839 $\pm$ 0.0000041 & 1.8020932 $\pm$ 0.0000012 & 3.985413 $\pm$ 0.000011 \\
$T_0$ \dotfill & Time of transit center (BJD$_{\mathrm{TDB}}$) \dotfill & $N$($t_{0,photom}$,0.1) \dotfill  &  2460661.83115 $\pm$ 0.00079 & 2460661.21859 $\pm$ 0.00047 & 2460071.37027 $\pm$ 0.00019 \\
$b$ \dotfill & Impact parameter of the orbit \dotfill & $U$(0,1) \dotfill & 0.29 $\pm$ 0.19 & 0.620 $\pm$ 0.027 & 0.386 $\pm$ 0.048 \\
$p = R_p/R_*$ \dotfill & Planet to star radius ratio \dotfill & $U$(0,1) \dotfill & 0.1132 $\pm$ 0.0018 & 0.08875 $\pm$ 0.00061 & 0.08843 $\pm$ 0.00052 \\
$\sqrt{e}sin(\omega)$ \dotfill & Parametrisation for e and $\omega$ \dotfill & $U$(-1,1) \dotfill & 0.28$^{+0.09}_{-0.17}$ & N.A. & N.A. \\
$\sqrt{e}cos(\omega)$ \dotfill & Parametrisation for e and $\omega$ \dotfill & $U$(-1,1) \dotfill & -0.01$^{+0.30}_{-0.33}$ & N.A. & N.A. \\
$K$ \dotfill & Radial velocity semi-amplitude (m/s) \dotfill & $U$(0,200) \dotfill & 51$^{+16}_{-15}$ & 128$^{+19}_{-21}$ & 80.3$^{+2.0}_{-2.1}$\\
$e$ \dotfill & Orbital eccentricity \dotfill & N.A. or \textit{fixed}\dotfill & 0.159 $\pm$ 0.074 & 0 & 0\\
$\omega$ \dotfill & Argument of periastron (deg) \dotfill & N.A. or \textit{fixed}\dotfill & 93$^{+53}_{-50}$ & 90 & 90\\
$i$ \dotfill & Inclination (deg) \dotfill & \dotfill & 87.8 $\pm$ 1.3 & 80.76 $\pm$ 0.38  & 86.77 $\pm$ 0.39\\
$a$ \dotfill & Semi-major axis (AU) \dotfill & \dotfill & 0.0431 $\pm$ 0.0010 & 0.03289 $\pm$ 0.00071  & 0.0542 $\pm$ 0.0012 \\
$R_p$ \dotfill & Planetary radius ($R_{\mathrm{Jup}}$) \dotfill & \dotfill & 1.222 $\pm$ 0.030 & 1.580 $\pm$ 0.020 & 1.463 $\pm$ 0.019 \\
$M_p$ \dotfill & Planetary mass ($M_{\mathrm{Jup}}$) \dotfill & \dotfill & 0.36 $\pm$ 0.12 & 0.87 $\pm$ 0.16 & 0.699 $\pm$ 0.066 \\
$\rho_p$ \dotfill & Planetary density (g/cm$^3$) \dotfill & \dotfill & 0.248 $\pm$ 0.081 & 0.275 $\pm$ 0.050 & 0.277 $\pm$ 0.028 \\
$T_{\mathrm{eq}}$ \dotfill & Equilibrium Temperature (K) \dotfill & \dotfill & 1232 $\pm$ 12 & 1907 $\pm$ 22 & 1445 $\pm$ 21 \\
$T_{\mathrm{dur}}$ \dotfill & Transit duration (hours) \dotfill & \dotfill & 3.22 $\pm$ 0.16 & 3.215 $\pm$ 0.086 & 4.53 $\pm$ 0.11 \\
\hline                                   
\\
\end{tabular}
}\\
*Notes: $U$(a,b) denotes a uniform prior between a and b, while $N$(a,b) denotes a normal distribution with mean of a and standard deviation b. Priors for $P$ and $T_0$ varied for each planet, situated around those computed in the photometric analysis described in section \ref{sec:phot_analysis}. Parameters without priors were derived from the fitted parameters and stellar parameters derived in section \ref{sec:stellar_params}, sampling from a normal distribution for each parameter based on their derived means and standard deviations. N.A. denotes `not applicable for this fit'. For the eccentric fit of TOI-3664\,b, $e$ and $\omega$ were derived from $\sqrt{e}sin(\omega)$ and $\sqrt{e}cos(\omega)$, while for TOI-4034\,b and TOI-6564\,b they were fixed at 0 and 90 deg respectively, with the reparametrisations unused.
\end{table*}

\subsection{Spectroscopic analysis}
\label{sec:spec_analysis}

Lomb-Scargle periodograms \citep{Lomb1976Least-squaresData,Scargle1982StudiesData} were also used to search for periodicity in the radial velocity measurements, yielding the results presented in Figures \ref{fig:rv_obs_toi-3664}, \ref{fig:rv_obs_toi-4034} and \ref{fig:rv_obs_toi-6564}. Because the CARMENES observing strategy employed two 22.5\,min exposures in series each observing night, for TOI-3664 and TOI-4034 the period search was trialled first with all individual measurements treated separately and secondly by binning data night by night. While both configurations yielded similar results, overall the unbinned RV measurements showed the strongest periodic signals despite the slightly decreased signal to noise of the individual spectra, so the unbinned data were used throughout this analysis.

Although the strength of the signal was limited by the available observing time in the small programmes, in all cases the first or second strongest peak in the periodogram corresponded to the same period alerted by the TOI release and found in the photometric analysis above (TOI-3664\,b = 3.30 days; TOI-4034\,b = 1.80 days; TOI-6564\,b = 3.99 days). In order to assess the statistical significance of these periodogram peaks, analytical False Alarm Probability (FAP) values were estimated for each target following the formalism of \citet{Baluev2008}. While TOI-6564\,b clearly passes even the 0.1\% FAP threshold, both TOI-3664\,b and TOI-4034\,b were found to cross the 10\% FAP threshold, and approach/cross the 1\% FAP threshold, validating the period seen in photometry. Note that after the removal of the best-fitting RV signal in each system, no additional FAP-crossing signals remained. Meanwhile, periodograms computed for common activity indicators such as the Full-Width Half-Maximum of the cross-correlation function (CCF), CCF bisector and CCF contrast across all spectrographs (and additionally the Ca IRT and H-alpha indicators for the CORALIE data) did not display notable peaks at these periods.

As an independent check of the spectroscopic analysis, all three systems were also analysed with the \texttt{kima} package \citep{Faria2018}. The \texttt{kima} package uses Bayesian inference to model a sum of Keplerian signals in the combined data across all available spectroscopic instruments. Despite the relatively small number of RV observations per target, the maximum likelihood period in the case of both TOI-3664\,b and TOI-6564\,b was found to be consistent with that seen in the \textit{TESS} photometry, with determined RV periods of P = 3.296 days for TOI-3664\,b and P = 3.987 days for TOI-6564\,b respectively. However, in the case of TOI-4034\,b, the photometric period of P$_{\mathrm{phot}}$ = 1.802 days actually corresponded to the second strongest period in the \texttt{kima} analysis (P$_{\mathrm{kima}}$ = 1.803 days), with the strongest period from the \texttt{kima} analysis occurring at 2.149 days. This is echoed by the Lomb-Scargle analysis, with the two strongest periods also falling at 2.15 and 1.80 days respectively for TOI-4034\,b (see Figure \ref{fig:rv_obs_toi-4034}). This may hint at a second periodic signal in the data (e.g. activity), but additional spectroscopic monitoring is required to confirm this.

\texttt{kima} also allows for a direct comparison between the evidence for different system architectures, i.e. the most likely number of planets in each system. For TOI-6564\,b, the candidate with the most RV follow-up, a 1-planet model was clearly preferred, with a 1-planet model having a log likelihood of log(Z) = -140.14 and a 0-planet model having a log likelihood of log(Z) = -176.19. However, the two other targets suffered from their small number of RV measurements and hence the 0-planet model was still preferred, despite many samples falling in the 1-planet (and in the case of TOI-4034, 2-planet) cases. Hence to confirm these latter two planets with radial velocities alone would require further observations, but when in support of the photometric data they validate the periods of the planetary systems quite effectively.

\subsection{Joint fit}
\label{sec:joint_fit}

Global modelling for all three systems was completed using the \texttt{juliet} package \citep{Espinoza2019Juliet:Systems}. \texttt{juliet} uses Bayesian inference to model planetary systems by combining transit fitting using \texttt{batman} \citep{Kreidberg2015Batman:Python}, RV fitting using \texttt{radvel} \citep{Fulton2018RadVel:Toolkit}, Gaussian process modelling with \texttt{george} \citep{celerite2:foremanmackey18} and \texttt{celerite} \citep{celerite2:foremanmackey18}, and nested sampling to simultaneously fit data from multiple photometric and spectroscopic datasets. Nested sampling was carried out using the \texttt{dynesty} package \citep{Speagle2019}, using $>N^2$ live points (where $N$ was the number of parameters in each fit), and in each case the fits were run until the estimated uncertainty in the log-evidence was less than 0.1.

Note that because of the significantly higher cadence of the short 2\,min vs long 30-min cadence data (and hence the much better coverage of individual transits), only the \textit{TESS} short-cadence (2\,min data) was included in the joint fit, but with Gaussian priors on t0 and planetary period drawn from the photometric analysis undertaken in Section \ref{sec:phot_analysis} using all sectors of available \textit{TESS} data. Both the radius ratio (p = R$_\mathrm{p}$/R$_*$) and impact parameter, b, were allowed to vary between zero and 1 uniformly.

A white-noise jitter term was added in quadrature to the error-bars of each photometric and RV dataset to account for underestimated uncertainties and/or additional white noise that was not captured by the model. In addition, to account for any leftover stellar activity and systematics caused by scattered light from the Earth/Moon and/or slight baseline shifts after data gaps in the \textit{TESS} data, a Gaussian Process (GP) model was included in the photometric models using the Matérn-3/2 kernel in \texttt{celerite} \citep{celerite2:foremanmackey18}. This GP is able to account from correlated noise from multiple sources, and ensured that a reliable baseline could be found for the out-of-transit data, enabling an accurately measured transit depth. 

Realistic priors on the limb-darkening parameters were found using the \texttt{LDCU} package.\footnote{\url{https://github.com/delinea/LDCU}} \texttt{LDCU} is a modified version of the python routine implemented by \citet{Espinoza2015LimbParameters} that computes the limb-darkening coefficients and their corresponding uncertainties using a set of stellar intensity profiles accounting for the uncertainties on the stellar parameters. The stellar intensity profiles are generated based on two libraries of synthetic stellar spectra: ATLAS \citep{Kurucz1979} and PHOENIX \citep{Husser2013}. For each target, quadratic coefficients were derived from the \texttt{LDCU} code, based on the q1 and q2 parameterisation of \citep{Kipping2013}, and used to set Gaussian priors on the limb-darkening parameters in the joint fit. For each set of \textit{TESS} data, the \textit{TESS} passband was used to calculate q1 and q2, while for the Brierfield Observatory the Bessel B-band was used and for PEST, uniform unconstrained priors of q1, q2 = [0,1] were used because LDCU did not contain a bandpass for the rp filter used by PEST. To account for any variations from the modelled limb-darkening in the real data, the errors found in LDCU were multiplied by five to give more conservative priors.

Because the Thoirum-Argon lamp in the CORALIE spectrograph was replaced in 2024, there are slight differences in both the baseline and properties of the pre-2024 and post-2024 data, so these were treated as two different datasets in the joint fit of TOI-6564\,b, denoted CORALIE14 and CORALIE24 respectively. Additionally, any spectroscopic measurements that fell within the expected times of transit were removed. 

For TOI-3664 and TOI-4034, the dilution value was fixed to one, as only the \textit{TESS} 2-min data was used, which is already corrected for dilution. As discussed by \citet{Han2025}, it is possible that additional systematic contamination caused by underestimation of stellar radii was not accounted for in this dilution correction, but tests with wide dilution priors of 0 to 1 were found to give unrealistically large radius values and much lower log evidence values, as was also cautioned by \citet{Espinoza2019Juliet:Systems}. Hence until higher resolution photometric follow-up is undertaken, fixing the dilution factor to one is the preferred approach for these two dimmer targets.

TOI-6564 however has data from three different photometric bands, of which only the \textit{TESS} 2\,min-cadence data is theoretically dilution corrected. The dilution value was thus allowed to vary uniformly between zero and one for the two follow-up observations carried out by Brierfield Observatory and PEST. Note that the end of the out-of-transit PEST data (t $>$ 2460390.35) was masked because it displayed an unexplained dip in flux which appeared systematic.


Circular and eccentric fits were trialled for each system and compared according to their Bayesian evidence. For the eccentric fits, the $\sqrt{e}\sin(\omega)$ and $\sqrt{e}\cos(\omega)$ allowed by \citet{Espinoza2015LimbParameters} was employed, with uniform priors between -1 and 1, and an eccentricity limit of 0.8 to aid convergence of the model. Following \citet{Espinoza2019Juliet:Systems} and \citet{Trotta2008}, different models were compared using the natural log of the Bayesian evidence (ln$Z$), with $\Delta$ln$Z$ = 2 the threshold between weak and moderate evidence for one model being preferred, and $\Delta$ln$Z>$  5 the threshold for strong evidence.

The final results of the joint photometric and spectroscopic fit, along with the priors used for all parameters, can be found in Tables \ref{planetary_table} and \ref{instrumental_table}, and discussed on a planet by planet basis in the next section.

\section{Results and Discussion}
\label{sec:results}

\subsection{Planetary fit results}

\subsubsection{TOI-3664\,b}

The joint fit of TOI-3664\,b revealed a close-in (P = 3.30 days), slightly inflated planet with a radius of 1.22 $\pm$ 0.03 R$_\mathrm{Jup}$ and mass of only 0.36 $\pm$ 0.12 M$_\mathrm{Jup}$, or roughly the mass of Saturn. This results in a reasonably low density of 0.25 $\pm$ 0.08 g/cm$^3$.

Because of the low eccentricity suggested by the spectroscopic-only fit, two different models were trialled: one with a circular orbit (e = 0, $\omega$ = 90 deg), and another with an eccentric orbit, using the discussed $\sqrt{e}sin(\omega)$ and $\sqrt{e}cos(\omega)$ reparametrisation. The comparative log-evidence values for these two fits were ln $Z_{\mathrm{circ}}$ = 140421.670 and ln $Z_{\mathrm{ecc}}$ = 140428.264, giving $\Delta$ln$Z$ = 6.59 in favour of the eccentric fit, which suggests strong evidence of the eccentric fit being preferred. However, with a calculated eccentricity of e = 0.16 $\pm$ 0.07, this eccentricity should be treated with a little caution. To get additional confirmation of the true eccentricity, additional RV monitoring with higher-precision instruments is recommended.

\begin{figure}
\centering
\includegraphics[width=\hsize]{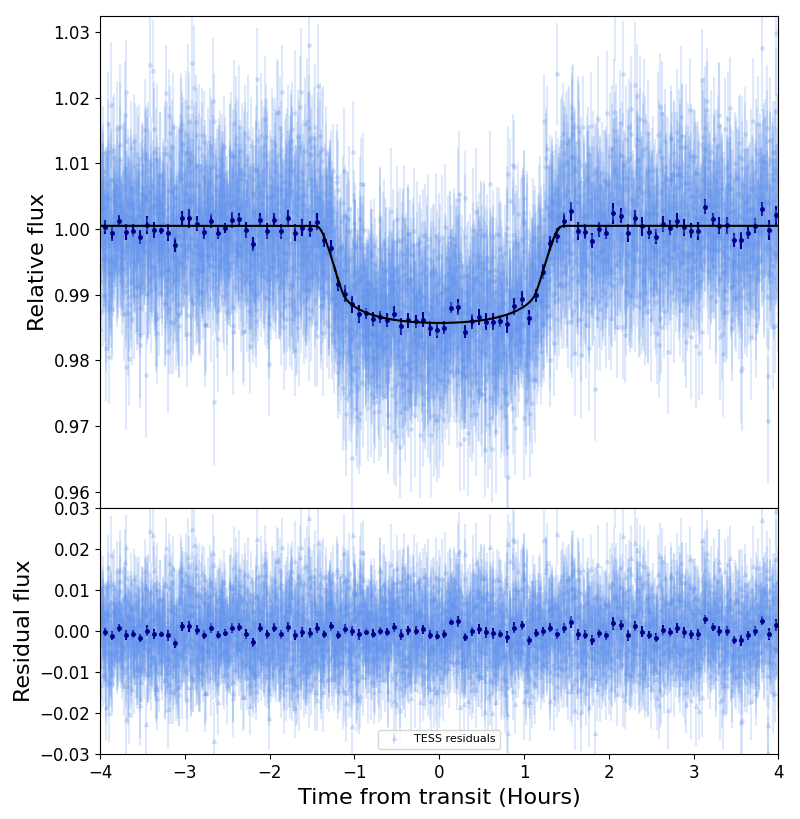}
  \caption{TOI-3664 \textit{TESS} data folded by the 3.30-day transit period. The full \textit{TESS} data is plotted in light blue, with the transit model from the joint fit over-plotted in black. Additional dark blue points show the data binned to 5\,min bins. Residuals from this fit are shown in the panel below.}
     \label{fig:toi3664_transit_fold}
\end{figure}

\begin{figure}
\centering
\includegraphics[width=\hsize]{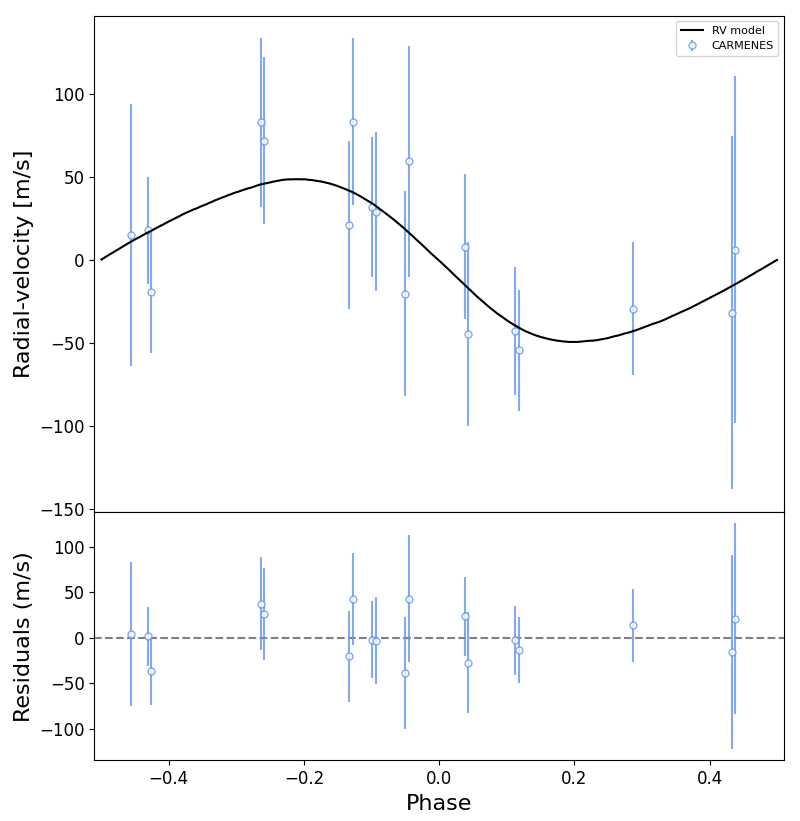}
  \caption{TOI-3664 radial velocity data folded by the 3.30-day planetary period. CARMENES\_VIS RVs are plotted in blue, and the fitted Keplerian planet model in black. One point which fell in the duration of the transit (near phase 0) has been removed. Residuals to the fit are shown in the lower panel.}
     \label{fig:toi3664_fitted_RVs}
\end{figure}

The folded photometric data showing the transit fit for TOI-3664\,b can be seen in Figure \ref{fig:toi3664_transit_fold}, with the fitted RVs in Figure \ref{fig:toi3664_fitted_RVs}.

\subsubsection{TOI-4034\,b}

The joint fit of TOI-4034\,b revealed a planet with an even closer orbit (P = 1.80 days), a radius of 1.58 $\pm$ 0.02 R$_\mathrm{Jup}$ and mass of 0.87 $\pm$ 0.16 M$_\mathrm{Jup}$, indicative of a more typical hot Jupiter exoplanet.

\begin{figure}
\centering
\includegraphics[width=\hsize]{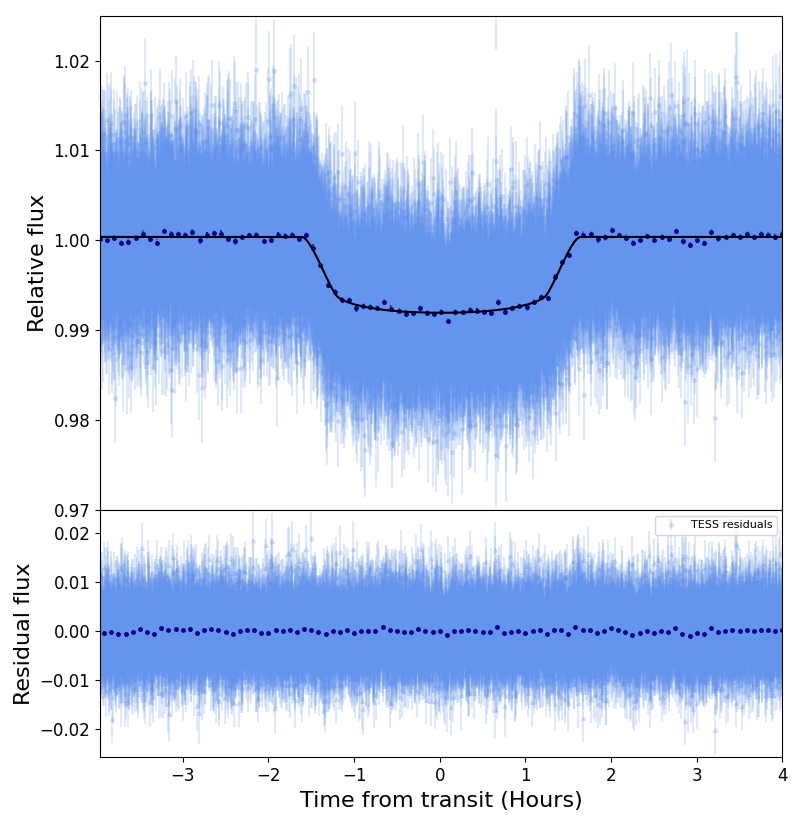}
  \caption{TOI-4034 \textit{TESS} data folded by the 1.80-day planetary period. The full \textit{TESS} data is shown in light blue, the \textit{TESS} data binned into 5\,min bins in dark blue and the joint fit transit model over-plotted in black. Note that the binned error-bars are smaller than the points themselves.}
     \label{fig:toi4034_transit_fold}
\end{figure}

\begin{figure}
\centering
\includegraphics[width=\hsize]{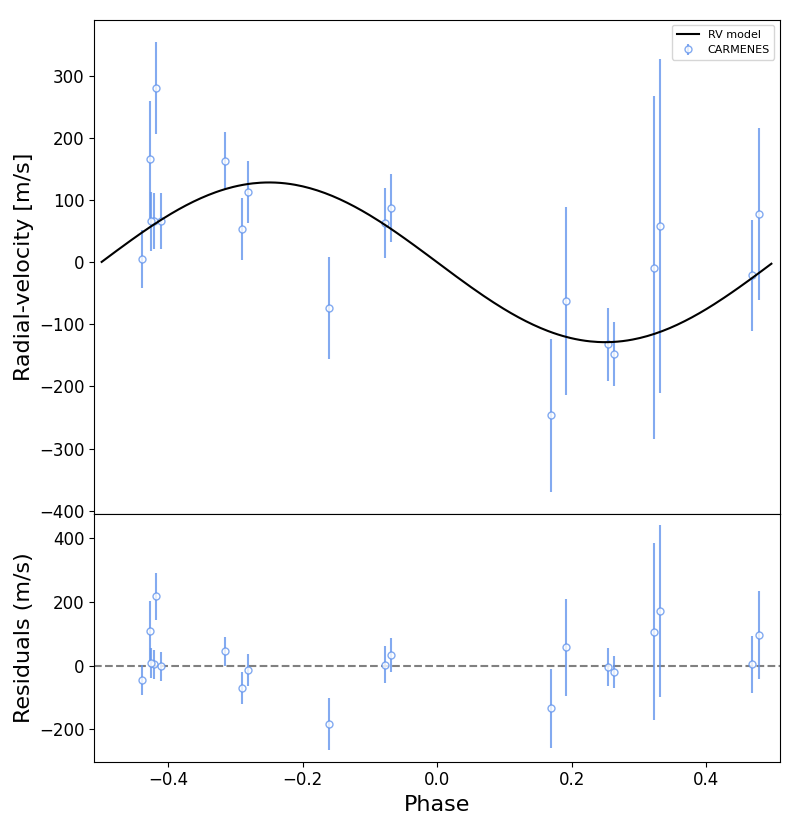}
  \caption{TOI-4034 radial velocity data folded by the 1.80-day planetary period. CARMENES\_VIS RVs are plotted in blue, and the fitted Keplerian planet model in black. Residuals to the fit are shown in the panel below. Two points which were taken at the time of transit were removed.}
     \label{fig:toi4034_fitted_RVs}
\end{figure}

In this case, the circular orbit was marginally preferred in the model comparison, with a ln $Z_{\mathrm{circ}}$ = 397184.588 and ln $Z_{\mathrm{ecc}}$ = 397182.586, giving $\Delta$ln$Z$ = 2.0 in favour of the circular fit. Given that for such a short orbital period the tidal circularisation timescale is also much shorter than the stellar age \citep[e.g.][c.f. $t_{\mathrm{circularisation}} \sim$ 820 Myr, age = 5.7$^{+0.5}_{-0.5}$ Gyr]{Jackson2008} we model the system as a simpler circular orbit. The folded photometric data showing the transit fit for TOI-4034\,b can be seen in Figure \ref{fig:toi4034_transit_fold}, with the fitted RVs in Figure \ref{fig:toi4034_fitted_RVs}. The CARMENES residuals to this fit are slightly less well behaved than those from the TOI-3664\,b fit, which could indicate the presence of additional signals in the RV data, however investigating this further would require additional RV monitoring of this system, which it outside the scope of the current analysis.

\subsubsection{TOI-6564\,b}

TOI-6564 hosts a hot planet with super-Jupiter radius ($R_p$ = 1.46 $\pm$ 0.02 R$_\mathrm{Jup}$) and sub-Jupiter mass ($M_p$ = 0.70 $\pm$ 0.07 M$_\mathrm{Jup}$). It is the furthest-out system of the three, with a period of 3.99 days.

For this system, a circular fit was significantly preferred to an eccentric one, with a $\Delta$ln$Z$ = 17.8 towards a circular fit. Furthermore, the eccentric fit resulted in an eccentricity value of only 0.015 $\pm$ 0.016, consistent with zero, so the simpler circular model was chosen. 

To test the effect of including the ground-based photometric data alongside the more extensive \textit{TESS} data, an initial fit was conducted using just the \textit{TESS} data and spectroscopic data, before additional fits were carried out by adding the two supporting photometric datasets (Brierfield and PEST) one by one. In each case, the addition of the ground-based data improved the overall log-evidence further and significantly improved the achieved precision in the transit ephemerides (t0 and P), so the fit including all photometric data is preferred.

\begin{figure*}
\centering
\includegraphics[width=\hsize]{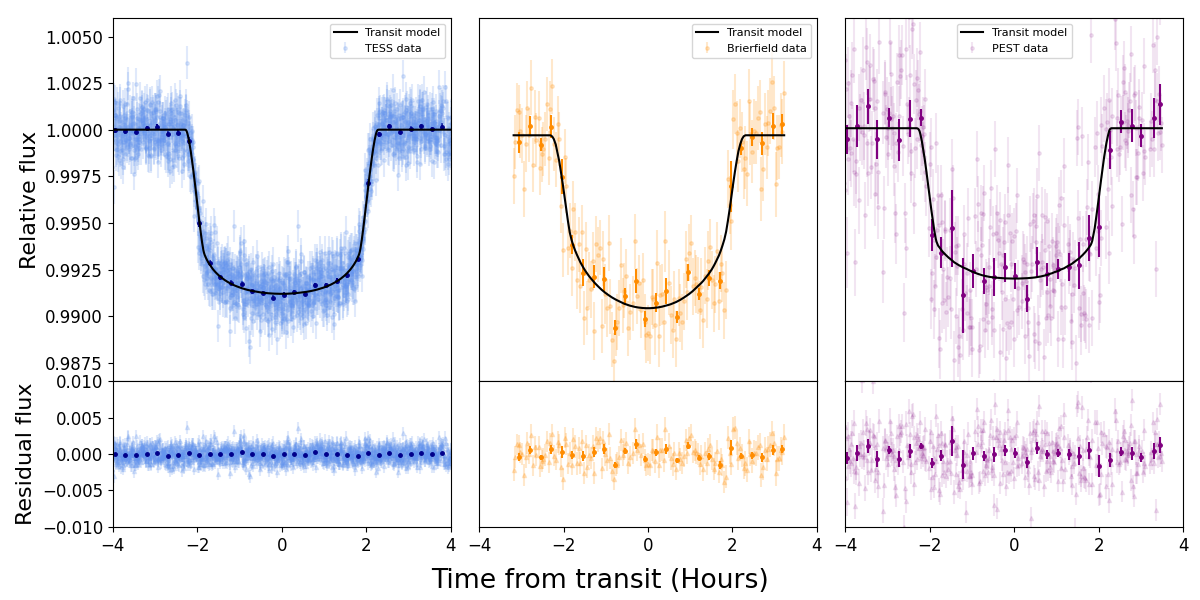}
  \caption{TOI-6564\,b phase-folded transits from \textit{TESS}, Brierfield Observatory and PEST. As in Figures \ref{fig:toi3664_transit_fold} and \ref{fig:toi4034_transit_fold}, the photometric data is shown in colour for each instrument, with the joint fit over-plotted in black and the residuals to the fit shown below. To aid clarity, the binned data-points are binned into 15\,min bins instead of 5\,min bins in this case.}
     \label{fig:toi6564_transit_fold}
\end{figure*}

\begin{figure}
\centering
\includegraphics[width=\hsize]{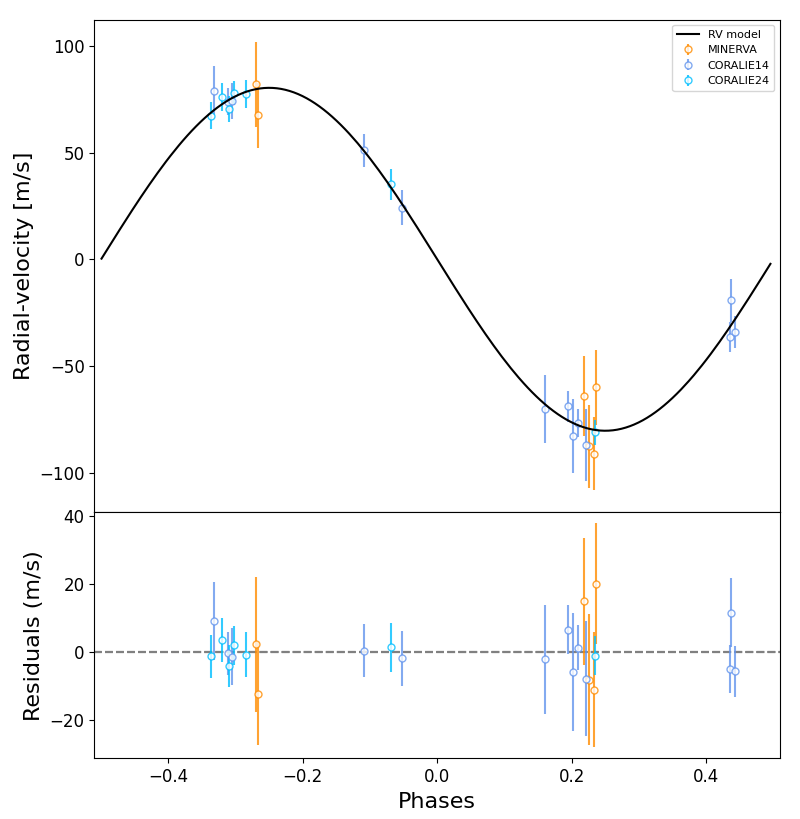}
  \caption{Radial velocity data for TOI-6564 folded by the best-fit period of 3.99 days. MINERVA-Australis, CORALIE (pre-2024) and CORALIE (post 2024) data is plotted in colour, with the radial velocity model from the joint fit shown in black. Residuals to the fit are shown in the panel below. One point from each instrument was removed due to occuring during the transit.}
     \label{fig:toi6564_fitted_RVs}
\end{figure}

The final photometric and spectroscopic fits can be seen in Figures \ref{fig:toi6564_transit_fold} and \ref{fig:toi6564_fitted_RVs} respectively. Note that the somewhat sparse RV sampling was caused by the near-integer day period (P = 3.99 days), which meant that many days were required to sample different regions of the phase curve.


\subsection{Comparison to the known population}

These three new planets are particularly interesting for two main reasons: their evolutionary phase near the Terminal Age Main Sequence and their relatively low densities. 

\begin{figure}
\centering
\includegraphics[width=\hsize]{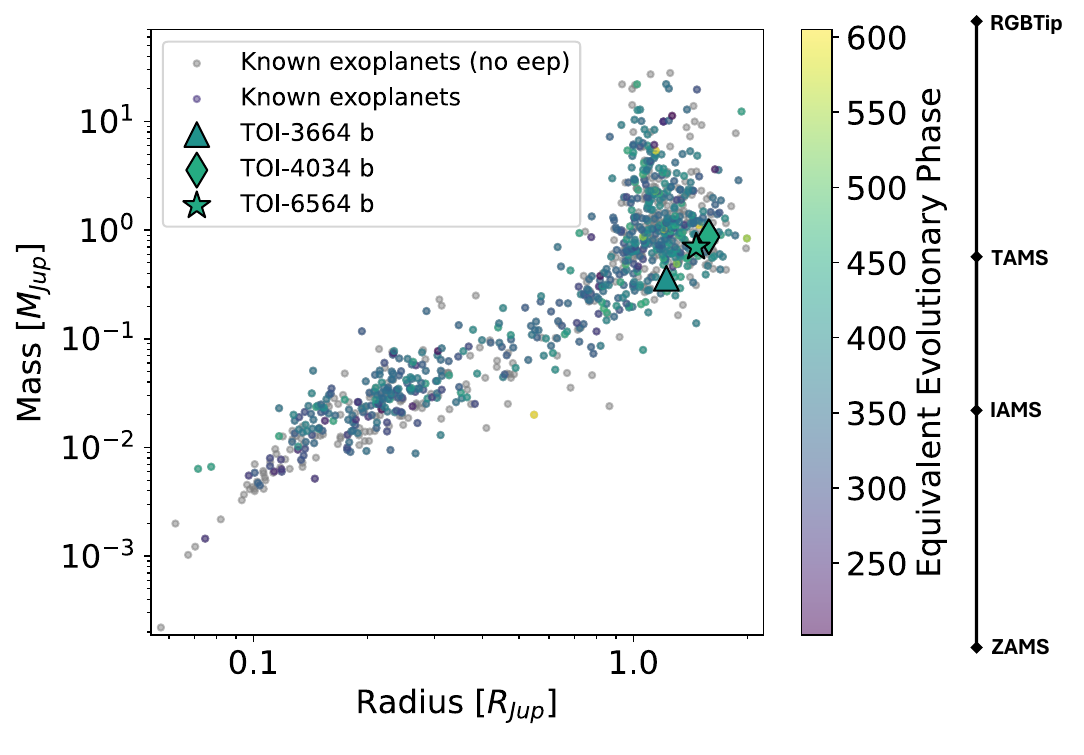}
  \caption{Mass-radius-EEP diagram showing all three new planets in the context of the known population of high precision exoplanets ($\Delta {\rm R}_{\rm p}/{\rm R}_{\rm p}$ <10\% and $\Delta \mathrm{M}_{\rm p}/\mathrm{M}_{\rm p}$ of <30\%). The planets confirmed in this work are plotted as symbols alongside the main population with black outlines for clarity. The equivalent evolutionary phase (EEP) of each host star is shown in colour, with the main evolutionary phase boundaries shown to the right at their respective EEPs. Grey points represent high precision known exoplanetary systems for which EEP values could not be calculated.}
     \label{fig:mass-radius}
\end{figure}

\begin{figure}
\centering
\includegraphics[width=\hsize]{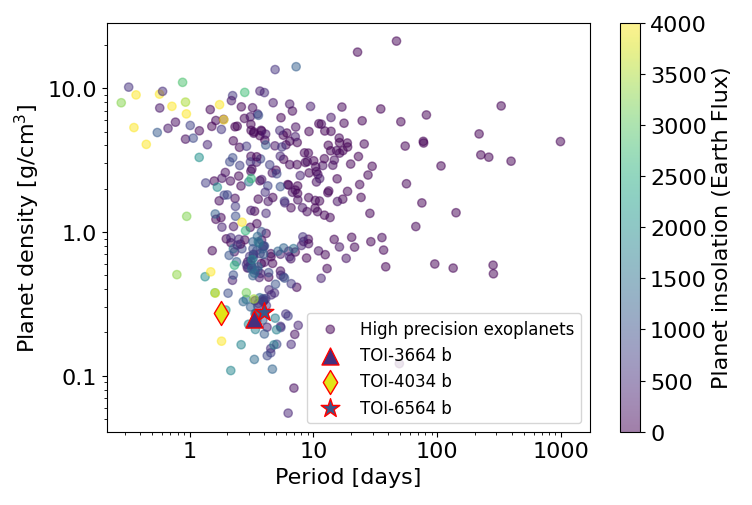}
  \caption{Period-density diagram showing the three new planets alongside the existing sample of high-precision exoplanets. The colour of each planet is plotted according to the planetary insolation flux. The three planets confirmed in this paper are outlined in red for clarity.}
     \label{fig:period_density}
\end{figure}

All three planets fall towards the upper end of the distribution of known exoplanets, with reasonably large radii but average to below-average masses compared to the rest of the gas-giant distribution, as shown in Figure \ref{fig:mass-radius}. Because of the masses and radii of the three exoplanets studied here, they all exhibit relatively small densities compared to the rest of the population, as illustrated in Figure \ref{fig:period_density}. Planets in this figure are coloured by their insolation flux, with TOI-4034\,b clearly possessing the largest planetary insolation due to its large host and close orbit (P=1.80 day). Such low densities are often seen for planets evolving off the main sequence, possibly due to re-inflation of planetary atmospheres as their host stars brighten \citep[e.g. see][]{Grunblatt2016,Wittenmyer2022}. 

\begin{figure}
\centering
\includegraphics[width=\hsize]{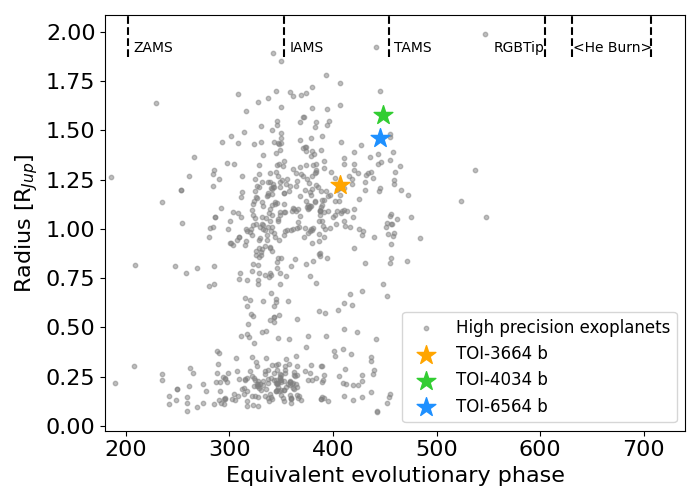}
  \caption{Radius-EEP plot for the host stars of the population of high precision exoplanets ($\Delta R_p/R_p$ <10\% and $\Delta M_p/M_p$ of <30\%) which also have characterised radii, metallicity and age (all required for the calculation of EEP). Known exoplanets are plotted in gray, while the new planets studied here are shown in colour. Dashed lines at the top of the plot denote the evolutionary stages which are associated with the numerical equivalent evolutionary phases and can be understood as follows: ZAMS = Zero Age Main Sequence; IAMS = Intermediate Age Main Sequence; TAMS = Terminal Age Main Sequence; RGBTip = the tip of the red giant branch. <He Burn> straddles the range of EEP values associated with Helium burning in the stellar atmosphere.}
     \label{fig:radius-eep_plot}
\end{figure}

Figure \ref{fig:mass-radius} also shows the equivalent evolutionary phase (EEP) number for every star which hosts a precisely characterised known exoplanet ($\Delta {\rm R}_{\rm p}/{\rm R}_{\rm p}$ <10\% and $\Delta \mathrm{M}_{\rm p}/\mathrm{M}_{\rm p}$ of <30\%) by way of the colour-bar. The EEP for each known exoplanet host star was calculated using the MESA isochrones \citep{Dotter2016,Choi2016} via the \texttt{isochrones} package\footnote{https://github.com/timothydmorton/isochrones} \citep{Morton2015} by interpolating from the closest MESA isochrones based on the stellar radius, metallicity and log(age) of each star from the Exoplanet Archive\footnote{https://exoplanetarchive.ipac.caltech.edu/} \citep{Akeson2013}. This plot suggests that these three new planets are around some of the most evolved stars currently in the high precision exoplanet sample, appearing lighter green than the bulk of the sample. This is corroborated by the planetary radius vs EEP plot in Figure \ref{fig:radius-eep_plot}, clearly illustrating that TOI-4034 and TOI-6564 sit right at the terminal age main sequence boundary and also host two of the largest radii exoplanets around high-EEP host stars. 

\subsection{Evolutionary state and planetary fate}

The evolutionary state of these three host stars, so close to the terminal age main sequence, raises an interesting question: what is the likely future evolution of these planetary systems as their hosts begin their journey onto the red giant branch?

As these planets' host stars leave the main sequence and enter the red giant branch, they will increase dramatically in radius, eventually expanding to well beyond the orbit of these short-period planets, leading to the engulfment of the planets \citep[e.g][]{Villaver2014,OConnor2023}. Before this engulfment, the planets will be subject to the competing effects of an inward tidal force \citep[e.g.][]{Rasio1996,Jackson2008,Villaver2014} and a mass-loss-induced orbital expansion caused by stellar winds \citep{Hadjidemetriou1963,Veras2011}. For close-in planets like hot Jupiters, the inward tidal force typically far outstrips the mass-loss-induced orbital expansion, leading to a decrease in orbital separation with time.

The fate of each planet can be investigated through equations 3-9 of \cite{Villaver2014} given time evolution profiles of stellar mass and radius from \cite{Hurley2000}. These equations are suitable for the subgiant and red giant branch phases, and hence are suitable for all three of these planets, because none of them will survive beyond the red giant branch phase. Each of these planets will be engulfed by the star and destroyed, as their masses are far too low to survive residence within a common envelope \citep{OConnor2023}. The equations from \cite{Villaver2014} allow us to simultaneously evolve the semi-major axis and eccentricity of the planets.

We find that in all three systems, the orbital radii at which the engulfment occurs ($\approx 0.025$~au for TOI-3664~b, $\approx 0.028$~au for TOI-4034~b and $\approx 0.033$~au for TOI-6564~b) are smaller then the planets' pericentre values at the terminal main sequence. Hence, all three planets are expected to be drawn into the star as the stellar radius was increases. The timescale for the planets to be engulfed, as measured from the end of the main sequence lifetime, are $\sim 1200$~Myr for TOI-3664~b, $\sim 650$~Myr for TOI-4034~b and $\sim 700$~Myr for TOI-6564~b. Further, the orbit of TOI-3664~b does not completely circularise before the planet is engulfed: at the moment of engulfment, the planet's eccentricity will have reduced to $\approx 0.05$.


The very short period of TOI-4034\,b (P = 1.80 days) may also accelerate this process through tidal destruction. \citet{Weinberg2024} found that for planets with periods of <2 days, non-linear damping of tidally driven g-modes in the host stars during the sub-giant phase could lead to orbital decay timescales of $\leq$ 10 Myr for stars around the size of TOI-4034 or larger ($M_\odot \geq 1.20$). 

Observationally, \citet{Bryant2025} also noted a steep drop-off in the occurrence rate of planets with periods <2 days around post-main sequence stars, even as early as the sub-giant phase. Hence as TOI-4034 passes the terminal age main sequence onto the sub-giant branch proper, its planet is likely to be swiftly destroyed by its host. Although this drop-off is less steep for planets with periods above 2 days, \citet{Bryant2025} also saw a notable decrease in the occurrence rate of planets around sub-giant and early red-giant phases of stellar evolution up to periods of 12.5 days, supporting the idea that the lifetimes of the short-period TOI-3664\,b and TOI-6564\,b planets will also be limited upon leaving the main sequence proper. 

The evolutionary state of these planets' host stars hence makes them very useful windows into the later stages of hot Jupiter evolution, and warrants follow-up over longer timescales to search for tell-tale signs of orbital decay.



\subsection{Prospects for atmospheric characterisation}

\begin{figure}
\centering
\includegraphics[width=\hsize]{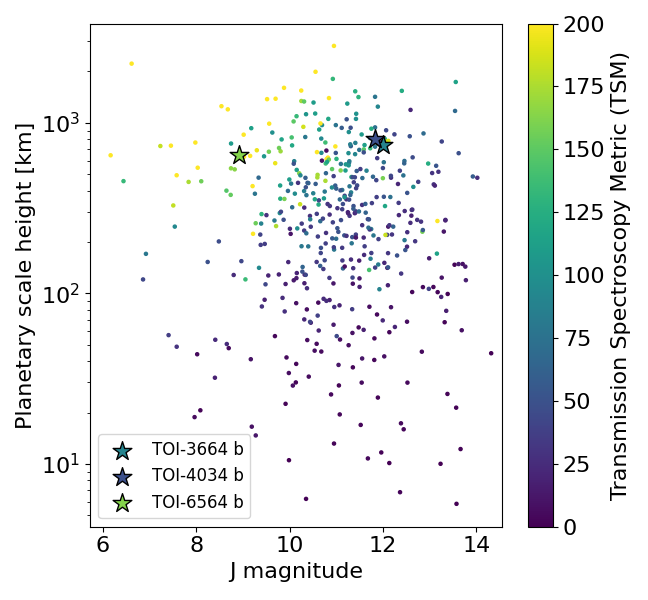}
  \caption{Atmospheric characterisation prospects for the three new planets. Alongside the planetary scale height and J-magnitude of their host stars, the Transmission Spectroscopy Metric (TSM) for each planet is shown in colour, curtailed at 200 to better differentiate between targets near the recommended follow-up (96) and first-quartile (159) cut-offs. TOI-6564\,b is particularly promising, with an atmospheric scale height of 645 km and TSM = 163.}
     \label{fig:scale_height}
\end{figure}

The interesting evolutionary stage, short periods and large radii of these three new planets make them intriguing targets for atmospheric characterisation. Two commonly used parameters to ascertain the suitability of different targets for atmospheric characterisation are the transmission spectroscopy metric  \citep[TSM,][]{Kempton2018ACharacterization} and the atmospheric scale height, $H_b$. The TSM for each target can be determined directly from the mass, radius and equilibrium of the planet, alongside the brightness and radius of the host star, while the atmospheric scale height can be calculated using the equation $H_b = kT_{eq}/(\mu g_b)$, where $k_b$ is the Boltzmann constant, $\mu$ is the mean molecular mass (assumed to be 2.3 amu here for a H/He atmosphere), and $g_b$ is the planet surface gravity based in the derived mass and radius. Because of its host's brightness, TOI-6564\,b was found to be particularly attractive for atmospheric characterisation, with TSM of 163 and an atmospheric scale height of 645km, roughly ten times that of Jupiter. For Jupiter-sized planets, \citet{Kempton2018ACharacterization} recommend a cut-off for follow-up efforts of TSM = 96, with TSM = 159 their cut-off for planets in the first quartile, meaning that TOI-6564\,b is a very high priority target for atmospheric characterisation. Indeed, by comparison to other giant planets in Figure \ref{fig:scale_height}, TOI-6564 can be seen to be one of the brightest targets with such a large scale height. On the other hand, TOI-3664\,b and TOI-4034\,b were found to have transmission spectroscopy metrics of 90 and 49 respectively (with $H_b$ = 745 km and 798km), which suggests that they may be of lower priority for follow-up despite having large scale heights, largely driven by their lower observed brightnesses.

\section{Conclusions}\label{sec:conclusion}

TOI-3664\,b, TOI-4034\,b and TOI-6564\,b are three new hot Jupiter planets orbiting host stars towards the very end of their main-sequence lifetimes. These planets, all with sub-Jupiter masses (0.36 $\pm$ 0.12, 0.87 $\pm$ 0.16 \& 0.70 $\pm$ 0.07 M$_\mathrm{Jup}$) and super-Jupiter radii (1.22 $\pm$ 0.03, 1.58 $\pm$ 0.02 \& 1.46 $\pm$ 0.02 R$_\mathrm{Jup}$) represent three key opportunities to explore the latter stages of planetary evolution, before their eventual engulfment as their host stars expand along the red giant branch. All three planets have reasonably low densities and moderate to high insolation flux due to their short orbital periods ($P =$ 3.30, 1.80 and 3.99 days respectively). TOI-6564\,b is also particularly promising for atmospheric characterisation, with a TSM of 163 and atmospheric scale height of 649\,km. 

Because of their position near the terminal main sequence, the ages of the systems are somewhat degenerate, so care is taken to determine the correct ages through consideration of a variety of age-dating methods. This resulted in ages of 9.0$^{+2.4}_{-2.1}$, 5.7 $\pm$ 0.5 Gyr and 4.0 $\pm$ 1.0 Gyr respectively, adding three important points to the evolutionary distribution of well characterised hot-Jupiters. Modelling of the evolution of the stellar envelopes and semi-major axes of the planets and their host stars suggests that all three systems will be engulfed by their hosts on the order of only 1 Gyr from the end of the main sequence. 

The results of a wider vetting effort of \textit{TESS} planet candidates conducted in the same CARMENES programme in which TOI-3664 and TOI-4034 were monitored are also presented to aid the community.

\section*{Acknowledgements}

The authors would like to thank the anonymous referee for their comments which improved the quality and robustness of this paper.

MPB and EG gratefully acknowledge support from UK Research and Innovation (UKRI) under the UK government’s Horizon Europe funding guarantee for an ERC starting grant [grant number EP/Z000890/1]. The contributions of MPB, SU-M, YF, FB, YC, MH, LP, and ST have been carried out within the framework of the NCCR PlanetS supported by the Swiss National Science Foundation under grants 51NF40\_182901 and 51NF40\_205606. ML acknowledges support of the Swiss National Science Foundation under grant number PCEFP2\_194576. MLafarga gratefully acknowledges support from UK Research and Innovation (UKRI) under the UK government’s Horizon Europe funding guarantee [grant number Grant EP/X027562/1].

This paper includes data collected with the TESS mission, obtained from the MAST data archive at the Space Telescope Science Institute (STScI). Funding for the TESS mission is provided by NASA’s Science Mission directorate. We acknowledge the use of public TOI Release data from pipelines at the TESS Science Office and at the TESS Science Processing Operations Center. STScI is operated by the Association of Universities for Research in Astronomy, Inc., under NASA contract NAS 5–26555. Resources supporting this work were provided by the NASA High-End Computing (HEC) Program through the NASA Advanced Supercomputing (NAS) Division at Ames Research Center for the production of the SPOC data products. This research has made use of the Exoplanet Follow-up Observation Program website, which is operated by the California Institute of Technology, under contract with the National Aeronautics and Space Administration under the Exoplanet Exploration Program. 

This paper was based on observations collected at Centro Astronómico Hispano en Andalucía (CAHA) at Calar Alto, operated jointly by Junta de Andalucía and Consejo Superior de Investigaciones Científicas (IAA-CSIC).

This publication makes use of The Data \& Analysis Center for Exoplanets (DACE), which is a facility based at the University of Geneva (CH) dedicated to extrasolar planets data visualisation, exchange and analysis. DACE is a platform of the Swiss National Centre of Competence in Research (NCCR) PlanetS, federating the Swiss expertise in Exoplanet research. The DACE platform is available at https://dace.unige.ch.

MINERVA-Australis is supported by Australian Research Council LIEF Grant LE160100001, Discovery Grants DP180100972 and DP220100365, Mount Cuba Astronomical Foundation, and institutional partners University of Southern Queensland, UNSW Sydney, MIT, Nanjing University, George Mason University, University of Louisville, University of California Riverside, University of Florida, and The University of Texas at Austin.

We respectfully acknowledge the traditional custodians of all lands throughout Australia, and recognise their continued cultural and spiritual connection to the land, waterways, cosmos, and community. We pay our deepest respects to all Elders, ancestors and descendants of the Giabal, Jarowair, and Kambuwal nations, upon whose lands the Minerva-Australis facility at Mt Kent is situated.

\section*{Data Availability}


\textit{TESS} data for all three systems are publicly available from the Mikulski Archive for Space Telescopes (MAST).\footnote{https://mast.stsci.edu/portal/Mashup/Clients/Mast/Portal.html} Additional photometry for TOI-6564 are also publicly available on the \textit{TESS} ExoFOP platform.\footnote{https://exofop.ipac.caltech.edu/tess/} Radial velocity measurements for each system are presented here in full, and are included in machine-readable form in the supplementary material alongside the online copy of this article. Additional intermediate data products underlying this article will be shared on reasonable request to the corresponding author.



\bibliographystyle{mnras}
\bibliography{references} 

@ARTICLE{Hurley2000,
       author = {{Hurley}, Jarrod R. and {Pols}, Onno R. and {Tout}, Christopher A.},
        title = "{Comprehensive analytic formulae for stellar evolution as a function of mass and metallicity}",
      journal = {\mnras},
         year = 2000,
        month = jul,
       volume = {315},
       number = {3},
        pages = {543-569},
          doi = {10.1046/j.1365-8711.2000.03426.x},
archivePrefix = {arXiv},
       eprint = {astro-ph/0001295},
 primaryClass = {astro-ph},
       adsurl = {https://ui.adsabs.harvard.edu/abs/2000MNRAS.315..543H}
}

@ARTICLE{Lafarga2020,
       author = {{Lafarga}, M. and {Ribas}, I. and {Lovis}, C. and {Perger}, M. and {Zechmeister}, M. and {Bauer}, F.~F. and {K{\"u}rster}, M. and {Cort{\'e}s-Contreras}, M. and {Morales}, J.~C. and {Herrero}, E. and {Rosich}, A. and {Baroch}, D. and {Reiners}, A. and {Caballero}, J.~A. and {Quirrenbach}, A. and {Amado}, P.~J. and {Alacid}, J.~M. and {B{\'e}jar}, V.~J.~S. and {Dreizler}, S. and {Hatzes}, A.~P. and {Henning}, T. and {Jeffers}, S.~V. and {Kaminski}, A. and {Montes}, D. and {Pedraz}, S. and {Rodr{\'\i}guez-L{\'o}pez}, C. and {Schmitt}, J.~H.~M.~M.},
        title = "{The CARMENES search for exoplanets around M dwarfs. Radial velocities and activity indicators from cross-correlation functions with weighted binary masks}",
      journal = {\aap},
         year = 2020,
        month = apr,
       volume = {636},
          eid = {A36},
        pages = {A36},
          doi = {10.1051/0004-6361/201937222},
archivePrefix = {arXiv},
       eprint = {2003.07471},
 primaryClass = {astro-ph.IM},
       adsurl = {https://ui.adsabs.harvard.edu/abs/2020A&A...636A..36L}
}

@ARTICLE{Kunimoto2022,
       author = {{Kunimoto}, Michelle and {Daylan}, Tansu and {Guerrero}, Natalia and {Fong}, William and {Bryson}, Steve and {Ricker}, George R. and {Fausnaugh}, Michael and {Huang}, Chelsea X. and {Sha}, Lizhou and {Shporer}, Avi and {Vanderburg}, Andrew and {Vanderspek}, Roland K. and {Yu}, Liang},
        title = "{The TESS Faint-star Search: 1617 TOIs from the TESS Primary Mission}",
      journal = {\apjs},
         year = 2022,
        month = apr,
       volume = {259},
       number = {2},
          eid = {33},
        pages = {33},
          doi = {10.3847/1538-4365/ac5688},
archivePrefix = {arXiv},
       eprint = {2112.02176},
 primaryClass = {astro-ph.EP},
       adsurl = {https://ui.adsabs.harvard.edu/abs/2022ApJS..259...33K}
}

@INPROCEEDINGS{Huang2019,
       author = {{Huang}, Xu and {Burt}, Jennifer and {Vanderburg}, Andrew and {Gunther}, Maximillian and {Shporer}, Avi and {Dittmann}, Jason and {Winn}, Joshua},
        title = "{A Quick look into the first discoveries of TESS}",
    booktitle = {American Astronomical Society Meeting Abstracts \#233},
         year = 2019,
       series = {American Astronomical Society Meeting Abstracts},
       volume = {233},
        month = jan,
          eid = {209.08},
        pages = {209.08},
       adsurl = {https://ui.adsabs.harvard.edu/abs/2019AAS...23320908H}
}

@ARTICLE{Huang2020a,
       author = {{Huang}, Chelsea X. and {Vanderburg}, Andrew and {P{\'a}l}, Andras and {Sha}, Lizhou and {Yu}, Liang and {Fong}, Willie and {Fausnaugh}, Michael and {Shporer}, Avi and {Guerrero}, Natalia and {Vanderspek}, Roland and {Ricker}, George},
        title = "{Photometry of 10 Million Stars from the First Two Years of TESS Full Frame Images: Part I}",
      journal = {Research Notes of the American Astronomical Society},
         year = 2020,
        month = nov,
       volume = {4},
       number = {11},
          eid = {204},
        pages = {204},
          doi = {10.3847/2515-5172/abca2e},
archivePrefix = {arXiv},
       eprint = {2011.06459},
 primaryClass = {astro-ph.EP},
       adsurl = {https://ui.adsabs.harvard.edu/abs/2020RNAAS...4..204H}
}

@ARTICLE{Huang2020b,
       author = {{Huang}, Chelsea X. and {Vanderburg}, Andrew and {P{\'a}l}, Andras and {Sha}, Lizhou and {Yu}, Liang and {Fong}, Willie and {Fausnaugh}, Michael and {Shporer}, Avi and {Guerrero}, Natalia and {Vanderspek}, Roland and {Ricker}, George},
        title = "{Photometry of 10 Million Stars from the First Two Years of TESS Full Frame Images: Part II}",
      journal = {Research Notes of the American Astronomical Society},
         year = 2020,
        month = nov,
       volume = {4},
       number = {11},
          eid = {206},
        pages = {206},
          doi = {10.3847/2515-5172/abca2d},
       adsurl = {https://ui.adsabs.harvard.edu/abs/2020RNAAS...4..206H}
}

@ARTICLE{Kunimoto2021,
       author = {{Kunimoto}, Michelle and {Huang}, Chelsea and {Tey}, Evan and {Fong}, Willie and {Hesse}, Katharine and {Shporer}, Avi and {Guerrero}, Natalia and {Fausnaugh}, Michael and {Vanderspek}, Roland and {Ricker}, George},
        title = "{Quick-look Pipeline Lightcurves for 9.1 Million Stars Observed over the First Year of the TESS Extended Mission}",
      journal = {Research Notes of the American Astronomical Society},
         year = 2021,
        month = oct,
       volume = {5},
       number = {10},
          eid = {234},
        pages = {234},
          doi = {10.3847/2515-5172/ac2ef0},
archivePrefix = {arXiv},
       eprint = {2110.05542},
 primaryClass = {astro-ph.EP},
       adsurl = {https://ui.adsabs.harvard.edu/abs/2021RNAAS...5..234K}
}

@ARTICLE{Dias2014,
       author = {{Dias}, W.~S. and {Monteiro}, H. and {Caetano}, T.~C. and {L{\'e}pine}, J.~R.~D. and {Assafin}, M. and {Oliveira}, A.~F.},
        title = "{Proper motions of the optically visible open clusters based on the UCAC4 catalog}",
      journal = {\aap},
         year = 2014,
        month = apr,
       volume = {564},
          eid = {A79},
        pages = {A79},
          doi = {10.1051/0004-6361/201323226},
       adsurl = {https://ui.adsabs.harvard.edu/abs/2014A&A...564A..79D}
}

@ARTICLE{Cantat-Gaudin_asterism,
       author = {{Cantat-Gaudin}, T. and {Anders}, F.},
        title = "{Clusters and mirages: cataloguing stellar aggregates in the Milky Way}",
      journal = {\aap},
         year = 2020,
        month = jan,
       volume = {633},
          eid = {A99},
        pages = {A99},
          doi = {10.1051/0004-6361/201936691},
archivePrefix = {arXiv},
       eprint = {1911.07075},
 primaryClass = {astro-ph.SR},
       adsurl = {https://ui.adsabs.harvard.edu/abs/2020A&A...633A..99C}
}

@article{GaiaCollaboration2016GaiaProperties,
    title = {{Gaia Data Release 1. Summary of the astrometric, photometric, and survey properties}},
    year = {2016},
    journal = {A{\&}A},
    author = {{Gaia Collaboration} and Brown, A. G. A. and Vallenari, A. and Prusti, T. and de Bruijne, J. and Mignard, F. and Drimmel, R. and co-authors, 585},
    number = {A2},
    pages = {A2},
    volume = {595},
    doi = {10.1051/0004-6361/201629512},
    pmid = {9010224},
    arxivId = {1609.04172}
}

@article{GaiaCollaboration2018GaiaProperties,
    title = {{Gaia Data Release 2. Summary of the contents and survey properties}},
    year = {2018},
    journal = {Astronomy and Astrophysics},
    author = {{Gaia Collaboration} and Brown, A. G. A. and Vallenari, A. and Prusti, T. and de Bruijne, J. H. J. and Babusiaux, C. and Bailer-Jones, C. A. L.},
    number = {A1},
    pages = {A1},
    volume = {616},
    doi = {10.1051/0004-6361/201833051},
    pmid = {9010224},
    arxivId = {1804.09365}
}

@article{Kovacs2002,
    title = {{A box-fitting algorithm in the search for periodic transits}},
    year = {2002},
    journal = {A{\&}A},
    author = {Kov{\'{a}}cs, G. and Zucker, S. and Mazeh, T.},
    number = {1},
    volume = {391},
    isbn = {0902009192},
    doi = {10.1051/0004-6361:20020802},
    pmid = {9010224},
    arxivId = {astro-ph/0206099}
}

@inproceedings{Jenkins2016,
    title = {{The TESS science processing operations center}},
    year = {2016},
    booktitle = {Software and Cyberinfrastructure for Astronomy IV},
    author = {Jenkins, Jon M. and Twicken, Joseph D. and McCauliff, Sean and Campbell, Jennifer and Sanderfer, Dwight and Lung, David and Mansouri-Samani, Masoud and Girouard, Forrest and Tenenbaum, Peter and Klaus, Todd and Smith, Jeffrey C. and Caldwell, Douglas A. and Chacon, A. D. and Henze, Christopher and Heiges, Cory and Latham, David W. and Morgan, Edward and Swade, Daryl and Rinehart, Stephen and Vanderspek, Roland},
    pages = {99133E},
    volume = {9913},
    isbn = {9781510602052},
    doi = {10.1117/12.2233418},
    issn = {1996756X}
}

@ARTICLE{SERVAL2018,
       author = {{Zechmeister}, M. and {Reiners}, A. and {Amado}, P.~J. and {Azzaro}, M. and {Bauer}, F.~F. and {B{\'e}jar}, V.~J.~S. and {Caballero}, J.~A. and {Guenther}, E.~W. and {Hagen}, H. -J. and {Jeffers}, S.~V. and {Kaminski}, A. and {K{\"u}rster}, M. and {Launhardt}, R. and {Montes}, D. and {Morales}, J.~C. and {Quirrenbach}, A. and {Reffert}, S. and {Ribas}, I. and {Seifert}, W. and {Tal-Or}, L. and {Wolthoff}, V.},
        title = "{Spectrum radial velocity analyser (SERVAL). High-precision radial velocities and two alternative spectral indicators}",
      journal = {\aap},
         year = 2018,
        month = jan,
       volume = {609},
          eid = {A12},
        pages = {A12},
          doi = {10.1051/0004-6361/201731483},
archivePrefix = {arXiv},
       eprint = {1710.10114},
 primaryClass = {astro-ph.IM},
       adsurl = {https://ui.adsabs.harvard.edu/abs/2018A&A...609A..12Z}
}

@ARTICLE{Vines2022_ARIADNE,
       author = {{Vines}, Jose I. and {Jenkins}, James S.},
        title = "{ARIADNE: Measuring accurate and precise stellar parameters through SED fitting}",
      journal = {\mnras},
         year = 2022,
        month = apr,
          doi = {10.1093/mnras/stac956},
archivePrefix = {arXiv},
       eprint = {2204.03769},
 primaryClass = {astro-ph.SR},
       adsurl = {https://ui.adsabs.harvard.edu/abs/2022MNRAS.tmp..920V}
}

@INPROCEEDINGS{Skilling2004_dynesty1,
       author = {{Skilling}, John},
        title = "{Nested Sampling}",
    booktitle = {American Institute of Physics Conference Series},
         year = "2004",
       editor = {{Fischer}, Rainer and {Preuss}, Roland and {Toussaint}, Udo Von},
       series = {American Institute of Physics Conference Series},
       volume = {735},
        month = "Nov",
        pages = {395-405},
          doi = {10.1063/1.1835238},
       adsurl = {https://ui.adsabs.harvard.edu/abs/2004AIPC..735..395S}
}

@article{Skilling2006_dynesty2,
       author = "Skilling, John",
          doi = "10.1214/06-BA127",
     fjournal = "Bayesian Analysis",
      journal = "Bayesian Anal.",
        month = "12",
       number = "4",
        pages = "833--859",
    publisher = "International Society for Bayesian Analysis",
        title = "Nested sampling for general Bayesian computation",
          url = "https://doi.org/10.1214/06-BA127",
       volume = "1",
         year = "2006"
}

@article{Husser2013,
   archivePrefix = {arXiv},
         arxivId = {1303.5632},
          author = {Husser, Tim-Oliver and von Berg, Sebastian Wende - and Dreizler, Stefan and Homeier, Derek and Reiners, Ansgar and Barman, Travis and Hauschildt, Peter H},
             doi = {10.1051/0004-6361/201219058},
          eprint = {1303.5632},
            issn = {0004-6361},
         journal = {Astronomy {\&} Astrophysics},
           pages = {A6},
           title = {{A new extensive library of PHOENIX stellar atmospheres and synthetic spectra}},
             url = {http://arxiv.org/abs/1303.5632{\%}0Ahttp://dx.doi.org/10.1051/0004-6361/201219058},
          volume = {553},
            year = {2013}
}

@article{Allard2012,
      author = {Allard, F. and Homeier, D. and Freytag, B.},
         doi = {10.1098/rsta.2011.0269},
        issn = {1364503X},
     journal = {Philosophical Transactions of the Royal Society A: Mathematical, Physical and Engineering Sciences},
      number = {1968},
       pages = {2765--2777},
       title = {{Models of very-low-mass stars, brown dwarfs and exoplanets}},
      volume = {370},
        year = {2012}
}

@article{Hauschildt1999,
      author = {Hauschildt, Peter H and Allard, France and Baron, E},
         doi = {10.1086/430754},
        issn = {0004-637X},
     journal = {The Astrophysical Journal},
      number = {2},
       pages = {865--872},
       title = {{THE NEXTGEN MODEL ATMOSPHERE GRID FOR 3000 {\textless}= Teff {\textless}= 10,000 K}},
      volume = {629},
        year = {1999}
}

@ARTICLE{Kurucz1993,
       author = {{Kurucz}, Robert},
        title = "{ATLAS9 Stellar Atmosphere Programs and 2 km/s grid.}",
      journal = {ATLAS9 Stellar Atmosphere Programs and 2 km/s grid. Kurucz CD-ROM No. 13. Cambridge},
         year = "1993",
        month = "Jan",
       volume = {13},
       adsurl = {https://ui.adsabs.harvard.edu/abs/1993KurCD..13.....K}
}

@INPROCEEDINGS{Castelli2004,
       author = {{Castelli}, F. and {Kurucz}, R.~L.},
        title = "{New Grids of ATLAS9 Model Atmospheres}",
    booktitle = {Modelling of Stellar Atmospheres},
         year = {2003},
       editor = {{Piskunov}, N. and {Weiss}, W.~W. and {Gray}, D.~F.},
       volume = {210},
        month = {jan},
        pages = {A20},
        series = {},
archivePrefix = {arXiv},
       eprint = {astro-ph/0405087},
 primaryClass = {astro-ph},
       adsurl = {https://ui.adsabs.harvard.edu/abs/2003IAUS..210P.A20C}
}

@ARTICLE{Speagle2019,
       author = {{Speagle}, Joshua S.},
        title = "{DYNESTY: a dynamic nested sampling package for estimating Bayesian posteriors and evidences}",
      journal = {\mnras},
         year = 2020,
        month = apr,
       volume = {493},
       number = {3},
        pages = {3132-3158},
          doi = {10.1093/mnras/staa278},
archivePrefix = {arXiv},
       eprint = {1904.02180},
 primaryClass = {astro-ph.IM},
       adsurl = {https://ui.adsabs.harvard.edu/abs/2020MNRAS.493.3132S}
}

@ARTICLE{Skrutskie2006_2MASS,
       author = {{Skrutskie}, M.~F. and {Cutri}, R.~M. and {Stiening}, R. and {Weinberg}, M.~D. and {Schneider}, S. and {Carpenter}, J.~M. and {Beichman}, C. and {Capps}, R. and {Chester}, T. and {Elias}, J. and {Huchra}, J. and {Liebert}, J. and {Lonsdale}, C. and {Monet}, D.~G. and {Price}, S. and {Seitzer}, P. and {Jarrett}, T. and {Kirkpatrick}, J.~D. and {Gizis}, J.~E. and {Howard}, E. and {Evans}, T. and {Fowler}, J. and {Fullmer}, L. and {Hurt}, R. and {Light}, R. and {Kopan}, E.~L. and {Marsh}, K.~A. and {McCallon}, H.~L. and {Tam}, R. and {Van Dyk}, S. and {Wheelock}, S.},
        title = "{The Two Micron All Sky Survey (2MASS)}",
      journal = {\aj},
         year = 2006,
        month = feb,
       volume = {131},
       number = {2},
        pages = {1163-1183},
          doi = {10.1086/498708},
       adsurl = {https://ui.adsabs.harvard.edu/abs/2006AJ....131.1163S}
}

@ARTICLE{Wright2010_WISE,
       author = {{Wright}, Edward L. and {Eisenhardt}, Peter R.~M. and {Mainzer}, Amy K. and {Ressler}, Michael E. and {Cutri}, Roc M. and {Jarrett}, Thomas and {Kirkpatrick}, J. Davy and {Padgett}, Deborah and {McMillan}, Robert S. and {Skrutskie}, Michael and {Stanford}, S.~A. and {Cohen}, Martin and {Walker}, Russell G. and {Mather}, John C. and {Leisawitz}, David and {Gautier}, III, Thomas N. and {McLean}, Ian and {Benford}, Dominic and {Lonsdale}, Carol J. and {Blain}, Andrew and {Mendez}, Bryan and {Irace}, William R. and {Duval}, Valerie and {Liu}, Fengchuan and {Royer}, Don and {Heinrichsen}, Ingolf and {Howard}, Joan and {Shannon}, Mark and {Kendall}, Martha and {Walsh}, Amy L. and {Larsen}, Mark and {Cardon}, Joel G. and {Schick}, Scott and {Schwalm}, Mark and {Abid}, Mohamed and {Fabinsky}, Beth and {Naes}, Larry and {Tsai}, Chao-Wei},
        title = "{The Wide-field Infrared Survey Explorer (WISE): Mission Description and Initial On-orbit Performance}",
      journal = {\aj},
         year = 2010,
        month = dec,
       volume = {140},
       number = {6},
        pages = {1868-1881},
          doi = {10.1088/0004-6256/140/6/1868},
archivePrefix = {arXiv},
       eprint = {1008.0031},
 primaryClass = {astro-ph.IM},
       adsurl = {https://ui.adsabs.harvard.edu/abs/2010AJ....140.1868W}
}

@ARTICLE{Stassun2018,
       author = {{Stassun}, Keivan G. and {Oelkers}, Ryan J. and {Pepper}, Joshua and {Paegert}, Martin and {De Lee}, Nathan and {Torres}, Guillermo and {Latham}, David W. and {Charpinet}, St{\'e}phane and {Dressing}, Courtney D. and {Huber}, Daniel and {Kane}, Stephen R. and {L{\'e}pine}, S{\'e}bastien and {Mann}, Andrew and {Muirhead}, Philip S. and {Rojas-Ayala}, B{\'a}rbara and {Silvotti}, Roberto and {Fleming}, Scott W. and {Levine}, Al and {Plavchan}, Peter},
        title = "{The TESS Input Catalog and Candidate Target List}",
      journal = {\aj},
         year = 2018,
        month = sep,
       volume = {156},
       number = {3},
          eid = {102},
        pages = {102},
          doi = {10.3847/1538-3881/aad050},
archivePrefix = {arXiv},
       eprint = {1706.00495},
 primaryClass = {astro-ph.EP},
       adsurl = {https://ui.adsabs.harvard.edu/abs/2018AJ....156..102S}
}

@ARTICLE{Stassun2019,
       author = {{Stassun}, Keivan G. and {Oelkers}, Ryan J. and {Paegert}, Martin and {Torres}, Guillermo and {Pepper}, Joshua and {De Lee}, Nathan and {Collins}, Kevin and {Latham}, David W. and {Muirhead}, Philip S. and {Chittidi}, Jay and {Rojas-Ayala}, B{\'a}rbara and {Fleming}, Scott W. and {Rose}, Mark E. and {Tenenbaum}, Peter and {Ting}, Eric B. and {Kane}, Stephen R. and {Barclay}, Thomas and {Bean}, Jacob L. and {Brassuer}, C.~E. and {Charbonneau}, David and {Ge}, Jian and {Lissauer}, Jack J. and {Mann}, Andrew W. and {McLean}, Brian and {Mullally}, Susan and {Narita}, Norio and {Plavchan}, Peter and {Ricker}, George R. and {Sasselov}, Dimitar and {Seager}, S. and {Sharma}, Sanjib and {Shiao}, Bernie and {Sozzetti}, Alessandro and {Stello}, Dennis and {Vanderspek}, Roland and {Wallace}, Geoff and {Winn}, Joshua N.},
        title = "{The Revised TESS Input Catalog and Candidate Target List}",
      journal = {\aj},
         year = 2019,
        month = oct,
       volume = {158},
       number = {4},
          eid = {138},
        pages = {138},
          doi = {10.3847/1538-3881/ab3467},
archivePrefix = {arXiv},
       eprint = {1905.10694},
 primaryClass = {astro-ph.SR},
       adsurl = {https://ui.adsabs.harvard.edu/abs/2019AJ....158..138S}
}

@ARTICLE{Hog2000_Tycho2,
       author = {{H{\o}g}, E. and {Fabricius}, C. and {Makarov}, V.~V. and {Urban}, S. and {Corbin}, T. and {Wycoff}, G. and {Bastian}, U. and {Schwekendiek}, P. and {Wicenec}, A.},
        title = "{The Tycho-2 catalogue of the 2.5 million brightest stars}",
      journal = {\aap},
         year = 2000,
        month = mar,
       volume = {355},
        pages = {L27-L30},
       adsurl = {https://ui.adsabs.harvard.edu/abs/2000A&A...355L..27H}
}

@ARTICLE{Kharchenko2001_ASCC,
       author = {{Kharchenko}, N.~V.},
        title = "{All-sky compiled catalogue of 2.5 million stars}",
      journal = {Kinematika i Fizika Nebesnykh Tel},
         year = 2001,
        month = oct,
       volume = {17},
       number = {5},
        pages = {409-423},
       adsurl = {https://ui.adsabs.harvard.edu/abs/2001KFNT...17..409K}
}

@ARTICLE{Chambers2016,
       author = {{Chambers}, K.~C. and {Magnier}, E.~A. and {Metcalfe}, N. and {Flewelling}, H.~A. and {Huber}, M.~E. and {Waters}, C.~Z. and {Denneau}, L. and {Draper}, P.~W. and {Farrow}, D. and {Finkbeiner}, D.~P. and {Holmberg}, C. and {Koppenhoefer}, J. and {Price}, P.~A. and {Rest}, A. and {Saglia}, R.~P. and {Schlafly}, E.~F. and {Smartt}, S.~J. and {Sweeney}, W. and {Wainscoat}, R.~J. and {Burgett}, W.~S. and {Chastel}, S. and {Grav}, T. and {Heasley}, J.~N. and {Hodapp}, K.~W. and {Jedicke}, R. and {Kaiser}, N. and {Kudritzki}, R. -P. and {Luppino}, G.~A. and {Lupton}, R.~H. and {Monet}, D.~G. and {Morgan}, J.~S. and {Onaka}, P.~M. and {Shiao}, B. and {Stubbs}, C.~W. and {Tonry}, J.~L. and {White}, R. and {Ba{\~n}ados}, E. and {Bell}, E.~F. and {Bender}, R. and {Bernard}, E.~J. and {Boegner}, M. and {Boffi}, F. and {Botticella}, M.~T. and {Calamida}, A. and {Casertano}, S. and {Chen}, W. -P. and {Chen}, X. and {Cole}, S. and {Deacon}, N. and {Frenk}, C. and {Fitzsimmons}, A. and {Gezari}, S. and {Gibbs}, V. and {Goessl}, C. and {Goggia}, T. and {Gourgue}, R. and {Goldman}, B. and {Grant}, P. and {Grebel}, E.~K. and {Hambly}, N.~C. and {Hasinger}, G. and {Heavens}, A.~F. and {Heckman}, T.~M. and {Henderson}, R. and {Henning}, T. and {Holman}, M. and {Hopp}, U. and {Ip}, W. -H. and {Isani}, S. and {Jackson}, M. and {Keyes}, C.~D. and {Koekemoer}, A.~M. and {Kotak}, R. and {Le}, D. and {Liska}, D. and {Long}, K.~S. and {Lucey}, J.~R. and {Liu}, M. and {Martin}, N.~F. and {Masci}, G. and {McLean}, B. and {Mindel}, E. and {Misra}, P. and {Morganson}, E. and {Murphy}, D.~N.~A. and {Obaika}, A. and {Narayan}, G. and {Nieto-Santisteban}, M.~A. and {Norberg}, P. and {Peacock}, J.~A. and {Pier}, E.~A. and {Postman}, M. and {Primak}, N. and {Rae}, C. and {Rai}, A. and {Riess}, A. and {Riffeser}, A. and {Rix}, H.~W. and {R{\"o}ser}, S. and {Russel}, R. and {Rutz}, L. and {Schilbach}, E. and {Schultz}, A.~S.~B. and {Scolnic}, D. and {Strolger}, L. and {Szalay}, A. and {Seitz}, S. and {Small}, E. and {Smith}, K.~W. and {Soderblom}, D.~R. and {Taylor}, P. and {Thomson}, R. and {Taylor}, A.~N. and {Thakar}, A.~R. and {Thiel}, J. and {Thilker}, D. and {Unger}, D. and {Urata}, Y. and {Valenti}, J. and {Wagner}, J. and {Walder}, T. and {Walter}, F. and {Watters}, S.~P. and {Werner}, S. and {Wood-Vasey}, W.~M. and {Wyse}, R.},
        title = "{The Pan-STARRS1 Surveys}",
      journal = {arXiv e-prints},
         year = 2016,
        month = dec,
          eid = {arXiv:1612.05560},
        pages = {arXiv:1612.05560},
          doi = {10.48550/arXiv.1612.05560},
archivePrefix = {arXiv},
       eprint = {1612.05560},
 primaryClass = {astro-ph.IM},
       adsurl = {https://ui.adsabs.harvard.edu/abs/2016arXiv161205560C}
}

@ARTICLE{Churchwell2009,
       author = {{Churchwell}, Ed and {Babler}, Brian L. and {Meade}, Marilyn R. and {Whitney}, Barbara A. and {Benjamin}, Robert and {Indebetouw}, Remy and {Cyganowski}, Claudia and {Robitaille}, Thomas P. and {Povich}, Matthew and {Watson}, Christer and {Bracker}, Steve},
        title = "{The Spitzer/GLIMPSE Surveys: A New View of the Milky Way}",
      journal = {\pasp},
         year = 2009,
        month = mar,
       volume = {121},
       number = {877},
        pages = {213},
          doi = {10.1086/597811},
       adsurl = {https://ui.adsabs.harvard.edu/abs/2009PASP..121..213C}
}

@ARTICLE{Bianchi2011_GALEX,
       author = {{Bianchi}, L. and {Herald}, J. and {Efremova}, B. and {Girardi}, L. and {Zabot}, A. and {Marigo}, P. and {Conti}, A. and {Shiao}, B.},
        title = "{GALEX catalogs of UV sources: statistical properties and sample science applications: hot white dwarfs in the Milky Way}",
      journal = {\apss},
         year = 2011,
        month = sep,
       volume = {335},
       number = {1},
        pages = {161-169},
          doi = {10.1007/s10509-010-0581-x},
       adsurl = {https://ui.adsabs.harvard.edu/abs/2011Ap&SS.335..161B}
}

@ARTICLE{Henden2014_APASS,
       author = {{Henden}, A. and {Munari}, U.},
        title = "{The APASS all-sky, multi-epoch BVgri photometric survey}",
      journal = {Contributions of the Astronomical Observatory Skalnate Pleso},
         year = 2014,
        month = mar,
       volume = {43},
       number = {3},
        pages = {518-522},
       adsurl = {https://ui.adsabs.harvard.edu/abs/2014CoSka..43..518H}
}

@ARTICLE{2015ApJS..219...12A,
       author = {{Alam}, Shadab and {Albareti}, Franco D. and {Allende Prieto}, Carlos and {Anders}, F. and {Anderson}, Scott F. and {Anderton}, Timothy and {Andrews}, Brett H. and {Armengaud}, Eric and {Aubourg}, {\'E}ric and {Bailey}, Stephen and {Basu}, Sarbani and {Bautista}, Julian E. and {Beaton}, Rachael L. and {Beers}, Timothy C. and {Bender}, Chad F. and {Berlind}, Andreas A. and {Beutler}, Florian and {Bhardwaj}, Vaishali and {Bird}, Jonathan C. and {Bizyaev}, Dmitry and {Blake}, Cullen H. and {Blanton}, Michael R. and {Blomqvist}, Michael and {Bochanski}, John J. and {Bolton}, Adam S. and {Bovy}, Jo and {Shelden Bradley}, A. and {Brandt}, W.~N. and {Brauer}, D.~E. and {Brinkmann}, J. and {Brown}, Peter J. and {Brownstein}, Joel R. and {Burden}, Angela and {Burtin}, Etienne and {Busca}, Nicol{\'a}s G. and {Cai}, Zheng and {Capozzi}, Diego and {Carnero Rosell}, Aurelio and {Carr}, Michael A. and {Carrera}, Ricardo and {Chambers}, K.~C. and {Chaplin}, William James and {Chen}, Yen-Chi and {Chiappini}, Cristina and {Chojnowski}, S. Drew and {Chuang}, Chia-Hsun and {Clerc}, Nicolas and {Comparat}, Johan and {Covey}, Kevin and {Croft}, Rupert A.~C. and {Cuesta}, Antonio J. and {Cunha}, Katia and {da Costa}, Luiz N. and {Da Rio}, Nicola and {Davenport}, James R.~A. and {Dawson}, Kyle S. and {De Lee}, Nathan and {Delubac}, Timoth{\'e}e and {Deshpande}, Rohit and {Dhital}, Saurav and {Dutra-Ferreira}, Let{\'\i}cia and {Dwelly}, Tom and {Ealet}, Anne and {Ebelke}, Garrett L. and {Edmondson}, Edward M. and {Eisenstein}, Daniel J. and {Ellsworth}, Tristan and {Elsworth}, Yvonne and {Epstein}, Courtney R. and {Eracleous}, Michael and {Escoffier}, Stephanie and {Esposito}, Massimiliano and {Evans}, Michael L. and {Fan}, Xiaohui and {Fern{\'a}ndez-Alvar}, Emma and {Feuillet}, Diane and {Filiz Ak}, Nurten and {Finley}, Hayley and {Finoguenov}, Alexis and {Flaherty}, Kevin and {Fleming}, Scott W. and {Font-Ribera}, Andreu and {Foster}, Jonathan and {Frinchaboy}, Peter M. and {Galbraith-Frew}, J.~G. and {Garc{\'\i}a}, Rafael A. and {Garc{\'\i}a-Hern{\'a}ndez}, D.~A. and {Garc{\'\i}a P{\'e}rez}, Ana E. and {Gaulme}, Patrick and {Ge}, Jian and {G{\'e}nova-Santos}, R. and {Georgakakis}, A. and {Ghezzi}, Luan and {Gillespie}, Bruce A. and {Girardi}, L{\'e}o and {Goddard}, Daniel and {Gontcho}, Satya Gontcho A. and {Gonz{\'a}lez Hern{\'a}ndez}, Jonay I. and {Grebel}, Eva K. and {Green}, Paul J. and {Grieb}, Jan Niklas and {Grieves}, Nolan and {Gunn}, James E. and {Guo}, Hong and {Harding}, Paul and {Hasselquist}, Sten and {Hawley}, Suzanne L. and {Hayden}, Michael and {Hearty}, Fred R. and {Hekker}, Saskia and {Ho}, Shirley and {Hogg}, David W. and {Holley-Bockelmann}, Kelly and {Holtzman}, Jon A. and {Honscheid}, Klaus and {Huber}, Daniel and {Huehnerhoff}, Joseph and {Ivans}, Inese I. and {Jiang}, Linhua and {Johnson}, Jennifer A. and {Kinemuchi}, Karen and {Kirkby}, David and {Kitaura}, Francisco and {Klaene}, Mark A. and {Knapp}, Gillian R. and {Kneib}, Jean-Paul and {Koenig}, Xavier P. and {Lam}, Charles R. and {Lan}, Ting-Wen and {Lang}, Dustin and {Laurent}, Pierre and {Le Goff}, Jean-Marc and {Leauthaud}, Alexie and {Lee}, Khee-Gan and {Lee}, Young Sun and {Licquia}, Timothy C. and {Liu}, Jian and {Long}, Daniel C. and {L{\'o}pez-Corredoira}, Mart{\'\i}n and {Lorenzo-Oliveira}, Diego and {Lucatello}, Sara and {Lundgren}, Britt and {Lupton}, Robert H. and {Mack}, III, Claude E. and {Mahadevan}, Suvrath and {Maia}, Marcio A.~G. and {Majewski}, Steven R. and {Malanushenko}, Elena and {Malanushenko}, Viktor and {Manchado}, A. and {Manera}, Marc and {Mao}, Qingqing and {Maraston}, Claudia and {Marchwinski}, Robert C. and {Margala}, Daniel and {Martell}, Sarah L. and {Martig}, Marie and {Masters}, Karen L. and {Mathur}, Savita and {McBride}, Cameron K. and {McGehee}, Peregrine M. and {McGreer}, Ian D. and {McMahon}, Richard G. and {M{\'e}nard}, Brice and {Menzel}, Marie-Luise and {Merloni}, Andrea and {M{\'e}sz{\'a}ros}, Szabolcs and {Miller}, Adam A. and {Miralda-Escud{\'e}}, Jordi and {Miyatake}, Hironao and {Montero-Dorta}, Antonio D. and {More}, Surhud and {Morganson}, Eric and {Morice-Atkinson}, Xan and {Morrison}, Heather L. and {Mosser}, Ben{\^o}it and {Muna}, Demitri and {Myers}, Adam D. and {Nandra}, Kirpal and {Newman}, Jeffrey A. and {Neyrinck}, Mark and {Nguyen}, Duy Cuong and {Nichol}, Robert C. and {Nidever}, David L. and {Noterdaeme}, Pasquier and {Nuza}, Sebasti{\'a}n E. and {O'Connell}, Julia E. and {O'Connell}, Robert W. and {O'Connell}, Ross and {Ogando}, Ricardo L.~C. and {Olmstead}, Matthew D. and {Oravetz}, Audrey E. and {Oravetz}, Daniel J. and {Osumi}, Keisuke and {Owen}, Russell and {Padgett}, Deborah L. and {Padmanabhan}, Nikhil and {Paegert}, Martin and {Palanque-Delabrouille}, Nathalie and {Pan}, Kaike},
        title = "{The Eleventh and Twelfth Data Releases of the Sloan Digital Sky Survey: Final Data from SDSS-III}",
      journal = {\apjs},
         year = 2015,
        month = jul,
       volume = {219},
       number = {1},
          eid = {12},
        pages = {12},
          doi = {10.1088/0067-0049/219/1/12},
archivePrefix = {arXiv},
       eprint = {1501.00963},
 primaryClass = {astro-ph.IM},
       adsurl = {https://ui.adsabs.harvard.edu/abs/2015ApJS..219...12A}
}

@ARTICLE{Paunzen2015,
       author = {{Paunzen}, E.},
        title = "{A new catalogue of Str{\"o}mgren-Crawford uvby{\ensuremath{\beta}} photometry}",
      journal = {\aap},
         year = 2015,
        month = aug,
       volume = {580},
          eid = {A23},
        pages = {A23},
          doi = {10.1051/0004-6361/201526413},
archivePrefix = {arXiv},
       eprint = {1506.04568},
 primaryClass = {astro-ph.SR},
       adsurl = {https://ui.adsabs.harvard.edu/abs/2015A&A...580A..23P}
}

@ARTICLE{Battley2020_YSD,
       author = {{Battley}, Matthew P. and {Pollacco}, Don and {Armstrong}, David J.},
        title = "{A search for young exoplanets in Sectors 1-5 of the TESS full-frame images}",
      journal = {\mnras},
         year = 2020,
        month = aug,
       volume = {496},
       number = {2},
        pages = {1197-1216},
          doi = {10.1093/mnras/staa1626},
archivePrefix = {arXiv},
       eprint = {2006.01721},
 primaryClass = {astro-ph.EP},
       adsurl = {https://ui.adsabs.harvard.edu/abs/2020MNRAS.496.1197B}
}

@MISC{Lightkurve2018,
   author = {{Lightkurve Collaboration} and {Cardoso}, J.~V.~d.~M. and
             {Hedges}, C. and {Gully-Santiago}, M. and {Saunders}, N. and
             {Cody}, A.~M. and {Barclay}, T. and {Hall}, O. and
             {Sagear}, S. and {Turtelboom}, E. and {Zhang}, J. and
             {Tzanidakis}, A. and {Mighell}, K. and {Coughlin}, J. and
             {Bell}, K. and {Berta-Thompson}, Z. and {Williams}, P. and
             {Dotson}, J. and {Barentsen}, G.},
    title = "{Lightkurve: Kepler and TESS time series analysis in Python}",
howpublished = {Astrophysics Source Code Library},
     year = 2018,
    month = dec,
archivePrefix = "ascl",
   eprint = {1812.013},
   adsurl = {http://adsabs.harvard.edu/abs/2018ascl.soft12013L},
}

@INPROCEEDINGS{Vogt1994_HIRES,
       author = {{Vogt}, S.~S. and {Allen}, S.~L. and {Bigelow}, B.~C. and {Bresee}, L. and {Brown}, B. and {Cantrall}, T. and {Conrad}, A. and {Couture}, M. and {Delaney}, C. and {Epps}, H.~W. and {Hilyard}, D. and {Hilyard}, D.~F. and {Horn}, E. and {Jern}, N. and {Kanto}, D. and {Keane}, M.~J. and {Kibrick}, R.~I. and {Lewis}, J.~W. and {Osborne}, J. and {Pardeilhan}, G.~H. and {Pfister}, T. and {Ricketts}, T. and {Robinson}, L.~B. and {Stover}, R.~J. and {Tucker}, D. and {Ward}, J. and {Wei}, M.~Z.},
        title = "{HIRES: the high-resolution echelle spectrometer on the Keck 10-m Telescope}",
    booktitle = {Instrumentation in Astronomy VIII},
         year = 1994,
       editor = {{Crawford}, David L. and {Craine}, Eric R.},
       series = {Society of Photo-Optical Instrumentation Engineers (SPIE) Conference Series},
       volume = {2198},
        month = jun,
        pages = {362},
          doi = {10.1117/12.176725},
       adsurl = {https://ui.adsabs.harvard.edu/abs/1994SPIE.2198..362V}
}

@ARTICLE{Panahi2022,
       author = {{Panahi}, Aviad and {Mazeh}, Tsevi and {Zucker}, Shay and {Latham}, David W. and {Collins}, Karen A. and {Rimoldini}, Lorenzo and {Evans}, Dafydd Wyn and {Eyer}, Laurent},
        title = "{Gaia-TESS synergy: improving the identification of transit candidates}",
      journal = {\aap},
         year = 2022,
        month = nov,
       volume = {667},
          eid = {A14},
        pages = {A14},
          doi = {10.1051/0004-6361/202244207},
archivePrefix = {arXiv},
       eprint = {2209.05845},
 primaryClass = {astro-ph.EP},
       adsurl = {https://ui.adsabs.harvard.edu/abs/2022A&A...667A..14P}
}

@ARTICLE{Safonov2017,
       author = {{Safonov}, B.~S. and {Lysenko}, P.~A. and {Dodin}, A.~V.},
        title = "{The speckle polarimeter of the 2.5-m telescope: Design and calibration}",
      journal = {Astronomy Letters},
         year = 2017,
        month = may,
       volume = {43},
       number = {5},
        pages = {344-364},
          doi = {10.1134/S1063773717050036},
       adsurl = {https://ui.adsabs.harvard.edu/abs/2017AstL...43..344S}
}

@article{Lomb1976Least-squaresData,
    title = {{Least-squares frequency analysis of unequally spaced data}},
    year = {1976},
    journal = {Astrophysics and Space Science},
    author = {Lomb, N.R.},
    pages = {447--462},
    volume = {39}
}

@article{Scargle1982StudiesData,
    title = {{Studies in Astronomical Time Series Analysis II. Statistical Aspects of Spectral Analysis of Unevenly Spaced Data}},
    year = {1982},
    journal = {Astrophysical Journal},
    author = {Scargle, J.D.},
    pages = {835--853},
    volume = {263}
}

@ARTICLE{Baluev2008,
       author = {{Baluev}, R.~V.},
        title = "{Assessing the statistical significance of periodogram peaks}",
      journal = {\mnras},
         year = 2008,
        month = apr,
       volume = {385},
       number = {3},
        pages = {1279-1285},
          doi = {10.1111/j.1365-2966.2008.12689.x},
archivePrefix = {arXiv},
       eprint = {0711.0330},
 primaryClass = {astro-ph},
       adsurl = {https://ui.adsabs.harvard.edu/abs/2008MNRAS.385.1279B}
}

@PHDTHESIS{Petigura2015,
       author = {{Petigura}, Erik Ardeshir},
        title = "{Prevalence of Earth-size Planets Orbiting Sun-like Stars}",
       school = {University of California, Berkeley},
         year = 2015,
        month = jan,
       adsurl = {https://ui.adsabs.harvard.edu/abs/2015PhDT........82P}
}

@article{Queloz2000,
    title = {{The CORALIE survey for southern extra-solar planets I. A planet orbiting the star Gliese 86}},
    year = {2000},
    journal = {Astronomy and Astrophysics},
    author = {Queloz, Didier and Mayor, M and Weber, Lizzy and Burnet, M and Confino, B and Naef, D and Pepe, F},
    month = {2},
    pages = {99--102},
    volume = {354}
}

@article{Pepe2002TheVII,
    title = {{The CORALIE survey for southern extra-solar planets VII}},
    year = {2002},
    journal = {Astronomy {\&} Astrophysics},
    author = {Pepe, F. and Mayor, M. and Galland, F. and Naef, D. and Queloz, D. and Santos, N. C. and Udry, S. and Burnet, M.},
    number = {2},
    month = {6},
    pages = {632--638},
    volume = {388},
    doi = {10.1051/0004-6361:20020433},
    issn = {0004-6361}
}

@ARTICLE{Blanco-Cuaresma14,
       author = {{Blanco-Cuaresma}, S. and {Soubiran}, C. and {Heiter}, U. and {Jofr{\'e}}, P.},
        title = "{Determining stellar atmospheric parameters and chemical abundances of FGK stars with iSpec}",
      journal = {\aap},
         year = 2014,
        month = sep,
       volume = {569},
          eid = {A111},
        pages = {A111},
          doi = {10.1051/0004-6361/201423945},
archivePrefix = {arXiv},
       eprint = {1407.2608},
 primaryClass = {astro-ph.IM},
       adsurl = {https://ui.adsabs.harvard.edu/abs/2014A&A...569A.111B}
}

@ARTICLE{Tabernero22,
       author = {{Tabernero}, H.~M. and {Marfil}, E. and {Montes}, D. and {Gonz{\'a}lez Hern{\'a}ndez}, J.~I.},
        title = "{STEPARSYN: A Bayesian code to infer stellar atmospheric parameters using spectral synthesis}",
      journal = {\aap},
         year = 2022,
        month = jan,
       volume = {657},
          eid = {A66},
        pages = {A66},
          doi = {10.1051/0004-6361/202141763},
archivePrefix = {arXiv},
       eprint = {2110.00444},
 primaryClass = {astro-ph.SR},
       adsurl = {https://ui.adsabs.harvard.edu/abs/2022A&A...657A..66T}
}

@ARTICLE{Addison2019_Minerva,
       author = {{Addison}, Brett and {Wright}, Duncan J. and {Wittenmyer}, Robert A. and {Horner}, Jonathan and {Mengel}, Matthew W. and {Johns}, Daniel and {Marti}, Connor and {Nicholson}, Belinda and {Soutter}, Jack and {Bowler}, Brendan and {Crossfield}, Ian and {Kane}, Stephen R. and {Kielkopf}, John and {Plavchan}, Peter and {Tinney}, C.~G. and {Zhang}, Hui and {Clark}, Jake T. and {Clerte}, Mathieu and {Eastman}, Jason D. and {Swift}, Jon and {Bottom}, Michael and {Muirhead}, Philip and {McCrady}, Nate and {Herzig}, Erich and {Hogstrom}, Kristina and {Wilson}, Maurice and {Sliski}, David and {Johnson}, Samson A. and {Wright}, Jason T. and {Johnson}, John Asher and {Blake}, Cullen and {Riddle}, Reed and {Lin}, Brian and {Cornachione}, Matthew and {Bedding}, Timothy R. and {Stello}, Dennis and {Huber}, Daniel and {Marsden}, Stephen and {Carter}, Bradley D.},
        title = "{Minerva-Australis. I. Design, Commissioning, and First Photometric Results}",
      journal = {\pasp},
         year = 2019,
        month = nov,
       volume = {131},
       number = {1005},
        pages = {115003},
          doi = {10.1088/1538-3873/ab03aa},
archivePrefix = {arXiv},
       eprint = {1901.11231},
 primaryClass = {astro-ph.IM},
       adsurl = {https://ui.adsabs.harvard.edu/abs/2019PASP..131k5003A}
}

@article{Rickeretal.2014TheSatellite,
    title = {{The Transiting Exoplanet Survey Satellite}},
    year = {2014},
    journal = {JATIS},
    author = {{Ricker et al.}},
    number = {1},
    pages = {014003 },
    volume = {1},
    arxivId = {1406.0151}
}

@ARTICLE{Collins2017,
       author = {{Collins}, Karen A. and {Kielkopf}, John F. and {Stassun}, Keivan G. and {Hessman}, Frederic V.},
        title = "{AstroImageJ: Image Processing and Photometric Extraction for Ultra-precise Astronomical Light Curves}",
      journal = {\aj},
         year = 2017,
        month = feb,
       volume = {153},
       number = {2},
          eid = {77},
        pages = {77},
          doi = {10.3847/1538-3881/153/2/77},
archivePrefix = {arXiv},
       eprint = {1701.04817},
 primaryClass = {astro-ph.IM},
       adsurl = {https://ui.adsabs.harvard.edu/abs/2017AJ....153...77C}
}

@article{astropy:2013,
Adsurl = {http://adsabs.harvard.edu/abs/2013A%26A...558A..33A},
Archiveprefix = {arXiv},
Author = {{Astropy Collaboration} and {Robitaille}, T.~P. and {Tollerud}, E.~J. and {Greenfield}, P. and {Droettboom}, M. and {Bray}, E. and {Aldcroft}, T. and {Davis}, M. and {Ginsburg}, A. and {Price-Whelan}, A.~M. and {Kerzendorf}, W.~E. and {Conley}, A. and {Crighton}, N. and {Barbary}, K. and {Muna}, D. and {Ferguson}, H. and {Grollier}, F. and {Parikh}, M.~M. and {Nair}, P.~H. and {Unther}, H.~M. and {Deil}, C. and {Woillez}, J. and {Conseil}, S. and {Kramer}, R. and {Turner}, J.~E.~H. and {Singer}, L. and {Fox}, R. and {Weaver}, B.~A. and {Zabalza}, V. and {Edwards}, Z.~I. and {Azalee Bostroem}, K. and {Burke}, D.~J. and {Casey}, A.~R. and {Crawford}, S.~M. and {Dencheva}, N. and {Ely}, J. and {Jenness}, T. and {Labrie}, K. and {Lim}, P.~L. and {Pierfederici}, F. and {Pontzen}, A. and {Ptak}, A. and {Refsdal}, B. and {Servillat}, M. and {Streicher}, O.},
Doi = {10.1051/0004-6361/201322068},
Eid = {A33},
Eprint = {1307.6212},
Journal = {\aap},
Month = oct,
Pages = {A33},
Primaryclass = {astro-ph.IM},
Title = {{Astropy: A community Python package for astronomy}},
Volume = 558,
Year = 2013}

@ARTICLE{astropy:2018,
       author = {{Astropy Collaboration} and {Price-Whelan}, A.~M. and
         {Sip{\H{o}}cz}, B.~M. and {G{\"u}nther}, H.~M. and {Lim}, P.~L. and
         {Crawford}, S.~M. and {Conseil}, S. and {Shupe}, D.~L. and
         {Craig}, M.~W. and {Dencheva}, N. and {Ginsburg}, A. and {Vand
        erPlas}, J.~T. and {Bradley}, L.~D. and {P{\'e}rez-Su{\'a}rez}, D. and
         {de Val-Borro}, M. and {Aldcroft}, T.~L. and {Cruz}, K.~L. and
         {Robitaille}, T.~P. and {Tollerud}, E.~J. and {Ardelean}, C. and
         {Babej}, T. and {Bach}, Y.~P. and {Bachetti}, M. and {Bakanov}, A.~V. and
         {Bamford}, S.~P. and {Barentsen}, G. and {Barmby}, P. and
         {Baumbach}, A. and {Berry}, K.~L. and {Biscani}, F. and {Boquien}, M. and
         {Bostroem}, K.~A. and {Bouma}, L.~G. and {Brammer}, G.~B. and
         {Bray}, E.~M. and {Breytenbach}, H. and {Buddelmeijer}, H. and
         {Burke}, D.~J. and {Calderone}, G. and {Cano Rodr{\'\i}guez}, J.~L. and
         {Cara}, M. and {Cardoso}, J.~V.~M. and {Cheedella}, S. and {Copin}, Y. and
         {Corrales}, L. and {Crichton}, D. and {D'Avella}, D. and {Deil}, C. and
         {Depagne}, {\'E}. and {Dietrich}, J.~P. and {Donath}, A. and
         {Droettboom}, M. and {Earl}, N. and {Erben}, T. and {Fabbro}, S. and
         {Ferreira}, L.~A. and {Finethy}, T. and {Fox}, R.~T. and
         {Garrison}, L.~H. and {Gibbons}, S.~L.~J. and {Goldstein}, D.~A. and
         {Gommers}, R. and {Greco}, J.~P. and {Greenfield}, P. and
         {Groener}, A.~M. and {Grollier}, F. and {Hagen}, A. and {Hirst}, P. and
         {Homeier}, D. and {Horton}, A.~J. and {Hosseinzadeh}, G. and {Hu}, L. and
         {Hunkeler}, J.~S. and {Ivezi{\'c}}, {\v{Z}}. and {Jain}, A. and
         {Jenness}, T. and {Kanarek}, G. and {Kendrew}, S. and {Kern}, N.~S. and
         {Kerzendorf}, W.~E. and {Khvalko}, A. and {King}, J. and {Kirkby}, D. and
         {Kulkarni}, A.~M. and {Kumar}, A. and {Lee}, A. and {Lenz}, D. and
         {Littlefair}, S.~P. and {Ma}, Z. and {Macleod}, D.~M. and
         {Mastropietro}, M. and {McCully}, C. and {Montagnac}, S. and
         {Morris}, B.~M. and {Mueller}, M. and {Mumford}, S.~J. and {Muna}, D. and
         {Murphy}, N.~A. and {Nelson}, S. and {Nguyen}, G.~H. and
         {Ninan}, J.~P. and {N{\"o}the}, M. and {Ogaz}, S. and {Oh}, S. and
         {Parejko}, J.~K. and {Parley}, N. and {Pascual}, S. and {Patil}, R. and
         {Patil}, A.~A. and {Plunkett}, A.~L. and {Prochaska}, J.~X. and
         {Rastogi}, T. and {Reddy Janga}, V. and {Sabater}, J. and
         {Sakurikar}, P. and {Seifert}, M. and {Sherbert}, L.~E. and
         {Sherwood-Taylor}, H. and {Shih}, A.~Y. and {Sick}, J. and
         {Silbiger}, M.~T. and {Singanamalla}, S. and {Singer}, L.~P. and
         {Sladen}, P.~H. and {Sooley}, K.~A. and {Sornarajah}, S. and
         {Streicher}, O. and {Teuben}, P. and {Thomas}, S.~W. and
         {Tremblay}, G.~R. and {Turner}, J.~E.~H. and {Terr{\'o}n}, V. and
         {van Kerkwijk}, M.~H. and {de la Vega}, A. and {Watkins}, L.~L. and
         {Weaver}, B.~A. and {Whitmore}, J.~B. and {Woillez}, J. and
         {Zabalza}, V. and {Astropy Contributors}},
        title = "{The Astropy Project: Building an Open-science Project and Status of the v2.0 Core Package}",
      journal = {\aj},
         year = 2018,
        month = sep,
       volume = {156},
       number = {3},
          eid = {123},
        pages = {123},
          doi = {10.3847/1538-3881/aabc4f},
archivePrefix = {arXiv},
       eprint = {1801.02634},
 primaryClass = {astro-ph.IM},
       adsurl = {https://ui.adsabs.harvard.edu/abs/2018AJ....156..123A}
}

@ARTICLE{astropy:2022,
       author = {{Astropy Collaboration} and {Price-Whelan}, Adrian M. and {Lim}, Pey Lian and {Earl}, Nicholas and {Starkman}, Nathaniel and {Bradley}, Larry and {Shupe}, David L. and {Patil}, Aarya A. and {Corrales}, Lia and {Brasseur}, C.~E. and {N{"o}the}, Maximilian and {Donath}, Axel and {Tollerud}, Erik and {Morris}, Brett M. and {Ginsburg}, Adam and {Vaher}, Eero and {Weaver}, Benjamin A. and {Tocknell}, James and {Jamieson}, William and {van Kerkwijk}, Marten H. and {Robitaille}, Thomas P. and {Merry}, Bruce and {Bachetti}, Matteo and {G{"u}nther}, H. Moritz and {Aldcroft}, Thomas L. and {Alvarado-Montes}, Jaime A. and {Archibald}, Anne M. and {B{'o}di}, Attila and {Bapat}, Shreyas and {Barentsen}, Geert and {Baz{'a}n}, Juanjo and {Biswas}, Manish and {Boquien}, M{'e}d{'e}ric and {Burke}, D.~J. and {Cara}, Daria and {Cara}, Mihai and {Conroy}, Kyle E. and {Conseil}, Simon and {Craig}, Matthew W. and {Cross}, Robert M. and {Cruz}, Kelle L. and {D'Eugenio}, Francesco and {Dencheva}, Nadia and {Devillepoix}, Hadrien A.~R. and {Dietrich}, J{"o}rg P. and {Eigenbrot}, Arthur Davis and {Erben}, Thomas and {Ferreira}, Leonardo and {Foreman-Mackey}, Daniel and {Fox}, Ryan and {Freij}, Nabil and {Garg}, Suyog and {Geda}, Robel and {Glattly}, Lauren and {Gondhalekar}, Yash and {Gordon}, Karl D. and {Grant}, David and {Greenfield}, Perry and {Groener}, Austen M. and {Guest}, Steve and {Gurovich}, Sebastian and {Handberg}, Rasmus and {Hart}, Akeem and {Hatfield-Dodds}, Zac and {Homeier}, Derek and {Hosseinzadeh}, Griffin and {Jenness}, Tim and {Jones}, Craig K. and {Joseph}, Prajwel and {Kalmbach}, J. Bryce and {Karamehmetoglu}, Emir and {Ka{l}uszy{'n}ski}, Miko{l}aj and {Kelley}, Michael S.~P. and {Kern}, Nicholas and {Kerzendorf}, Wolfgang E. and {Koch}, Eric W. and {Kulumani}, Shankar and {Lee}, Antony and {Ly}, Chun and {Ma}, Zhiyuan and {MacBride}, Conor and {Maljaars}, Jakob M. and {Muna}, Demitri and {Murphy}, N.~A. and {Norman}, Henrik and {O'Steen}, Richard and {Oman}, Kyle A. and {Pacifici}, Camilla and {Pascual}, Sergio and {Pascual-Granado}, J. and {Patil}, Rohit R. and {Perren}, Gabriel I. and {Pickering}, Timothy E. and {Rastogi}, Tanuj and {Roulston}, Benjamin R. and {Ryan}, Daniel F. and {Rykoff}, Eli S. and {Sabater}, Jose and {Sakurikar}, Parikshit and {Salgado}, Jes{'u}s and {Sanghi}, Aniket and {Saunders}, Nicholas and {Savchenko}, Volodymyr and {Schwardt}, Ludwig and {Seifert-Eckert}, Michael and {Shih}, Albert Y. and {Jain}, Anany Shrey and {Shukla}, Gyanendra and {Sick}, Jonathan and {Simpson}, Chris and {Singanamalla}, Sudheesh and {Singer}, Leo P. and {Singhal}, Jaladh and {Sinha}, Manodeep and {Sip{H{o}}cz}, Brigitta M. and {Spitler}, Lee R. and {Stansby}, David and {Streicher}, Ole and {{{S}}umak}, Jani and {Swinbank}, John D. and {Taranu}, Dan S. and {Tewary}, Nikita and {Tremblay}, Grant R. and {Val-Borro}, Miguel de and {Van Kooten}, Samuel J. and {Vasovi{'c}}, Zlatan and {Verma}, Shresth and {de Miranda Cardoso}, Jos{'e} Vin{'i}cius and {Williams}, Peter K.~G. and {Wilson}, Tom J. and {Winkel}, Benjamin and {Wood-Vasey}, W.~M. and {Xue}, Rui and {Yoachim}, Peter and {Zhang}, Chen and {Zonca}, Andrea and {Astropy Project Contributors}},
        title = "{The Astropy Project: Sustaining and Growing a Community-oriented Open-source Project and the Latest Major Release (v5.0) of the Core Package}",
      journal = {\apj},
         year = 2022,
        month = aug,
       volume = {935},
       number = {2},
          eid = {167},
        pages = {167},
          doi = {10.3847/1538-4357/ac7c74},
archivePrefix = {arXiv},
       eprint = {2206.14220},
 primaryClass = {astro-ph.IM},
       adsurl = {https://ui.adsabs.harvard.edu/abs/2022ApJ...935..167A}
}

@ARTICLE{Boyle2025_TESS_rotation,
       author = {{Boyle}, Andrew W. and {Mann}, Andrew W. and {Bush}, Jonathan},
        title = "{Quantifying the Limits of TESS Stellar Rotation Measurements with the K2-TESS Overlap}",
      journal = {\apj},
         year = 2025,
        month = jun,
       volume = {985},
       number = {2},
          eid = {233},
        pages = {233},
          doi = {10.3847/1538-4357/adcecc},
archivePrefix = {arXiv},
       eprint = {2504.13262},
 primaryClass = {astro-ph.SR},
       adsurl = {https://ui.adsabs.harvard.edu/abs/2025ApJ...985..233B}
}

@ARTICLE{Dawson2018,
       author = {{Dawson}, Rebekah I. and {Johnson}, John Asher},
        title = "{Origins of Hot Jupiters}",
      journal = {\araa},
         year = 2018,
        month = sep,
       volume = {56},
        pages = {175-221},
          doi = {10.1146/annurev-astro-081817-051853},
archivePrefix = {arXiv},
       eprint = {1801.06117},
 primaryClass = {astro-ph.EP},
       adsurl = {https://ui.adsabs.harvard.edu/abs/2018ARA&A..56..175D}
}

@ARTICLE{Nielsen2019,
       author = {{Nielsen}, L.~D. and {Bouchy}, F. and {Turner}, O. and {Giles}, H. and {Su{\'a}rez Mascare{\~n}o}, A. and {Lovis}, C. and {Marmier}, M. and {Pepe}, F. and {S{\'e}gransan}, D. and {Udry}, S. and {Otegi}, J.~F. and {Ottoni}, G. and {Stalport}, M. and {Ricker}, G. and {Vanderspek}, R. and {Latham}, D.~W. and {Seager}, S. and {Winn}, J.~N. and {Jenkins}, J.~M. and {Kane}, S.~R. and {Wittenmyer}, R.~A. and {Bowler}, B. and {Crossfield}, I. and {Horner}, J. and {Kielkopf}, J. and {Morton}, T. and {Plavchan}, P. and {Tinney}, C.~G. and {Zhang}, Hui and {Wright}, D.~J. and {Mengel}, M.~W. and {Clark}, J.~T. and {Okumura}, J. and {Addison}, B. and {Caldwell}, D.~A. and {Cartwright}, S.~M. and {Collins}, K.~A. and {Francis}, J. and {Guerrero}, N. and {Huang}, C.~X. and {Matthews}, E.~C. and {Pepper}, J. and {Rose}, M. and {Villase{\~n}or}, J. and {Wohler}, B. and {Stassun}, K. and {Howell}, S. and {Ciardi}, D. and {Gonzales}, E. and {Matson}, R. and {Beichman}, C. and {Schlieder}, J.},
        title = "{A Jovian planet in an eccentric 11.5 day orbit around HD 1397 discovered by TESS}",
      journal = {\aap},
         year = 2019,
        month = mar,
       volume = {623},
          eid = {A100},
        pages = {A100},
          doi = {10.1051/0004-6361/201834577},
archivePrefix = {arXiv},
       eprint = {1811.01882},
 primaryClass = {astro-ph.EP},
       adsurl = {https://ui.adsabs.harvard.edu/abs/2019A&A...623A.100N}
}

@ARTICLE{Addison2021,
       author = {{Addison}, Brett C. and {Wright}, Duncan J. and {Nicholson}, Belinda A. and {Cale}, Bryson and {Mocnik}, Teo and {Huber}, Daniel and {Plavchan}, Peter and {Wittenmyer}, Robert A. and {Vanderburg}, Andrew and {Chaplin}, William J. and {Chontos}, Ashley and {Clark}, Jake T. and {Eastman}, Jason D. and {Ziegler}, Carl and {Brahm}, Rafael and {Carter}, Bradley D. and {Clerte}, Mathieu and {Espinoza}, N{\'e}stor and {Horner}, Jonathan and {Bentley}, John and {Jord{\'a}n}, Andr{\'e}s and {Kane}, Stephen R. and {Kielkopf}, John F. and {Laychock}, Emilie and {Mengel}, Matthew W. and {Okumura}, Jack and {Stassun}, Keivan G. and {Bedding}, Timothy R. and {Bowler}, Brendan P. and {Burnelis}, Andrius and {Blanco-Cuaresma}, Sergi and {Collins}, Michaela and {Crossfield}, Ian and {Davis}, Allen B. and {Evensberget}, Dag and {Heitzmann}, Alexis and {Howell}, Steve B. and {Law}, Nicholas and {Mann}, Andrew W. and {Marsden}, Stephen C. and {Matson}, Rachel A. and {O'Connor}, James H. and {Shporer}, Avi and {Stevens}, Catherine and {Tinney}, C.~G. and {Tylor}, Christopher and {Wang}, Songhu and {Zhang}, Hui and {Henning}, Thomas and {Kossakowski}, Diana and {Ricker}, George and {Sarkis}, Paula and {Schlecker}, Martin and {Torres}, Pascal and {Vanderspek}, Roland and {Latham}, David W. and {Seager}, Sara and {Winn}, Joshua N. and {Jenkins}, Jon M. and {Mireles}, Ismael and {Rowden}, Pam and {Pepper}, Joshua and {Daylan}, Tansu and {Schlieder}, Joshua E. and {Collins}, Karen A. and {Collins}, Kevin I. and {Tan}, Thiam-Guan and {Ball}, Warrick H. and {Basu}, Sarbani and {Buzasi}, Derek L. and {Campante}, Tiago L. and {Corsaro}, Enrico and {Gonz{\'a}lez-Cuesta}, L. and {Davies}, Guy R. and {de Almeida}, Leandro and {do Nascimento}, Jr., Jose-Dias and {Garc{\'\i}a}, Rafael A. and {Guo}, Zhao and {Handberg}, Rasmus and {Hekker}, Saskia and {Hey}, Daniel R. and {Kallinger}, Thomas and {Kawaler}, Steven D. and {Kayhan}, Cenk and {Kuszlewicz}, James S. and {Lund}, Mikkel N. and {Lyttle}, Alexander and {Mathur}, Savita and {Miglio}, Andrea and {Mosser}, Benoit and {Nielsen}, Martin B. and {Serenelli}, Aldo M. and {Aguirre}, Victor Silva and {Theme{\ss}l}, Nathalie},
        title = "{TOI-257b (HD 19916b): a warm sub-saturn orbiting an evolved F-type star}",
      journal = {\mnras},
         year = 2021,
        month = apr,
       volume = {502},
       number = {3},
        pages = {3704-3722},
          doi = {10.1093/mnras/staa3960},
archivePrefix = {arXiv},
       eprint = {2001.07345},
 primaryClass = {astro-ph.EP},
       adsurl = {https://ui.adsabs.harvard.edu/abs/2021MNRAS.502.3704A}
}

@ARTICLE{Wittenmyer2022,
       author = {{Wittenmyer}, Robert A. and {Clark}, Jake T. and {Trifonov}, Trifon and {Addison}, Brett C. and {Wright}, Duncan J. and {Stassun}, Keivan G. and {Horner}, Jonathan and {Lowson}, Nataliea and {Kielkopf}, John and {Kane}, Stephen R. and {Plavchan}, Peter and {Shporer}, Avi and {Zhang}, Hui and {Bowler}, Brendan P. and {Mengel}, Matthew W. and {Okumura}, Jack and {Rabus}, Markus and {Johnson}, Marshall C. and {Harbeck}, Daniel and {Tronsgaard}, Ren{\'e} and {Buchhave}, Lars A. and {Collins}, Karen A. and {Collins}, Kevin I. and {Gan}, Tianjun and {Jensen}, Eric L.~N. and {Howell}, Steve B. and {Furlan}, E. and {Gnilka}, Crystal L. and {Lester}, Kathryn V. and {Matson}, Rachel A. and {Scott}, Nicholas J. and {Ricker}, George R. and {Vanderspek}, Roland and {Latham}, David W. and {Seager}, S. and {Winn}, Joshua N. and {Jenkins}, Jon M. and {Rudat}, Alexander and {Quintana}, Elisa V. and {Rodriguez}, David R. and {Caldwell}, Douglas A. and {Quinn}, Samuel N. and {Essack}, Zahra and {Bouma}, Luke G.},
        title = "{TOI-1842b: A Transiting Warm Saturn Undergoing Reinflation around an Evolving Subgiant}",
      journal = {\aj},
         year = 2022,
        month = feb,
       volume = {163},
       number = {2},
          eid = {82},
        pages = {82},
          doi = {10.3847/1538-3881/ac3f39},
archivePrefix = {arXiv},
       eprint = {2112.00198},
 primaryClass = {astro-ph.EP},
       adsurl = {https://ui.adsabs.harvard.edu/abs/2022AJ....163...82W}
}

@ARTICLE{GaiaDR3_2023,
       author = {{Gaia Collaboration} and {Vallenari}, A. and {Brown}, A.~G.~A. and {Prusti}, T. and {de Bruijne}, J.~H.~J. and {Arenou}, F. and {Babusiaux}, C. and {Biermann}, M. and {Creevey}, O.~L. and {Ducourant}, C. and {Evans}, D.~W. and {Eyer}, L. and {Guerra}, R. and {Hutton}, A. and {Jordi}, C. and {Klioner}, S.~A. and {Lammers}, U.~L. and {Lindegren}, L. and {Luri}, X. and {Mignard}, F. and {Panem}, C. and {Pourbaix}, D. and {Randich}, S. and {Sartoretti}, P. and {Soubiran}, C. and {Tanga}, P. and {Walton}, N.~A. and {Bailer-Jones}, C.~A.~L. and {Bastian}, U. and {Drimmel}, R. and {Jansen}, F. and {Katz}, D. and {Lattanzi}, M.~G. and {van Leeuwen}, F. and {Bakker}, J. and {Cacciari}, C. and {Casta{\~n}eda}, J. and {De Angeli}, F. and {Fabricius}, C. and {Fouesneau}, M. and {Fr{\'e}mat}, Y. and {Galluccio}, L. and {Guerrier}, A. and {Heiter}, U. and {Masana}, E. and {Messineo}, R. and {Mowlavi}, N. and {Nicolas}, C. and {Nienartowicz}, K. and {Pailler}, F. and {Panuzzo}, P. and {Riclet}, F. and {Roux}, W. and {Seabroke}, G.~M. and {Sordo}, R. and {Th{\'e}venin}, F. and {Gracia-Abril}, G. and {Portell}, J. and {Teyssier}, D. and {Altmann}, M. and {Andrae}, R. and {Audard}, M. and {Bellas-Velidis}, I. and {Benson}, K. and {Berthier}, J. and {Blomme}, R. and {Burgess}, P.~W. and {Busonero}, D. and {Busso}, G. and {C{\'a}novas}, H. and {Carry}, B. and {Cellino}, A. and {Cheek}, N. and {Clementini}, G. and {Damerdji}, Y. and {Davidson}, M. and {de Teodoro}, P. and {Nu{\~n}ez Campos}, M. and {Delchambre}, L. and {Dell'Oro}, A. and {Esquej}, P. and {Fern{\'a}ndez-Hern{\'a}ndez}, J. and {Fraile}, E. and {Garabato}, D. and {Garc{\'\i}a-Lario}, P. and {Gosset}, E. and {Haigron}, R. and {Halbwachs}, J. -L. and {Hambly}, N.~C. and {Harrison}, D.~L. and {Hern{\'a}ndez}, J. and {Hestroffer}, D. and {Hodgkin}, S.~T. and {Holl}, B. and {Jan{\ss}en}, K. and {Jevardat de Fombelle}, G. and {Jordan}, S. and {Krone-Martins}, A. and {Lanzafame}, A.~C. and {L{\"o}ffler}, W. and {Marchal}, O. and {Marrese}, P.~M. and {Moitinho}, A. and {Muinonen}, K. and {Osborne}, P. and {Pancino}, E. and {Pauwels}, T. and {Recio-Blanco}, A. and {Reyl{\'e}}, C. and {Riello}, M. and {Rimoldini}, L. and {Roegiers}, T. and {Rybizki}, J. and {Sarro}, L.~M. and {Siopis}, C. and {Smith}, M. and {Sozzetti}, A. and {Utrilla}, E. and {van Leeuwen}, M. and {Abbas}, U. and {{\'A}brah{\'a}m}, P. and {Abreu Aramburu}, A. and {Aerts}, C. and {Aguado}, J.~J. and {Ajaj}, M. and {Aldea-Montero}, F. and {Altavilla}, G. and {{\'A}lvarez}, M.~A. and {Alves}, J. and {Anders}, F. and {Anderson}, R.~I. and {Anglada Varela}, E. and {Antoja}, T. and {Baines}, D. and {Baker}, S.~G. and {Balaguer-N{\'u}{\~n}ez}, L. and {Balbinot}, E. and {Balog}, Z. and {Barache}, C. and {Barbato}, D. and {Barros}, M. and {Barstow}, M.~A. and {Bartolom{\'e}}, S. and {Bassilana}, J. -L. and {Bauchet}, N. and {Becciani}, U. and {Bellazzini}, M. and {Berihuete}, A. and {Bernet}, M. and {Bertone}, S. and {Bianchi}, L. and {Binnenfeld}, A. and {Blanco-Cuaresma}, S. and {Blazere}, A. and {Boch}, T. and {Bombrun}, A. and {Bossini}, D. and {Bouquillon}, S. and {Bragaglia}, A. and {Bramante}, L. and {Breedt}, E. and {Bressan}, A. and {Brouillet}, N. and {Brugaletta}, E. and {Bucciarelli}, B. and {Burlacu}, A. and {Butkevich}, A.~G. and {Buzzi}, R. and {Caffau}, E. and {Cancelliere}, R. and {Cantat-Gaudin}, T. and {Carballo}, R. and {Carlucci}, T. and {Carnerero}, M.~I. and {Carrasco}, J.~M. and {Casamiquela}, L. and {Castellani}, M. and {Castro-Ginard}, A. and {Chaoul}, L. and {Charlot}, P. and {Chemin}, L. and {Chiaramida}, V. and {Chiavassa}, A. and {Chornay}, N. and {Comoretto}, G. and {Contursi}, G. and {Cooper}, W.~J. and {Cornez}, T. and {Cowell}, S. and {Crifo}, F. and {Cropper}, M. and {Crosta}, M. and {Crowley}, C. and {Dafonte}, C. and {Dapergolas}, A. and {David}, M. and {David}, P. and {de Laverny}, P. and {De Luise}, F. and {De March}, R.},
        title = "{Gaia Data Release 3. Summary of the content and survey properties}",
      journal = {\aap},
         year = 2023,
        month = jun,
       volume = {674},
          eid = {A1},
        pages = {A1},
          doi = {10.1051/0004-6361/202243940},
archivePrefix = {arXiv},
       eprint = {2208.00211},
 primaryClass = {astro-ph.GA},
       adsurl = {https://ui.adsabs.harvard.edu/abs/2023A&A...674A...1G}
}

@ARTICLE{Dotter2016,
       author = {{Dotter}, Aaron},
        title = "{MESA Isochrones and Stellar Tracks (MIST) 0: Methods for the Construction of Stellar Isochrones}",
      journal = {\apjs},
         year = 2016,
        month = jan,
       volume = {222},
       number = {1},
          eid = {8},
        pages = {8},
          doi = {10.3847/0067-0049/222/1/8},
archivePrefix = {arXiv},
       eprint = {1601.05144},
 primaryClass = {astro-ph.SR},
       adsurl = {https://ui.adsabs.harvard.edu/abs/2016ApJS..222....8D}
}

@ARTICLE{Choi2016,
       author = {{Choi}, Jieun and {Dotter}, Aaron and {Conroy}, Charlie and {Cantiello}, Matteo and {Paxton}, Bill and {Johnson}, Benjamin D.},
        title = "{Mesa Isochrones and Stellar Tracks (MIST). I. Solar-scaled Models}",
      journal = {\apj},
         year = 2016,
        month = jun,
       volume = {823},
       number = {2},
          eid = {102},
        pages = {102},
          doi = {10.3847/0004-637X/823/2/102},
archivePrefix = {arXiv},
       eprint = {1604.08592},
 primaryClass = {astro-ph.SR},
       adsurl = {https://ui.adsabs.harvard.edu/abs/2016ApJ...823..102C}
}

@INPROCEEDINGS{ASAS-SN1,
       author = {{Shappee}, Benjamin and {Prieto}, J. and {Stanek}, K.~Z. and {Kochanek}, C.~S. and {Holoien}, T. and {Jencson}, J. and {Basu}, U. and {Beacom}, J.~F. and {Szczygiel}, D. and {Pojmanski}, G. and {Brimacombe}, J. and {Dubberley}, M. and {Elphick}, M. and {Foale}, S. and {Hawkins}, E. and {Mullins}, D. and {Rosing}, W. and {Ross}, R. and {Walker}, Z.},
        title = "{All Sky Automated Survey for SuperNovae (ASAS-SN or ``Assassin'')}",
    booktitle = {American Astronomical Society Meeting Abstracts \#223},
         year = 2014,
       series = {American Astronomical Society Meeting Abstracts},
       volume = {223},
        month = jan,
          eid = {236.03},
        pages = {236.03},
       adsurl = {https://ui.adsabs.harvard.edu/abs/2014AAS...22323603S}
}

@ARTICLE{ASAS-SN2,
       author = {{Kochanek}, C.~S. and {Shappee}, B.~J. and {Stanek}, K.~Z. and {Holoien}, T.~W. -S. and {Thompson}, Todd A. and {Prieto}, J.~L. and {Dong}, Subo and {Shields}, J.~V. and {Will}, D. and {Britt}, C. and {Perzanowski}, D. and {Pojma{\'n}ski}, G.},
        title = "{The All-Sky Automated Survey for Supernovae (ASAS-SN) Light Curve Server v1.0}",
      journal = {\pasp},
         year = 2017,
        month = oct,
       volume = {129},
       number = {980},
        pages = {104502},
          doi = {10.1088/1538-3873/aa80d9},
archivePrefix = {arXiv},
       eprint = {1706.07060},
 primaryClass = {astro-ph.SR},
       adsurl = {https://ui.adsabs.harvard.edu/abs/2017PASP..129j4502K}
}

@ARTICLE{Faria2018,
       author = {{Faria}, J.~P. and {Santos}, N.~C. and {Figueira}, P. and {Brewer}, B.~J.},
        title = "{kima: Exoplanet detection in radial velocities}",
      journal = {The Journal of Open Source Software},
         year = 2018,
        month = jun,
       volume = {3},
       number = {26},
        pages = {487},
          doi = {10.21105/joss.00487},
archivePrefix = {arXiv},
       eprint = {1806.08305},
 primaryClass = {astro-ph.IM},
       adsurl = {https://ui.adsabs.harvard.edu/abs/2018JOSS....3..487F}
}

@article{Espinoza2015LimbParameters,
    title = {{Limb darkening and exoplanets: testing stellar model atmospheres and identifying biases in transit parameters}},
    year = {2015},
    journal = {Monthly Notices of the Royal Astronomical Society},
    author = {Espinoza, Néstor and Jord{\'{a}}n, Andrés},
    number = {2},
    month = {6},
    pages = {1879--1899},
    volume = {450},
    doi = {10.1093/mnras/stv744},
    issn = {1365-2966}
}

@article{Kreidberg2015Batman:Python,
    title = {{batman: BAsic Transit Model cAlculatioN in Python}},
    year = {2015},
    journal = {Publications of the Astronomical Society of the Pacific},
    author = {Kreidberg, Laura},
    pages = {1161--1165},
    volume = {127},
    url = {https://github.com/lkreidberg/batman.}
}

@article{Fulton2018RadVel:Toolkit,
    title = {{RadVel: The Radial Velocity Modeling Toolkit}},
    year = {2018},
    journal = {Publications of the Astronomical Society of the Pacific},
    author = {Fulton, Benjamin J. and Petigura, Erik A. and Blunt, Sarah and Sinukoff, Evan},
    number = {986},
    month = {4},
    pages = {044504},
    volume = {130},
    doi = {10.1088/1538-3873/aaaaa8},
    issn = {0004-6280}
}

@article{celerite2:foremanmackey18,
    title = {{Scalable Backpropagation for Gaussian Processes using Celerite}},
    year = {2018},
    journal = {Research Notes of the American Astronomical Society},
    author = {Foreman-Mackey, D},
    number = {1},
    month = {2},
    pages = {31},
    volume = {2},
    doi = {10.3847/2515-5172/aaaf6c}
}

@article{Espinoza2019Juliet:Systems,
    title = {{juliet: a versatile modelling tool for transiting and non-transiting exoplanetary systems}},
    year = {2019},
    journal = {Monthly Notices of the Royal Astronomical Society},
    author = {Espinoza, Néstor and Kossakowski, Diana and Brahm, Rafael},
    number = {2},
    month = {12},
    pages = {2262--2283},
    volume = {490},
    doi = {10.1093/mnras/stz2688},
    issn = {0035-8711}
}

@ARTICLE{Kipping2013,
       author = {{Kipping}, David M.},
        title = "{Efficient, uninformative sampling of limb darkening coefficients for two-parameter laws}",
      journal = {\mnras},
         year = 2013,
        month = nov,
       volume = {435},
       number = {3},
        pages = {2152-2160},
          doi = {10.1093/mnras/stt1435},
archivePrefix = {arXiv},
       eprint = {1308.0009},
 primaryClass = {astro-ph.SR},
       adsurl = {https://ui.adsabs.harvard.edu/abs/2013MNRAS.435.2152K}
}

@ARTICLE{Kurucz1979,
       author = {{Kurucz}, R.~L.},
        title = "{Model atmospheres for G, F, A, B, and O stars.}",
      journal = {\apjs},
         year = 1979,
        month = may,
       volume = {40},
        pages = {1-340},
          doi = {10.1086/190589},
       adsurl = {https://ui.adsabs.harvard.edu/abs/1979ApJS...40....1K}
}

@ARTICLE{Trotta2008,
       author = {{Trotta}, Roberto},
        title = "{Bayes in the sky: Bayesian inference and model selection in cosmology}",
      journal = {Contemporary Physics},
         year = 2008,
        month = mar,
       volume = {49},
       number = {2},
        pages = {71-104},
          doi = {10.1080/00107510802066753},
archivePrefix = {arXiv},
       eprint = {0803.4089},
 primaryClass = {astro-ph},
       adsurl = {https://ui.adsabs.harvard.edu/abs/2008ConPh..49...71T}
}

@ARTICLE{Damiani2019,
       author = {{Damiani}, F. and {Prisinzano}, L. and {Pillitteri}, I. and {Micela}, G. and {Sciortino}, S.},
        title = "{Stellar population of Sco OB2 revealed by Gaia DR2 data}",
      journal = {\aap},
         year = 2019,
        month = mar,
       volume = {623},
          eid = {A112},
        pages = {A112},
          doi = {10.1051/0004-6361/201833994},
archivePrefix = {arXiv},
       eprint = {1807.11884},
 primaryClass = {astro-ph.SR},
       adsurl = {https://ui.adsabs.harvard.edu/abs/2019A&A...623A.112D}
}

@ARTICLE{Tofflemire2021,
       author = {{Tofflemire}, Benjamin M. and {Rizzuto}, Aaron C. and {Newton}, Elisabeth R. and {Kraus}, Adam L. and {Mann}, Andrew W. and {Vanderburg}, Andrew and {Nelson}, Tyler and {Hawkins}, Keith and {Wood}, Mackenna L. and {Zhou}, George and {Quinn}, Samuel N. and {Howell}, Steve B. and {Collins}, Karen A. and {Schwarz}, Richard P. and {Stassun}, Keivan G. and {Bouma}, Luke G. and {Essack}, Zahra and {Osborn}, Hugh and {Boyd}, Patricia T. and {F{\H{u}}r{\'e}sz}, G{\'a}bor and {Glidden}, Ana and {Twicken}, Joseph D. and {Wohler}, Bill and {McLean}, Brian and {Ricker}, George R. and {Vanderspek}, Roland and {Latham}, David W. and {Seager}, S. and {Winn}, Joshua N. and {Jenkins}, Jon M.},
        title = "{TESS Hunt for Young and Maturing Exoplanets (THYME). V. A Sub-Neptune Transiting a Young Star in a Newly Discovered 250 Myr Association}",
      journal = {\aj},
         year = 2021,
        month = apr,
       volume = {161},
       number = {4},
          eid = {171},
        pages = {171},
          doi = {10.3847/1538-3881/abdf53},
archivePrefix = {arXiv},
       eprint = {2102.06066},
 primaryClass = {astro-ph.EP},
       adsurl = {https://ui.adsabs.harvard.edu/abs/2021AJ....161..171T}
}

@ARTICLE{Bouma2023,
       author = {{Bouma}, Luke G. and {Palumbo}, Elsa K. and {Hillenbrand}, Lynne A.},
        title = "{The Empirical Limits of Gyrochronology}",
      journal = {\apjl},
         year = 2023,
        month = apr,
       volume = {947},
       number = {1},
          eid = {L3},
        pages = {L3},
          doi = {10.3847/2041-8213/acc589},
archivePrefix = {arXiv},
       eprint = {2303.08830},
 primaryClass = {astro-ph.SR},
       adsurl = {https://ui.adsabs.harvard.edu/abs/2023ApJ...947L...3B}
}

@ARTICLE{Jeffries2023,
       author = {{Jeffries}, R.~D. and {Jackson}, R.~J. and {Wright}, Nicholas J. and {Weaver}, G. and {Gilmore}, G. and {Randich}, S. and {Bragaglia}, A. and {Korn}, A.~J. and {Smiljanic}, R. and {Biazzo}, K. and {Casey}, A.~R. and {Frasca}, A. and {Gonneau}, A. and {Guiglion}, G. and {Morbidelli}, L. and {Prisinzano}, L. and {Sacco}, G.~G. and {Tautvai{\v{s}}ien{\.{e}}}, G. and {Worley}, C.~C. and {Zaggia}, S.},
        title = "{The Gaia-ESO Survey: empirical estimates of stellar ages from lithium equivalent widths (EAGLES)}",
      journal = {\mnras},
         year = 2023,
        month = jul,
       volume = {523},
       number = {1},
        pages = {802-824},
          doi = {10.1093/mnras/stad1293},
archivePrefix = {arXiv},
       eprint = {2304.12197},
 primaryClass = {astro-ph.SR},
       adsurl = {https://ui.adsabs.harvard.edu/abs/2023MNRAS.523..802J}
}

@software{Morton2015,
       author = {{Morton}, Timothy D.},
        title = "{isochrones: Stellar model grid package}",
 howpublished = {Astrophysics Source Code Library, record ascl:1503.010},
         year = 2015,
        month = mar,
          eid = {ascl:1503.010},
archivePrefix = {ascl},
       eprint = {1503.010},
       adsurl = {https://ui.adsabs.harvard.edu/abs/2015ascl.soft03010M}
}

@ARTICLE{Mayor1995,
       author = {{Mayor}, Michel and {Queloz}, Didier},
        title = "{A Jupiter-mass companion to a solar-type star}",
      journal = {\nat},
         year = 1995,
        month = nov,
       volume = {378},
       number = {6555},
        pages = {355-359},
          doi = {10.1038/378355a0},
       adsurl = {https://ui.adsabs.harvard.edu/abs/1995Natur.378..355M}
}

@ARTICLE{Boss1997,
       author = {{Boss}, A.~P.},
        title = "{Giant planet formation by gravitational instability.}",
      journal = {Science},
         year = 1997,
        month = jan,
       volume = {276},
        pages = {1836-1839},
          doi = {10.1126/science.276.5320.1836},
       adsurl = {https://ui.adsabs.harvard.edu/abs/1997Sci...276.1836B}
}

@ARTICLE{Perri1974,
       author = {{Perri}, F. and {Cameron}, A.~G.~W.},
        title = "{Hydrodynamic Instability of the Solar Nebula in the Presence of a Planetary Core}",
      journal = {\icarus},
         year = 1974,
        month = aug,
       volume = {22},
       number = {4},
        pages = {416-425},
          doi = {10.1016/0019-1035(74)90074-8},
       adsurl = {https://ui.adsabs.harvard.edu/abs/1974Icar...22..416P}
}

@ARTICLE{Pollack1996,
       author = {{Pollack}, James B. and {Hubickyj}, Olenka and {Bodenheimer}, Peter and {Lissauer}, Jack J. and {Podolak}, Morris and {Greenzweig}, Yuval},
        title = "{Formation of the Giant Planets by Concurrent Accretion of Solids and Gas}",
      journal = {\icarus},
         year = 1996,
        month = nov,
       volume = {124},
       number = {1},
        pages = {62-85},
          doi = {10.1006/icar.1996.0190},
       adsurl = {https://ui.adsabs.harvard.edu/abs/1996Icar..124...62P}
}

@ARTICLE{Rafikov2005,
       author = {{Rafikov}, Roman R.},
        title = "{Can Giant Planets Form by Direct Gravitational Instability?}",
      journal = {\apjl},
         year = 2005,
        month = mar,
       volume = {621},
       number = {1},
        pages = {L69-L72},
          doi = {10.1086/428899},
archivePrefix = {arXiv},
       eprint = {astro-ph/0406469},
 primaryClass = {astro-ph},
       adsurl = {https://ui.adsabs.harvard.edu/abs/2005ApJ...621L..69R}
}

@ARTICLE{Goldreich1980,
       author = {{Goldreich}, P. and {Tremaine}, S.},
        title = "{Disk-satellite interactions.}",
      journal = {\apj},
         year = 1980,
        month = oct,
       volume = {241},
        pages = {425-441},
          doi = {10.1086/158356},
       adsurl = {https://ui.adsabs.harvard.edu/abs/1980ApJ...241..425G}
}

@ARTICLE{Ida2004,
       author = {{Ida}, Shigeru and {Lin}, D.~N.~C.},
        title = "{Toward a Deterministic Model of Planetary Formation. II. The Formation and Retention of Gas Giant Planets around Stars with a Range of Metallicities}",
      journal = {\apj},
         year = 2004,
        month = nov,
       volume = {616},
       number = {1},
        pages = {567-572},
          doi = {10.1086/424830},
archivePrefix = {arXiv},
       eprint = {astro-ph/0408019},
 primaryClass = {astro-ph},
       adsurl = {https://ui.adsabs.harvard.edu/abs/2004ApJ...616..567I}
}

@ARTICLE{Lin1996,
       author = {{Lin}, D.~N.~C. and {Bodenheimer}, P. and {Richardson}, D.~C.},
        title = "{Orbital migration of the planetary companion of 51 Pegasi to its present location}",
      journal = {\nat},
         year = 1996,
        month = apr,
       volume = {380},
       number = {6575},
        pages = {606-607},
          doi = {10.1038/380606a0},
       adsurl = {https://ui.adsabs.harvard.edu/abs/1996Natur.380..606L}
}

@ARTICLE{Jackson2008,
       author = {{Jackson}, Brian and {Greenberg}, Richard and {Barnes}, Rory},
        title = "{Tidal Evolution of Close-in Extrasolar Planets}",
      journal = {\apj},
         year = 2008,
        month = may,
       volume = {678},
       number = {2},
        pages = {1396-1406},
          doi = {10.1086/529187},
archivePrefix = {arXiv},
       eprint = {0802.1543},
 primaryClass = {astro-ph},
       adsurl = {https://ui.adsabs.harvard.edu/abs/2008ApJ...678.1396J}
}

@ARTICLE{Papaloizou2000,
       author = {{Papaloizou}, J.~C.~B. and {Larwood}, J.~D.},
        title = "{On the orbital evolution and growth of protoplanets embedded in a gaseous disc}",
      journal = {\mnras},
         year = 2000,
        month = jul,
       volume = {315},
       number = {4},
        pages = {823-833},
          doi = {10.1046/j.1365-8711.2000.03466.x},
archivePrefix = {arXiv},
       eprint = {astro-ph/9911431},
 primaryClass = {astro-ph},
       adsurl = {https://ui.adsabs.harvard.edu/abs/2000MNRAS.315..823P}
}

@ARTICLE{Rasio1996,
       author = {{Rasio}, Frederic A. and {Ford}, Eric B.},
        title = "{Dynamical instabilities and the formation of extrasolar planetary systems}",
      journal = {Science},
         year = 1996,
        month = nov,
       volume = {274},
        pages = {954-956},
          doi = {10.1126/science.274.5289.954},
       adsurl = {https://ui.adsabs.harvard.edu/abs/1996Sci...274..954R}
}

@ARTICLE{Weidenschilling1996,
       author = {{Weidenschilling}, Stuart J. and {Marzari}, Francesco},
        title = "{Gravitational scattering as a possible origin for giant planets at small stellar distances}",
      journal = {\nat},
         year = 1996,
        month = dec,
       volume = {384},
       number = {6610},
        pages = {619-621},
          doi = {10.1038/384619a0},
       adsurl = {https://ui.adsabs.harvard.edu/abs/1996Natur.384..619W}
}

@ARTICLE{Chatterjee2008,
       author = {{Chatterjee}, Sourav and {Ford}, Eric B. and {Matsumura}, Soko and {Rasio}, Frederic A.},
        title = "{Dynamical Outcomes of Planet-Planet Scattering}",
      journal = {\apj},
         year = 2008,
        month = oct,
       volume = {686},
       number = {1},
        pages = {580-602},
          doi = {10.1086/590227},
archivePrefix = {arXiv},
       eprint = {astro-ph/0703166},
 primaryClass = {astro-ph},
       adsurl = {https://ui.adsabs.harvard.edu/abs/2008ApJ...686..580C}
}

@article{Kozai1962SecularEccentricity,
    title = {{Secular perturbations of asteroids with high inclination and eccentricity}},
    year = {1962},
    journal = {The Astronomical Journal},
    author = {Kozai, Yoshihide},
    month = {11},
    pages = {591},
    volume = {67},
    doi = {10.1086/108790},
    issn = {00046256}
}

@article{Lidov1962TheBodies,
    title = {{The evolution of orbits of artificial satellites of planets under the action of gravitational perturbations of external bodies}},
    year = {1962},
    journal = {Planetary and Space Science},
    author = {Lidov, M.L.},
    number = {10},
    month = {10},
    pages = {719--759},
    volume = {9},
    doi = {10.1016/0032-0633(62)90129-0},
    issn = {00320633}
}

@ARTICLE{Newton2019,
       author = {{Newton}, Elisabeth R. and {Mann}, Andrew W. and {Tofflemire}, Benjamin M. and {Pearce}, Logan and {Rizzuto}, Aaron C. and {Vanderburg}, Andrew and {Martinez}, Raquel A. and {Wang}, Jason J. and {Ruffio}, Jean-Baptiste and {Kraus}, Adam L. and {Johnson}, Marshall C. and {Thao}, Pa Chia and {Wood}, Mackenna L. and {Rampalli}, Rayna and {Nielsen}, Eric L. and {Collins}, Karen A. and {Dragomir}, Diana and {Hellier}, Coel and {Anderson}, D.~R. and {Barclay}, Thomas and {Brown}, Carolyn and {Feiden}, Gregory and {Hart}, Rhodes and {Isopi}, Giovanni and {Kielkopf}, John F. and {Mallia}, Franco and {Nelson}, Peter and {Rodriguez}, Joseph E. and {Stockdale}, Chris and {Waite}, Ian A. and {Wright}, Duncan J. and {Lissauer}, Jack J. and {Ricker}, George R. and {Vanderspek}, Roland and {Latham}, David W. and {Seager}, S. and {Winn}, Joshua N. and {Jenkins}, Jon M. and {Bouma}, Luke G. and {Burke}, Christopher J. and {Davies}, Misty and {Fausnaugh}, Michael and {Li}, Jie and {Morris}, Robert L. and {Mukai}, Koji and {Villase{\~n}or}, Joel and {Villeneuva}, Steven and {De Rosa}, Robert J. and {Macintosh}, Bruce and {Mengel}, Matthew W. and {Okumura}, Jack and {Wittenmyer}, Robert A.},
        title = "{TESS Hunt for Young and Maturing Exoplanets (THYME): A Planet in the 45 Myr Tucana-Horologium Association}",
      journal = {\apjl},
         year = 2019,
        month = jul,
       volume = {880},
       number = {1},
          eid = {L17},
        pages = {L17},
          doi = {10.3847/2041-8213/ab2988},
archivePrefix = {arXiv},
       eprint = {1906.10703},
 primaryClass = {astro-ph.EP},
       adsurl = {https://ui.adsabs.harvard.edu/abs/2019ApJ...880L..17N}
}

@ARTICLE{Vach2024,
       author = {{Vach}, Sydney and {Zhou}, George and {Huang}, Chelsea X. and {Rogers}, James G. and {Bouma}, L.~G. and {Douglas}, Stephanie T. and {Kunimoto}, Michelle and {Mann}, Andrew W. and {Barber}, Madyson G. and {Quinn}, Samuel N. and {Latham}, David W. and {Bieryla}, Allyson and {Collins}, Karen},
        title = "{The Occurrence of Small, Short-period Planets Younger than 200 Myr with TESS}",
      journal = {\aj},
         year = 2024,
        month = may,
       volume = {167},
       number = {5},
          eid = {210},
        pages = {210},
          doi = {10.3847/1538-3881/ad3108},
archivePrefix = {arXiv},
       eprint = {2403.03261},
 primaryClass = {astro-ph.EP},
       adsurl = {https://ui.adsabs.harvard.edu/abs/2024AJ....167..210V}
}

@ARTICLE{Mantovan2024,
       author = {{Mantovan}, G. and {Malavolta}, L. and {Desidera}, S. and {Zingales}, T. and {Borsato}, L. and {Piotto}, G. and {Maggio}, A. and {Locci}, D. and {Polychroni}, D. and {Turrini}, D. and {Baratella}, M. and {Biazzo}, K. and {Nardiello}, D. and {Stassun}, K. and {Nascimbeni}, V. and {Benatti}, S. and {John}, A. Anna and {Watkins}, C. and {Bieryla}, A. and {Lissauer}, J.~J. and {Twicken}, J.~D. and {Lanza}, A.~F. and {Winn}, J.~N. and {Messina}, S. and {Montalto}, M. and {Sozzetti}, A. and {Boffin}, H. and {Cheryasov}, D. and {Strakhov}, I. and {Murgas}, F. and {D'Arpa}, M. and {Barkaoui}, K. and {Benni}, P. and {Bignamini}, A. and {Bonomo}, A.~S. and {Borsa}, F. and {Cabona}, L. and {Cameron}, A.~C. and {Claudi}, R. and {Cochran}, W. and {Collins}, K.~A. and {Damasso}, M. and {Dong}, J. and {Endl}, M. and {Fukui}, A. and {F{\H{u}}r{\'e}sz}, G. and {Gandolfi}, D. and {Ghedina}, A. and {Jenkins}, J. and {Kab{\'a}th}, P. and {Latham}, D.~W. and {Lorenzi}, V. and {Luque}, R. and {Maldonado}, J. and {McLeod}, K. and {Molinaro}, M. and {Narita}, N. and {Nowak}, G. and {Orell-Miquel}, J. and {Pall{\'e}}, E. and {Parviainen}, H. and {Pedani}, M. and {Quinn}, S.~N. and {Relles}, H. and {Rowden}, P. and {Scandariato}, G. and {Schwarz}, R. and {Seager}, S. and {Shporer}, A. and {Vanderburg}, A. and {Wilson}, T.~G.},
        title = "{The GAPS programme at TNG. XLIX. TOI-5398, the youngest compact multi-planet system composed of an inner sub-Neptune and an outer warm Saturn}",
      journal = {\aap},
         year = 2024,
        month = feb,
       volume = {682},
          eid = {A129},
        pages = {A129},
          doi = {10.1051/0004-6361/202347472},
archivePrefix = {arXiv},
       eprint = {2310.16888},
 primaryClass = {astro-ph.EP},
       adsurl = {https://ui.adsabs.harvard.edu/abs/2024A&A...682A.129M}
}

@ARTICLE{Nardiello2019,
       author = {{Nardiello}, D. and {Borsato}, L. and {Piotto}, G. and {Colombo}, L.~S. and {Manthopoulou}, E.~E. and {Bedin}, L.~R. and {Granata}, V. and {Lacedelli}, G. and {Libralato}, M. and {Malavolta}, L. and {Montalto}, M. and {Nascimbeni}, V.},
        title = "{A PSF-based Approach to TESS High quality data Of Stellar clusters (PATHOS) - I. Search for exoplanets and variable stars in the field of 47 Tuc}",
      journal = {\mnras},
         year = 2019,
        month = dec,
       volume = {490},
       number = {3},
        pages = {3806-3823},
          doi = {10.1093/mnras/stz2878},
archivePrefix = {arXiv},
       eprint = {1910.03592},
 primaryClass = {astro-ph.SR},
       adsurl = {https://ui.adsabs.harvard.edu/abs/2019MNRAS.490.3806N}
}

@ARTICLE{Pereira2024,
       author = {{Pereira}, Filipe and {Grunblatt}, Samuel K. and {Psaridi}, Angelica and {Campante}, Tiago L. and {Cunha}, Margarida S. and {Santos}, Nuno C. and {Bossini}, Diego and {Thorngren}, Daniel and {Hellier}, Coel and {Bouchy}, Fran{\c{c}}ois and {Lendl}, Monika and {Mounzer}, Dany and {Udry}, St{\'e}phane and {Beard}, Corey and {Brinkman}, Casey L. and {Isaacson}, Howard and {Quinn}, Samuel N. and {Tyler}, Dakotah and {Zhou}, George and {Howell}, Steve B. and {Howard}, Andrew W. and {Jenkins}, Jon M. and {Seager}, Sara and {Vanderspek}, Roland K. and {Winn}, Joshua N. and {Saunders}, Nicholas and {Huber}, Daniel},
        title = "{TESS giants transiting giants V - two hot Jupiters orbiting red giant hosts}",
      journal = {\mnras},
         year = 2024,
        month = jan,
       volume = {527},
       number = {3},
        pages = {6332-6345},
          doi = {10.1093/mnras/stad3449},
archivePrefix = {arXiv},
       eprint = {2311.06678},
 primaryClass = {astro-ph.EP},
       adsurl = {https://ui.adsabs.harvard.edu/abs/2024MNRAS.527.6332P}
}

@ARTICLE{Villaver2009,
       author = {{Villaver}, Eva and {Livio}, Mario},
        title = "{The Orbital Evolution of Gas Giant Planets Around Giant Stars}",
      journal = {\apjl},
         year = 2009,
        month = nov,
       volume = {705},
       number = {1},
        pages = {L81-L85},
          doi = {10.1088/0004-637X/705/1/L81},
archivePrefix = {arXiv},
       eprint = {0910.2396},
 primaryClass = {astro-ph.EP},
       adsurl = {https://ui.adsabs.harvard.edu/abs/2009ApJ...705L..81V}
}

@ARTICLE{Schlaufman2013,
       author = {{Schlaufman}, Kevin C. and {Winn}, Joshua N.},
        title = "{Evidence for the Tidal Destruction of Hot Jupiters by Subgiant Stars}",
      journal = {\apj},
         year = 2013,
        month = aug,
       volume = {772},
       number = {2},
          eid = {143},
        pages = {143},
          doi = {10.1088/0004-637X/772/2/143},
archivePrefix = {arXiv},
       eprint = {1306.0567},
 primaryClass = {astro-ph.EP},
       adsurl = {https://ui.adsabs.harvard.edu/abs/2013ApJ...772..143S}
}

@ARTICLE{Bryant2020,
       author = {{Bryant}, Edward M. and {Bayliss}, Daniel and {Nielsen}, Louise D. and {Veras}, Dimitri and {Acton}, Jack S. and {Anderson}, David R. and {Armstrong}, David J. and {Bouchy}, Fran{\c{c}}ois and {Briegal}, Joshua T. and {Burleigh}, Matthew R. and {Cabrera}, Juan and {Casewell}, Sarah L. and {Chaushev}, Alexander and {Cooke}, Benjamin F. and {Csizmadia}, Szil{\'a}rd and {Eigm{\"u}ller}, Philipp and {Erikson}, Anders and {Gill}, Samuel and {Gillen}, Edward and {Goad}, Michael R. and {Grieves}, Nolan and {G{\"u}nther}, Maximilian N. and {Henderson}, Beth and {Hogan}, Aleisha and {Jenkins}, James S. and {Lendl}, Monika and {McCormac}, James and {Moyano}, Maximiliano and {Queloz}, Didier and {Rauer}, Heike and {Raynard}, Liam and {Smith}, Alexis M.~S. and {Tilbrook}, Rosanna H. and {Udry}, St{\'e}phane and {Vines}, Jose I. and {Watson}, Christopher A. and {West}, Richard G. and {Wheatley}, Peter J.},
        title = "{NGTS-12b: A sub-Saturn mass transiting exoplanet in a 7.53 day orbit}",
      journal = {\mnras},
         year = 2020,
        month = dec,
       volume = {499},
       number = {3},
        pages = {3139-3148},
          doi = {10.1093/mnras/staa2976},
archivePrefix = {arXiv},
       eprint = {2009.10620},
 primaryClass = {astro-ph.EP},
       adsurl = {https://ui.adsabs.harvard.edu/abs/2020MNRAS.499.3139B}
}

@ARTICLE{Sha2021,
       author = {{Sha}, Lizhou and {Huang}, Chelsea X. and {Shporer}, Avi and {Rodriguez}, Joseph E. and {Vanderburg}, Andrew and {Brahm}, Rafael and {Hagelberg}, Janis and {Matthews}, Elisabeth C. and {Ziegler}, Carl and {Livingston}, John H. and {Stassun}, Keivan G. and {Wright}, Duncan J. and {Crane}, Jeffrey D. and {Espinoza}, N{\'e}stor and {Bouchy}, Fran{\c{c}}ois and {Bakos}, G{\'a}sp{\'a}r {\'A}. and {Collins}, Karen A. and {Zhou}, George and {Bieryla}, Allyson and {Hartman}, Joel D. and {Wittenmyer}, Robert A. and {Nielsen}, Louise D. and {Plavchan}, Peter and {Bayliss}, Daniel and {Sarkis}, Paula and {Tan}, Thiam-Guan and {Cloutier}, Ryan and {Mancini}, Luigi and {Jord{\'a}n}, Andr{\'e}s and {Wang}, Sharon and {Henning}, Thomas and {Narita}, Norio and {Penev}, Kaloyan and {Teske}, Johanna K. and {Kane}, Stephen R. and {Mann}, Andrew W. and {Addison}, Brett C. and {Tamura}, Motohide and {Horner}, Jonathan and {Barbieri}, Mauro and {Burt}, Jennifer A. and {D{\'\i}az}, Mat{\'\i}as R. and {Crossfield}, Ian J.~M. and {Dragomir}, Diana and {Drass}, Holger and {Feinstein}, Adina D. and {Zhang}, Hui and {Hart}, Rhodes and {Kielkopf}, John F. and {Jensen}, Eric L.~N. and {Montet}, Benjamin T. and {Ottoni}, Ga{\"e}l and {Schwarz}, Richard P. and {Rojas}, Felipe and {Nespral}, David and {Torres}, Pascal and {Mengel}, Matthew W. and {Udry}, St{\'e}phane and {Zapata}, Abner and {Snoddy}, Erin and {Okumura}, Jack and {Ricker}, George R. and {Vanderspek}, Roland K. and {Latham}, David W. and {Winn}, Joshua N. and {Seager}, Sara and {Jenkins}, Jon M. and {Col{\'o}n}, Knicole D. and {Henze}, Christopher E. and {Krishnamurthy}, Akshata and {Ting}, Eric B. and {Vezie}, Michael and {Villanueva}, Steven},
        title = "{TOI-954 b and K2-329 b: Short-period Saturn-mass Planets that Test whether Irradiation Leads to Inflation}",
      journal = {\aj},
         year = 2021,
        month = feb,
       volume = {161},
       number = {2},
          eid = {82},
        pages = {82},
          doi = {10.3847/1538-3881/abd187},
archivePrefix = {arXiv},
       eprint = {2010.14436},
 primaryClass = {astro-ph.EP},
       adsurl = {https://ui.adsabs.harvard.edu/abs/2021AJ....161...82S}
}

@ARTICLE{Johnson2006,
       author = {{Johnson}, John Asher and {Marcy}, Geoffrey W. and {Fischer}, Debra A. and {Henry}, Gregory W. and {Wright}, Jason T. and {Isaacson}, Howard and {McCarthy}, Chris},
        title = "{An Eccentric Hot Jupiter Orbiting the Subgiant HD 185269}",
      journal = {\apj},
         year = 2006,
        month = dec,
       volume = {652},
       number = {2},
        pages = {1724-1728},
          doi = {10.1086/508255},
archivePrefix = {arXiv},
       eprint = {astro-ph/0608035},
 primaryClass = {astro-ph},
       adsurl = {https://ui.adsabs.harvard.edu/abs/2006ApJ...652.1724J}
}

@ARTICLE{Lam2017,
       author = {{Lam}, K.~W.~F. and {Faedi}, F. and {Brown}, D.~J.~A. and {Anderson}, D.~R. and {Delrez}, L. and {Gillon}, M. and {H{\'e}brard}, G. and {Lendl}, M. and {Mancini}, L. and {Southworth}, J. and {Smalley}, B. and {Triaud}, A.~H.~M. and {Turner}, O.~D. and {Hay}, K.~L. and {Armstrong}, D.~J. and {Barros}, S.~C.~C. and {Bonomo}, A.~S. and {Bouchy}, F. and {Boumis}, P. and {Collier Cameron}, A. and {Doyle}, A.~P. and {Hellier}, C. and {Henning}, T. and {Jehin}, E. and {King}, G. and {Kirk}, J. and {Louden}, T. and {Maxted}, P.~F.~L. and {McCormac}, J.~J. and {Osborn}, H.~P. and {Palle}, E. and {Pepe}, F. and {Pollacco}, D. and {Prieto-Arranz}, J. and {Queloz}, D. and {Rey}, J. and {S{\'e}gransan}, D. and {Udry}, S. and {Walker}, S. and {West}, R.~G. and {Wheatley}, P.~J.},
        title = "{From dense hot Jupiter to low-density Neptune: The discovery of WASP-127b, WASP-136b, and WASP-138b}",
      journal = {\aap},
         year = 2017,
        month = mar,
       volume = {599},
          eid = {A3},
        pages = {A3},
          doi = {10.1051/0004-6361/201629403},
archivePrefix = {arXiv},
       eprint = {1607.07859},
 primaryClass = {astro-ph.EP},
       adsurl = {https://ui.adsabs.harvard.edu/abs/2017A&A...599A...3L}
}

@article{Kempton2018ACharacterization,
    title = {{A framework for prioritizing the TESS planetary candidates most amenable to atmospheric characterization}},
    year = {2018},
    journal = {Publications of the Astronomical Society of the Pacific},
    author = {Kempton, Eliza M.R. and Bean, Jacob L. and Louie, Dana R. and Deming, Drake and Koll, Daniel D.B. and Mansfield, Megan and Christiansen, Jessie L. and L{\'{o}}pez-Morales, Mercedes and Swain, Mark R. and Zellem, Robert T. and Ballard, Sarah and Barclay, Thomas and Barstow, Joanna K. and Batalha, Natasha E. and Beatty, Thomas G. and Berta-Thompson, Zach and Birkby, Jayne and Buchhave, Lars A. and Charbonneau, David and Cowan, Nicolas B. and Crossfield, Ian and De Val-Borro, Miguel and Dragomir, Diana and Heng, Kevin and Hu, Renyu and Kane, Stephen R. and Kreidberg, Laura and Mallonn, Matthias and Morley, Caroline V. and Narita, Norio and Nascimbeni, Valerio and Pall{\'{e}}, Enric and Quintana, Elisa V. and Rauscher, Emily and Seager, Sara and Shkolnik, Evgenya L. and Sing, David K. and Sozzetti, Alessandro and Stassun, Keivan G. and Essen, Carolina Von and Valenti, Jeff A.},
    number = {993},
    volume = {130},
    doi = {10.1088/1538-3873/aadf6f},
    issn = {00046280}
}

@ARTICLE{Villaver2014,
       author = {{Villaver}, Eva and {Livio}, Mario and {Mustill}, Alexander J. and {Siess}, Lionel},
        title = "{Hot Jupiters and Cool Stars}",
      journal = {\apj},
         year = 2014,
        month = oct,
       volume = {794},
       number = {1},
          eid = {3},
        pages = {3},
          doi = {10.1088/0004-637X/794/1/3},
archivePrefix = {arXiv},
       eprint = {1407.7879},
 primaryClass = {astro-ph.EP},
       adsurl = {https://ui.adsabs.harvard.edu/abs/2014ApJ...794....3V}
}

@ARTICLE{OConnor2023,
       author = {{O'Connor}, Christopher E. and {Bildsten}, Lars and {Cantiello}, Matteo and {Lai}, Dong},
        title = "{Giant Planet Engulfment by Evolved Giant Stars: Light Curves, Asteroseismology, and Survivability}",
      journal = {\apj},
         year = 2023,
        month = jun,
       volume = {950},
       number = {2},
          eid = {128},
        pages = {128},
          doi = {10.3847/1538-4357/acd2d4},
archivePrefix = {arXiv},
       eprint = {2304.09882},
 primaryClass = {astro-ph.EP},
       adsurl = {https://ui.adsabs.harvard.edu/abs/2023ApJ...950..128O}
}

@ARTICLE{Santos2013,
       author = {{Santos}, N.~C. and {Sousa}, S.~G. and {Mortier}, A. and {Neves}, V. and {Adibekyan}, V. and {Tsantaki}, M. and {Delgado Mena}, E. and {Bonfils}, X. and {Israelian}, G. and {Mayor}, M. and {Udry}, S.},
        title = "{SWEET-Cat: A catalogue of parameters for Stars With ExoplanETs. I. New atmospheric parameters and masses for 48 stars with planets}",
      journal = {\aap},
         year = 2013,
        month = aug,
       volume = {556},
          eid = {A150},
        pages = {A150},
          doi = {10.1051/0004-6361/201321286},
archivePrefix = {arXiv},
       eprint = {1307.0354},
 primaryClass = {astro-ph.SR},
       adsurl = {https://ui.adsabs.harvard.edu/abs/2013A&A...556A.150S}
}

@INCOLLECTION{Sousa2014,
       author = {{Sousa}, S{\'e}rgio G.},
        title = "{ARES + MOOG: A Practical Overview of an Equivalent Width (EW) Method to Derive Stellar Parameters}",
    booktitle = {Determination of Atmospheric Parameters of B},
    publisher = {Springer},
         year = 2014,
        pages = {297-310},
          doi = {10.1007/978-3-319-06956-2_26},
       adsurl = {https://ui.adsabs.harvard.edu/abs/2014dapb.book..297S}
}

@ARTICLE{Sousa2021,
       author = {{Sousa}, S.~G. and {Adibekyan}, V. and {Delgado-Mena}, E. and {Santos}, N.~C. and {Rojas-Ayala}, B. and {Soares}, B.~M.~T.~B. and {Legoinha}, H. and {Ulmer-Moll}, S. and {Camacho}, J.~D. and {Barros}, S.~C.~C. and {Demangeon}, O.~D.~S. and {Hoyer}, S. and {Israelian}, G. and {Mortier}, A. and {Tsantaki}, M. and {Monteiro}, M.~A.},
        title = "{SWEET-Cat 2.0: The Cat just got SWEETer. Higher quality spectra and precise parallaxes from Gaia eDR3}",
      journal = {\aap},
         year = 2021,
        month = dec,
       volume = {656},
          eid = {A53},
        pages = {A53},
          doi = {10.1051/0004-6361/202141584},
archivePrefix = {arXiv},
       eprint = {2109.04781},
 primaryClass = {astro-ph.EP},
       adsurl = {https://ui.adsabs.harvard.edu/abs/2021A&A...656A..53S}
}

@ARTICLE{Sousa2007,
       author = {{Sousa}, S.~G. and {Santos}, N.~C. and {Israelian}, G. and {Mayor}, M. and {Monteiro}, M.~J.~P.~F.~G.},
        title = "{A new code for automatic determination of equivalent widths: Automatic Routine for line Equivalent widths in stellar Spectra (ARES)}",
      journal = {\aap},
         year = 2007,
        month = jul,
       volume = {469},
       number = {2},
        pages = {783-791},
          doi = {10.1051/0004-6361:20077288},
archivePrefix = {arXiv},
       eprint = {astro-ph/0703696},
 primaryClass = {astro-ph},
       adsurl = {https://ui.adsabs.harvard.edu/abs/2007A&A...469..783S}
}

@ARTICLE{Sousa2015,
       author = {{Sousa}, S.~G. and {Santos}, N.~C. and {Adibekyan}, V. and {Delgado-Mena}, E. and {Israelian}, G.},
        title = "{ARES v2: new features and improved performance}",
      journal = {\aap},
         year = 2015,
        month = may,
       volume = {577},
          eid = {A67},
        pages = {A67},
          doi = {10.1051/0004-6361/201425463},
archivePrefix = {arXiv},
       eprint = {1504.02725},
 primaryClass = {astro-ph.IM},
       adsurl = {https://ui.adsabs.harvard.edu/abs/2015A&A...577A..67S}
}

@ARTICLE{Sousa2008,
       author = {{Sousa}, S.~G. and {Santos}, N.~C. and {Mayor}, M. and {Udry}, S. and {Casagrande}, L. and {Israelian}, G. and {Pepe}, F. and {Queloz}, D. and {Monteiro}, M.~J.~P.~F.~G.},
        title = "{Spectroscopic parameters for 451 stars in the HARPS GTO planet search program. Stellar [Fe/H] and the frequency of exo-Neptunes}",
      journal = {\aap},
         year = 2008,
        month = aug,
       volume = {487},
       number = {1},
        pages = {373-381},
          doi = {10.1051/0004-6361:200809698},
archivePrefix = {arXiv},
       eprint = {0805.4826},
 primaryClass = {astro-ph},
       adsurl = {https://ui.adsabs.harvard.edu/abs/2008A&A...487..373S}
}

@PHDTHESIS{Sneden1973,
       author = {{Sneden}, Christopher Alan},
        title = "{Carbon and Nitrogen Abundances in Metal-Poor Stars.}",
       school = {University of Texas, Austin},
         year = 1973,
        month = jan,
       adsurl = {https://ui.adsabs.harvard.edu/abs/1973PhDT.......180S}
}

@ARTICLE{BattleyYOUNGSTER,
       author = {{Battley}, Matthew P. and {Armstrong}, David J. and {Pollacco}, Don},
        title = "{YOUNG Star detrending for Transiting Exoplanet Recovery (YOUNGSTER) - II. Using self-organizing maps to explore young star variability in sectors 1-13 of TESS data}",
      journal = {\mnras},
         year = 2022,
        month = apr,
       volume = {511},
       number = {3},
        pages = {4285-4304},
          doi = {10.1093/mnras/stac278},
archivePrefix = {arXiv},
       eprint = {2202.00031},
 primaryClass = {astro-ph.EP},
       adsurl = {https://ui.adsabs.harvard.edu/abs/2022MNRAS.511.4285B}
}

@INPROCEEDINGS{Quirrenbach2020,
       author = {{Quirrenbach}, Andreas and {CARMENES Consortium} and {Amado}, P.~J. and {Ribas}, I. and {Reiners}, A. and {Caballero}, J.~A. and {Aceituno}, J. and {Alacid}, J.~M. and {Alonso-Floriano}, F.~J. and {Anglada-Escud{\'e}}, G. and {Azzaro}, M. and {Baroch}, D. and {Bauer}, F.~F. and {Becerril}, S. and {B{\'e}jar}, V.~J.~S. and {Bluhm}, P. and {Calvo Ortega}, R. and {Cardona Guill{\'e}n}, C. and {Casasayas-Barris}, N. and {Chaturvedi}, P. and {Cifuentes}, C. and {Colom{\'e}}, J. and {Conte}, D. and {Cort{\'e}s-Contreras}, M. and {Czesla}, S. and {D{\'\i}ez-Alonso}, E. and {Dom{\'\i}nguez Fern{\'a}ndez}, A.~J. and {Dreizler}, S. and {Duque-Arribas}, C. and {Espinoza}, N. and {Fuhrmeister}, B. and {Galad{\'\i}-Enr{\'\i}quez}, D. and {Gar{\textasciiacute}a Quintana}, E. and {Gonz{\'a}lez-Alvare}, E. and {Gonz{\'a}lez Cuesta}, z. L. and {Gonz{\'a}lez Hern{\'a}ndez}, J.~I. and {Guenther}, E.~W. and {de Guindos}, E. and {Hatzes}, A.~P. and {Henning}, T. and {Herbort}, O. and {Herrero}, E. and {Hintz}, D. and {Iglesias-P{\'a}ra}, J. and {Jeffers}, S.~V. and {Johnson}, E.~N. and {de Juan}, E. and {Kaminski}, A. and {Kemmer}, J. and {Khaimova}, J. and {Khalafinejad}, S. and {Klahr}, H. and {Kossakowski}, D. and {Kreidberg}, L. and {K{\"u}rster}, M. and {Labarga}, F. and {Lafarga}, M. and {Lamp{\'o}n}, M. and {Lara}, L.~M. and {Lillo-Box}, J. and {Lodieu}, N. and {L{\'o}pez Gallifa}, A. and {L{\'o}pez Gonz{\'a}lez}, M.~J. and {L{\'o}pez-Puertas}, M. and {Luque}, R. and {Marfil}, E. and {Mart{\'\i}n-Ruiz}, S. and {Matth{\'e}}, C. and {Molaverdikhani}, K. and {Montes}, D. and {Morales}, J.~C. and {Morales-Calder{\'o}on}, M. and {Nagel}, E. and {Nortmann}, L. and {Nowak}, G. and {Ofir}, A. and {Oshaghi}, M. and {Pall{\'e}}, E. and {Passegger}, V.~M. and {Pavlov}, A. and {Pedraz}, S. and {Perdelwitz}, V. and {Perger}, M. and {Reffert}, S. and {Revilla}, D. and {Rodr{\'\i}guez}, E. and {Rodr{\'\i}guez L{\'o}pez}, C. and {Sabotta}, S. and {Sadegi}, S. and {Sairam}, L. and {Salz}, M. and {S{\'a}nchez-L{\'o}pez}, A. and {Sanz-Forcada}, J. and {Sarkis}, P. and {Sch{\"a}fer}, S. and {Schiller}, J. and {Schlecker}, M. and {Schmitt}, J.~H.~M.~M. and {Sch{\"o}fer}, P. and {Schweitzer}, A. and {Seiferta}, W. and {Shan}, Y. and {Shulyak}, D. and {Skrzypinski}, S.~L. and {Solano}, E. and {Soto}, M.~G. and {Stahl}, O. and {Stangret}, M. and {Stock}, S.~A. and {Strachan}, J.~B.~P. and {Stuber}, T. and {St{\"u}rmer}, J. and {Tabernero}, H.~M. and {Tal-Or}, L. and {Tala-Pinto}, M. and {Trifonov}, T. and {Vanaverbeke}, S. and {Yan}, F. and {Zapatero Osorio}, M.~R. and {Zechmeister}, M.},
        title = "{The CARMENES M-dwarf planet survey}",
    booktitle = {Ground-based and Airborne Instrumentation for Astronomy VIII},
         year = 2020,
       editor = {{Evans}, Christopher J. and {Bryant}, Julia J. and {Motohara}, Kentaro},
       series = {Society of Photo-Optical Instrumentation Engineers (SPIE) Conference Series},
       volume = {11447},
        month = dec,
          eid = {114473C},
        pages = {114473C},
          doi = {10.1117/12.2561380},
       adsurl = {https://ui.adsabs.harvard.edu/abs/2020SPIE11447E..3CQ}
}

@INPROCEEDINGS{Caballero2016,
       author = {{Caballero}, J.~A. and {Gu{\`a}rdia}, J. and {L{\'o}pez del Fresno}, M. and {Zechmeister}, M. and {de Juan}, E. and {Alonso-Floriano}, F.~J. and {Amado}, P.~J. and {Colom{\'e}}, J. and {Cort{\'e}s-Contreras}, M. and {Garc{\'\i}a-Piquer}, {\'A}. and {Gesa}, L. and {de Guindos}, E. and {Hagen}, H.-J. and {Helmling}, J. and {Hern{\'a}ndez Casta{\~n}o}, L. and {K{\"u}rster}, M. and {L{\'o}pez-Santiago}, J. and {Montes}, D. and {Morales Mu{\~n}oz}, R. and {Pavlov}, A. and {Quirrenbach}, A. and {Reiners}, A. and {Ribas}, I. and {Seifert}, W. and {Solano}, E.},
        title = "{CARMENES: data flow}",
    booktitle = {Observatory Operations: Strategies, Processes, and Systems VI},
         year = 2016,
       editor = {{Peck}, Alison B. and {Seaman}, Robert L. and {Benn}, Chris R.},
       series = {Society of Photo-Optical Instrumentation Engineers (SPIE) Conference Series},
       volume = {9910},
        month = jul,
          eid = {99100E},
        pages = {99100E},
          doi = {10.1117/12.2233574},
       adsurl = {https://ui.adsabs.harvard.edu/abs/2016SPIE.9910E..0EC}
}

@ARTICLE{Bryant2025,
       author = {{Bryant}, Edward M. and {Van Eylen}, Vincent},
        title = "{Determining the impact of post-main sequence stellar evolution on the transiting giant planet population}",
      journal = {\mnras},
         year = 2025,
        month = oct,
          doi = {10.1093/mnras/staf1771},
archivePrefix = {arXiv},
       eprint = {2511.02896},
 primaryClass = {astro-ph.EP},
       adsurl = {https://ui.adsabs.harvard.edu/abs/2025MNRAS.tmp.1665B}
}

@ARTICLE{Wang2019,
       author = {{Wang}, Songhu and {Jones}, Matias and {Shporer}, Avi and {Fulton}, Benjamin J. and {Paredes}, Leonardo A. and {Trifonov}, Trifon and {Kossakowski}, Diana and {Eastman}, Jason and {Redfield}, Seth and {G{\"u}nther}, Maximilian N. and {Kreidberg}, Laura and {Huang}, Chelsea X. and {Millholland}, Sarah and {Seligman}, Darryl and {Fischer}, Debra and {Brahm}, Rafael and {Wang}, Xian-Yu and {Cruz}, Bryndis and {Henry}, Todd and {James}, Hodari-Sadiki and {Addison}, Brett and {Liang}, En-Si and {Davis}, Allen B. and {Tronsgaard}, Ren{\'e} and {Worku}, Keduse and {Brewer}, John M. and {K{\"u}rster}, Martin and {Zhang}, Hui and {Beichman}, Charles A. and {Bieryla}, Allyson and {Brown}, Timothy M. and {Christiansen}, Jessie L. and {Ciardi}, David R. and {Collins}, Karen A. and {Esquerdo}, Gilbert A. and {Howard}, Andrew W. and {Isaacson}, Howard and {Latham}, David W. and {Mazeh}, Tsevi and {Petigura}, Erik A. and {Quinn}, Samuel N. and {Shahaf}, Sahar and {Siverd}, Robert J. and {Rodler}, Florian and {Reffert}, Sabine and {Zakhozhay}, Olga and {Ricker}, George R. and {Vanderspek}, Roland and {Seager}, Sara and {Winn}, Joshua N. and {Jenkins}, Jon M. and {Boyd}, Patricia T. and {F{\H{u}}r{\'e}sz}, G{\'a}bor and {Henze}, Christopher and {Levine}, Alen M. and {Morris}, Robert and {Paegert}, Martin and {Stassun}, Keivan G. and {Ting}, Eric B. and {Vezie}, Michael and {Laughlin}, Gregory},
        title = "{HD 202772A b: A Transiting Hot Jupiter around a Bright, Mildly Evolved Star in a Visual Binary Discovered by TESS}",
      journal = {\aj},
         year = 2019,
        month = feb,
       volume = {157},
       number = {2},
          eid = {51},
        pages = {51},
          doi = {10.3847/1538-3881/aaf1b7},
archivePrefix = {arXiv},
       eprint = {1810.02341},
 primaryClass = {astro-ph.EP},
       adsurl = {https://ui.adsabs.harvard.edu/abs/2019AJ....157...51W}
}

@ARTICLE{Jones2019,
       author = {{Jones}, Mat{\'\i}as I. and {Brahm}, Rafael and {Espinoza}, Nestor and {Wang}, Songhu and {Shporer}, Avi and {Henning}, Thomas and {Jord{\'a}n}, Andr{\'e}s and {Sarkis}, Paula and {Paredes}, Leonardo A. and {Hodari-Sadiki}, James and {Henry}, Todd and {Cruz}, Bryndis and {Nielsen}, Louise D. and {Bouchy}, Fran{\c{c}}ois and {Pepe}, Francesco and {S{\'e}gransan}, Damien and {Turner}, Oliver and {Udry}, St{\'e}phane and {Marmier}, Maxime and {Lovis}, Christophe and {Bakos}, Gaspar and {Osip}, David and {Suc}, Vincent and {Ziegler}, Carl and {Tokovinin}, Andrei and {Law}, Nick M. and {Mann}, Andrew W. and {Relles}, Howard and {Collins}, Karen A. and {Bayliss}, Daniel and {Sedaghati}, Elyar and {Latham}, David W. and {Seager}, Sara and {Winn}, Joshua N. and {Jenkins}, Jon M. and {Smith}, Jeffrey C. and {Davies}, Misty and {Tenenbaum}, Peter and {Dittmann}, Jason and {Vanderburg}, Andrew and {Christiansen}, Jessie L. and {Haworth}, Kari and {Doty}, John and {Fur{\'e}sz}, Gabor and {Laughlin}, Greg and {Matthews}, Elisabeth and {Crossfield}, Ian and {Howell}, Steve and {Ciardi}, David and {Gonzales}, Erica and {Matson}, Rachel and {Beichman}, Charles and {Schlieder}, Joshua},
        title = "{HD 2685 b: a hot Jupiter orbiting an early F-type star detected by TESS}",
      journal = {\aap},
         year = 2019,
        month = may,
       volume = {625},
          eid = {A16},
        pages = {A16},
          doi = {10.1051/0004-6361/201834640},
archivePrefix = {arXiv},
       eprint = {1811.05518},
 primaryClass = {astro-ph.EP},
       adsurl = {https://ui.adsabs.harvard.edu/abs/2019A&A...625A..16J}
}

@ARTICLE{Pecaut2013,
       author = {{Pecaut}, Mark J. and {Mamajek}, Eric E.},
        title = "{Intrinsic Colors, Temperatures, and Bolometric Corrections of Pre-main-sequence Stars}",
      journal = {\apjs},
         year = 2013,
        month = sep,
       volume = {208},
       number = {1},
          eid = {9},
        pages = {9},
          doi = {10.1088/0067-0049/208/1/9},
archivePrefix = {arXiv},
       eprint = {1307.2657},
 primaryClass = {astro-ph.SR},
       adsurl = {https://ui.adsabs.harvard.edu/abs/2013ApJS..208....9P}
}

@ARTICLE{Han2025,
       author = {{Han}, Te and {Robertson}, Paul and {Brandt}, Timothy D. and {Kanodia}, Shubham and {Ca{\~n}as}, Caleb and {Shporer}, Avi and {Ricker}, George and {Beard}, Corey},
        title = "{Hundreds of TESS Exoplanets Might Be Larger than We Thought}",
      journal = {\apjl},
         year = 2025,
        month = jul,
       volume = {988},
       number = {1},
          eid = {L4},
        pages = {L4},
          doi = {10.3847/2041-8213/ade794},
archivePrefix = {arXiv},
       eprint = {2506.19985},
 primaryClass = {astro-ph.EP},
       adsurl = {https://ui.adsabs.harvard.edu/abs/2025ApJ...988L...4H}
}

@ARTICLE{Grunblatt2016,
       author = {{Grunblatt}, Samuel K. and {Huber}, Daniel and {Gaidos}, Eric J. and {Lopez}, Eric D. and {Fulton}, Benjamin J. and {Vanderburg}, Andrew and {Barclay}, Thomas and {Fortney}, Jonathan J. and {Howard}, Andrew W. and {Isaacson}, Howard T. and {Mann}, Andrew W. and {Petigura}, Erik and {Silva Aguirre}, Victor and {Sinukoff}, Evan J.},
        title = "{K2-97b: A (Re-?)Inflated Planet Orbiting a Red Giant Star}",
      journal = {\aj},
         year = 2016,
        month = dec,
       volume = {152},
       number = {6},
          eid = {185},
        pages = {185},
          doi = {10.3847/0004-6256/152/6/185},
archivePrefix = {arXiv},
       eprint = {1606.05818},
 primaryClass = {astro-ph.EP},
       adsurl = {https://ui.adsabs.harvard.edu/abs/2016AJ....152..185G}
}

@ARTICLE{Grunblatt2017,
       author = {{Grunblatt}, Samuel K. and {Huber}, Daniel and {Gaidos}, Eric and {Lopez}, Eric D. and {Howard}, Andrew W. and {Isaacson}, Howard T. and {Sinukoff}, Evan and {Vanderburg}, Andrew and {Nofi}, Larissa and {Yu}, Jie and {North}, Thomas S.~H. and {Chaplin}, William and {Foreman-Mackey}, Daniel and {Petigura}, Erik and {Ansdell}, Megan and {Weiss}, Lauren and {Fulton}, Benjamin and {Lin}, Douglas N.~C.},
        title = "{Seeing Double with K2: Testing Re-inflation with Two Remarkably Similar Planets around Red Giant Branch Stars}",
      journal = {\aj},
         year = 2017,
        month = dec,
       volume = {154},
       number = {6},
          eid = {254},
        pages = {254},
          doi = {10.3847/1538-3881/aa932d},
archivePrefix = {arXiv},
       eprint = {1706.05865},
 primaryClass = {astro-ph.EP},
       adsurl = {https://ui.adsabs.harvard.edu/abs/2017AJ....154..254G}
}

@ARTICLE{He22_USco,
       author = {{He}, Zhihong and {Wang}, Kun and {Luo}, Yangping and {Li}, Jing and {Liu}, Xiaochen and {Jiang}, Qingquan},
        title = "{A Blind All-sky Search for Star Clusters in Gaia EDR3: 886 Clusters within 1.2 kpc of the Sun}",
      journal = {\apjs},
         year = 2022,
        month = sep,
       volume = {262},
       number = {1},
          eid = {7},
        pages = {7},
          doi = {10.3847/1538-4365/ac7c17},
archivePrefix = {arXiv},
       eprint = {2206.12170},
 primaryClass = {astro-ph.GA},
       adsurl = {https://ui.adsabs.harvard.edu/abs/2022ApJS..262....7H}
}

@ARTICLE{Akeson2013,
       author = {{Akeson}, R.~L. and {Chen}, X. and {Ciardi}, D. and {Crane}, M. and {Good}, J. and {Harbut}, M. and {Jackson}, E. and {Kane}, S.~R. and {Laity}, A.~C. and {Leifer}, S. and {Lynn}, M. and {McElroy}, D.~L. and {Papin}, M. and {Plavchan}, P. and {Ram{\'\i}rez}, S.~V. and {Rey}, R. and {von Braun}, K. and {Wittman}, M. and {Abajian}, M. and {Ali}, B. and {Beichman}, C. and {Beekley}, A. and {Berriman}, G.~B. and {Berukoff}, S. and {Bryden}, G. and {Chan}, B. and {Groom}, S. and {Lau}, C. and {Payne}, A.~N. and {Regelson}, M. and {Saucedo}, M. and {Schmitz}, M. and {Stauffer}, J. and {Wyatt}, P. and {Zhang}, A.},
        title = "{The NASA Exoplanet Archive: Data and Tools for Exoplanet Research}",
      journal = {\pasp},
         year = 2013,
        month = aug,
       volume = {125},
       number = {930},
        pages = {989},
          doi = {10.1086/672273},
archivePrefix = {arXiv},
       eprint = {1307.2944},
 primaryClass = {astro-ph.IM},
       adsurl = {https://ui.adsabs.harvard.edu/abs/2013PASP..125..989A}
}

@ARTICLE{Weinberg2024,
       author = {{Weinberg}, Nevin N. and {Davachi}, Niyousha and {Essick}, Reed and {Yu}, Hang and {Arras}, Phil and {Belland}, Brent},
        title = "{Orbital Decay of Hot Jupiters due to Weakly Nonlinear Tidal Dissipation}",
      journal = {\apj},
         year = 2024,
        month = jan,
       volume = {960},
       number = {1},
          eid = {50},
        pages = {50},
          doi = {10.3847/1538-4357/ad05c9},
archivePrefix = {arXiv},
       eprint = {2305.11974},
 primaryClass = {astro-ph.EP},
       adsurl = {https://ui.adsabs.harvard.edu/abs/2024ApJ...960...50W}
}

@ARTICLE{Hadjidemetriou1963,
       author = {{Hadjidemetriou}, John D.},
        title = "{Two-body problem with variable mass: A new approach}",
      journal = {\icarus},
         year = 1963,
        month = jan,
       volume = {2},
        pages = {440-451},
          doi = {10.1016/0019-1035(63)90072-1},
       adsurl = {https://ui.adsabs.harvard.edu/abs/1963Icar....2..440H}
}

@ARTICLE{Veras2011,
       author = {{Veras}, Dimitri and {Wyatt}, Mark C. and {Mustill}, Alexander J. and {Bonsor}, Amy and {Eldridge}, John J.},
        title = "{The great escape: how exoplanets and smaller bodies desert dying stars}",
      journal = {\mnras},
         year = 2011,
        month = nov,
       volume = {417},
       number = {3},
        pages = {2104-2123},
          doi = {10.1111/j.1365-2966.2011.19393.x},
archivePrefix = {arXiv},
       eprint = {1107.1239},
 primaryClass = {astro-ph.EP},
       adsurl = {https://ui.adsabs.harvard.edu/abs/2011MNRAS.417.2104V}
}

@ARTICLE{Stumpe2012,
       author = {{Stumpe}, Martin C. and {Smith}, Jeffrey C. and {Van Cleve}, Jeffrey E. and {Twicken}, Joseph D. and {Barclay}, Thomas S. and {Fanelli}, Michael N. and {Girouard}, Forrest R. and {Jenkins}, Jon M. and {Kolodziejczak}, Jeffery J. and {McCauliff}, Sean D. and {Morris}, Robert L.},
        title = "{Kepler Presearch Data Conditioning I{\textemdash}Architecture and Algorithms for Error Correction in Kepler Light Curves}",
      journal = {\pasp},
         year = 2012,
        month = sep,
       volume = {124},
       number = {919},
        pages = {985},
          doi = {10.1086/667698},
archivePrefix = {arXiv},
       eprint = {1203.1382},
 primaryClass = {astro-ph.IM},
       adsurl = {https://ui.adsabs.harvard.edu/abs/2012PASP..124..985S}
}

@ARTICLE{Stumpe2014,
       author = {{Stumpe}, Martin C. and {Smith}, Jeffrey C. and {Catanzarite}, Joseph H. and {Van Cleve}, Jeffrey E. and {Jenkins}, Jon M. and {Twicken}, Joseph D. and {Girouard}, Forrest R.},
        title = "{Multiscale Systematic Error Correction via Wavelet-Based Bandsplitting in Kepler Data}",
      journal = {\pasp},
         year = 2014,
        month = jan,
       volume = {126},
       number = {935},
        pages = {100},
          doi = {10.1086/674989},
       adsurl = {https://ui.adsabs.harvard.edu/abs/2014PASP..126..100S}
}

@ARTICLE{Smith2012,
       author = {{Smith}, Jeffrey C. and {Stumpe}, Martin C. and {Van Cleve}, Jeffrey E. and {Jenkins}, Jon M. and {Barclay}, Thomas S. and {Fanelli}, Michael N. and {Girouard}, Forrest R. and {Kolodziejczak}, Jeffery J. and {McCauliff}, Sean D. and {Morris}, Robert L. and {Twicken}, Joseph D.},
        title = "{Kepler Presearch Data Conditioning II - A Bayesian Approach to Systematic Error Correction}",
      journal = {\pasp},
         year = 2012,
        month = sep,
       volume = {124},
       number = {919},
        pages = {1000},
          doi = {10.1086/667697},
archivePrefix = {arXiv},
       eprint = {1203.1383},
 primaryClass = {astro-ph.IM},
       adsurl = {https://ui.adsabs.harvard.edu/abs/2012PASP..124.1000S}
}

@ARTICLE{Santos2002,
       author = {{Santos}, N.~C. and {Mayor}, M. and {Naef}, D. and {Pepe}, F. and {Queloz}, D. and {Udry}, S. and {Burnet}, M. and {Clausen}, J.~V. and {Helt}, B.~E. and {Olsen}, E.~H. and {Pritchard}, J.~D.},
        title = "{The CORALIE survey for southern extra-solar planets. IX. A 1.3-day period brown dwarf disguised as a planet}",
      journal = {\aap},
         year = 2002,
        month = sep,
       volume = {392},
        pages = {215-229},
          doi = {10.1051/0004-6361:20020876},
archivePrefix = {arXiv},
       eprint = {astro-ph/0206213},
 primaryClass = {astro-ph},
       adsurl = {https://ui.adsabs.harvard.edu/abs/2002A&A...392..215S}
}

@ARTICLE{Doyle2014,
       author = {{Doyle}, Amanda P. and {Davies}, Guy R. and {Smalley}, Barry and {Chaplin}, William J. and {Elsworth}, Yvonne},
        title = "{Determining stellar macroturbulence using asteroseismic rotational velocities from Kepler}",
      journal = {\mnras},
         year = 2014,
        month = nov,
       volume = {444},
       number = {4},
        pages = {3592-3602},
          doi = {10.1093/mnras/stu1692},
archivePrefix = {arXiv},
       eprint = {1408.3988},
 primaryClass = {astro-ph.SR},
       adsurl = {https://ui.adsabs.harvard.edu/abs/2014MNRAS.444.3592D}
}

@ARTICLE{Bailer-Jones2023,
       author = {{Bailer-Jones}, C.~A.~L.},
        title = "{Estimating Distances from Parallaxes. VI. A Method for Inferring Distances and Transverse Velocities from Parallaxes and Proper Motions Demonstrated on Gaia Data Release 3}",
      journal = {\aj},
         year = 2023,
        month = dec,
       volume = {166},
       number = {6},
          eid = {269},
        pages = {269},
          doi = {10.3847/1538-3881/ad08bb},
archivePrefix = {arXiv},
       eprint = {2311.00374},
 primaryClass = {astro-ph.GA},
       adsurl = {https://ui.adsabs.harvard.edu/abs/2023AJ....166..269B}
}

@ARTICLE{Schlegel1998,
       author = {{Schlegel}, David J. and {Finkbeiner}, Douglas P. and {Davis}, Marc},
        title = "{Maps of Dust Infrared Emission for Use in Estimation of Reddening and Cosmic Microwave Background Radiation Foregrounds}",
      journal = {\apj},
         year = 1998,
        month = jun,
       volume = {500},
       number = {2},
        pages = {525-553},
          doi = {10.1086/305772},
archivePrefix = {arXiv},
       eprint = {astro-ph/9710327},
 primaryClass = {astro-ph},
       adsurl = {https://ui.adsabs.harvard.edu/abs/1998ApJ...500..525S}
}

@ARTICLE{Schlafly2011,
       author = {{Schlafly}, Edward F. and {Finkbeiner}, Douglas P.},
        title = "{Measuring Reddening with Sloan Digital Sky Survey Stellar Spectra and Recalibrating SFD}",
      journal = {\apj},
         year = 2011,
        month = aug,
       volume = {737},
       number = {2},
          eid = {103},
        pages = {103},
          doi = {10.1088/0004-637X/737/2/103},
archivePrefix = {arXiv},
       eprint = {1012.4804},
 primaryClass = {astro-ph.GA},
       adsurl = {https://ui.adsabs.harvard.edu/abs/2011ApJ...737..103S}
}

@ARTICLE{Bakos2011,
       author = {{Bakos}, G. {\'A}. and {Hartman}, J. and {Torres}, G. and {Latham}, D.~W. and {Kov{\'a}cs}, G{\'e}za and {Noyes}, R.~W. and {Fischer}, D.~A. and {Johnson}, J.~A. and {Marcy}, G.~W. and {Howard}, A.~W. and {Kipping}, D. and {Esquerdo}, G.~A. and {Shporer}, A. and {B{\'e}ky}, B. and {Buchhave}, L.~A. and {Perumpilly}, G. and {Everett}, M. and {Sasselov}, D.~D. and {Stefanik}, R.~P. and {L{\'a}z{\'a}r}, J. and {Papp}, I. and {S{\'a}ri}, P.},
        title = "{HAT-P-20b-HAT-P-23b: Four Massive Transiting Extrasolar Planets}",
      journal = {\apj},
         year = 2011,
        month = dec,
       volume = {742},
       number = {2},
          eid = {116},
        pages = {116},
          doi = {10.1088/0004-637X/742/2/116},
archivePrefix = {arXiv},
       eprint = {1008.3388},
 primaryClass = {astro-ph.EP},
       adsurl = {https://ui.adsabs.harvard.edu/abs/2011ApJ...742..116B}
}

@ARTICLE{Santerne2011,
       author = {{Santerne}, A. and {Bonomo}, A.~S. and {H{\'e}brard}, G. and {Deleuil}, M. and {Moutou}, C. and {Almenara}, J.-M. and {Bouchy}, F. and {D{\'\i}az}, R.~F.},
        title = "{SOPHIE velocimetry of Kepler transit candidates. IV. KOI-196b: a non-inflated hot Jupiter with a high albedo}",
      journal = {\aap},
         year = 2011,
        month = dec,
       volume = {536},
          eid = {A70},
        pages = {A70},
          doi = {10.1051/0004-6361/201117807},
archivePrefix = {arXiv},
       eprint = {1108.0550},
 primaryClass = {astro-ph.EP},
       adsurl = {https://ui.adsabs.harvard.edu/abs/2011A&A...536A..70S}
}

@ARTICLE{Bonomo2015,
       author = {{Bonomo}, A.~S. and {Sozzetti}, A. and {Santerne}, A. and {Deleuil}, M. and {Almenara}, J.-M. and {Bruno}, G. and {D{\'\i}az}, R.~F. and {H{\'e}brard}, G. and {Moutou}, C.},
        title = "{Improved parameters of seven Kepler giant companions characterized with SOPHIE and HARPS-N}",
      journal = {\aap},
         year = 2015,
        month = mar,
       volume = {575},
          eid = {A85},
        pages = {A85},
          doi = {10.1051/0004-6361/201323042},
archivePrefix = {arXiv},
       eprint = {1501.02653},
 primaryClass = {astro-ph.EP},
       adsurl = {https://ui.adsabs.harvard.edu/abs/2015A&A...575A..85B}
}

@ARTICLE{Tejada2021,
       author = {{Tejada Arevalo}, Roberto A. and {Winn}, Joshua N. and {Anderson}, Kassandra R.},
        title = "{Further Evidence for Tidal Spin-up of Hot Jupiter Host Stars}",
      journal = {\apj},
         year = 2021,
        month = oct,
       volume = {919},
       number = {2},
          eid = {138},
        pages = {138},
          doi = {10.3847/1538-4357/ac1429},
archivePrefix = {arXiv},
       eprint = {2107.05759},
 primaryClass = {astro-ph.EP},
       adsurl = {https://ui.adsabs.harvard.edu/abs/2021ApJ...919..138T}
}

@ARTICLE{Poppenhaeger2014,
       author = {{Poppenhaeger}, K. and {Wolk}, S.~J.},
        title = "{Indications for an influence of hot Jupiters on the rotation and activity of their host stars}",
      journal = {\aap},
         year = 2014,
        month = may,
       volume = {565},
          eid = {L1},
        pages = {L1},
          doi = {10.1051/0004-6361/201423454},
archivePrefix = {arXiv},
       eprint = {1404.1073},
 primaryClass = {astro-ph.SR},
       adsurl = {https://ui.adsabs.harvard.edu/abs/2014A&A...565L...1P}
}

@ARTICLE{Brown2014,
       author = {{Brown}, D.~J.~A.},
        title = "{Discrepancies between isochrone fitting and gyrochronology for exoplanet host stars?}",
      journal = {\mnras},
         year = 2014,
        month = aug,
       volume = {442},
       number = {2},
        pages = {1844-1862},
          doi = {10.1093/mnras/stu950},
archivePrefix = {arXiv},
       eprint = {1406.4402},
 primaryClass = {astro-ph.EP},
       adsurl = {https://ui.adsabs.harvard.edu/abs/2014MNRAS.442.1844B}
}

@ARTICLE{Maxted2015,
       author = {{Maxted}, P.~F.~L. and {Serenelli}, A.~M. and {Southworth}, J.},
        title = "{Comparison of gyrochronological and isochronal age estimates for transiting exoplanet host stars}",
      journal = {\aap},
         year = 2015,
        month = may,
       volume = {577},
          eid = {A90},
        pages = {A90},
          doi = {10.1051/0004-6361/201525774},
archivePrefix = {arXiv},
       eprint = {1503.09111},
 primaryClass = {astro-ph.EP},
       adsurl = {https://ui.adsabs.harvard.edu/abs/2015A&A...577A..90M}
}

@ARTICLE{Magrini2021,
       author = {{Magrini}, L. and {Smiljanic}, R. and {Franciosini}, E. and {Pasquini}, L. and {Randich}, S. and {Casali}, G. and {Viscasillas V{\'a}zquez}, C. and {Bragaglia}, A. and {Spina}, L. and {Biazzo}, K. and {Tautvai{\v{s}}ien{\.{e}}}, G. and {Masseron}, T. and {Van der Swaelmen}, M. and {Pancino}, E. and {Jim{\'e}nez-Esteban}, F. and {Guiglion}, G. and {Martell}, S. and {Bensby}, T. and {D'Orazi}, V. and {Baratella}, M. and {Korn}, A. and {Jofre}, P. and {Gilmore}, G. and {Worley}, C. and {Hourihane}, A. and {Gonneau}, A. and {Sacco}, G.~G. and {Morbidelli}, L.},
        title = "{Gaia-ESO survey: Lithium abundances in open cluster Red Clump stars}",
      journal = {\aap},
         year = 2021,
        month = nov,
       volume = {655},
          eid = {A23},
        pages = {A23},
          doi = {10.1051/0004-6361/202141275},
archivePrefix = {arXiv},
       eprint = {2108.11677},
 primaryClass = {astro-ph.SR},
       adsurl = {https://ui.adsabs.harvard.edu/abs/2021A&A...655A..23M}
}

@ARTICLE{Sun2025,
       author = {{Sun}, Qinghui and {Deliyannis}, Constantine P. and {Twarog}, Bruce A. and {Anthony-Twarog}, Barbara J.},
        title = "{Lithium Abundances of Main-sequence Stars in the Old Open Cluster NGC 188: Probes of Stellar Evolution beyond the Solar Age}",
      journal = {\apj},
         year = 2025,
        month = oct,
       volume = {992},
       number = {1},
          eid = {75},
        pages = {75},
          doi = {10.3847/1538-4357/ae032a},
archivePrefix = {arXiv},
       eprint = {2509.02931},
 primaryClass = {astro-ph.SR},
       adsurl = {https://ui.adsabs.harvard.edu/abs/2025ApJ...992...75S}
}



\appendix


\section{CARMENES vetting}
\label{sec:CARM_vetting_append}

\begin{table*}
	\centering
	\caption{Overview of CARMENES vetting results. Note that the RV values were taken as close to phases 0.25 and 0.75 in the expected orbit in order to probe the approximate magnitude of the radial velocity signal. N.b. EB = Eclipsing Binary and SB = Spectroscopic Binary}
	\label{tab:vetting}
	\begin{tabular}{ccccccc} 
		\hline
		Name & Time 0.25 [BJD] & RV 0.25 [km/s] & Time 0.75 [BJD] & RV 0.75 [km/s] & $\Delta$RV [m/s] & Vetting Result\\
		\hline
		TOI-2538 & 2460312.5278260 & 762.817 $\pm$ 0.044 & 2460311.4602998 & 277.610 $\pm$ 0.042 & 485206.81 & Broad CCFs \\
            TOI-3571 & 2460311.3410344 & 29.759 $\pm$ 0.085  & 2460313.2636574 & 29.469  $\pm$ 0.078 & 289.87 & Anti-phase\\
            TOI-3664 & 2460312.3571130 & -16.915 $\pm$ 0.032 & 2460317.3912960 & -16.969 $\pm$ 0.060 & 54.43 & \textbf{Passed}\\
            TOI-4034 & 2460311.3235430 & -6.648 $\pm$ 0.056  & 2460317.3242420 & -6.820  $\pm$ 0.081 & 172.45 &\textbf{Passed}\\
            TOI-4688 & Missed          & Missed              & 2460311.3837339 & 76.684  $\pm$ 0.051 & N/A & Missed\\
            TOI-5459 & 2460312.4723960 & -86.415 $\pm$ 0.121 & 2460313.4496347 & -81.991 $\pm$ 0.079 & -4423.80 & SB\\
            TOI-6248 & 2460311.2984069 & 54.995 $\pm$ 0.093  & 2460313.2844601 & 71.419  $\pm$ 0.099 & -16423.87 & EB \\
		TOI-6382 & 2460325.4066204 & Fit failed          & 2460311.4844659 & 36.353  $\pm$ 0.097 & N/A & Likely SB\\
		\hline
	\end{tabular}
\end{table*}

In addition to the CARMENES follow-up carried out above for TOI-3644 and TOI-4034, six additional TOIs with evidence of youth in a variety of literature catalogues were vetted in the same programme. The results of this vetting are presented here to support the community's wider efforts. 

Each of the eight targets was vetted by taking one spectrum at each of the expected maximum and minimum radial velocity epochs, or as close to phases 0.25 and 0.75 in the orbit, according to the photometric ephemeris presented on ExoFOP.\footnote{https://exofop.ipac.caltech.edu/tess/}

With only two exceptions, vetting observations were carried out between 1-7 January 2024, in order to quickly decide which targets warranted further monitoring. Following each observation, the raw data was analysed using the \texttt{raccoon} pipeline \citep{Lafarga2020}, in order to check the spectra, extract measurements of the radial velocity, and to inspect the Cross Correlation Functions (CCFs).

Two simple vetting tests were carried out to evaluate their suitability to further monitoring: 1) Inspecting the shape of the extracted CCFs and 2) evaluating the approximate amplitude of the radial velocity signal by calculating the difference between the measurements at phases 0.25 and 0.75. 

During this process TOI-5459 and TOI-6382 were found to have CCFs with more than one dip, suggestive of spectroscopic binaries, while TOI-2538 was found to have untenably wide spectral lines (likely due to fast stellar rotation), preventing precise and reliable radial velocity measurements from being extracted. Similarly, in the case of TOI-6248, the difference between the RV measurements at phases 0.25 and 0.75 was in excess of 16 km/s, suggesting that this candidate signal was caused by an eclipsing binary, not an exoplanet. Meanwhile, although the RV amplitude measured for TOI-3571 was reasonable for a Jupiter-sized planet around its host star, the two collected radial velocity measurements actually occurred at opposite times to what was expected from the TESS ephemeris, suggestive of either an eclipsing binary system or complex stellar activity, so this target too was dropped from the programme. Unfortunately due to scheduling and weather issues, the observation of TOI-4688 at phase 0.25 could not be carried out in the programme, so nothing can be said definitively about this target.

A full overview of the collected vetting observations and their final dispositions is presented in Table \ref{tab:vetting}. 

\section{Radial Velocity Measurements}\label{sec:appendix_rvs}

The full radial-velocity measurements analysed in this paper are presented in Tables \ref{rv_table_TOI-3664}, \ref{rv_table_TOI-4034} and \ref{rv_table_TOI-6564} and are also available in machine-reading format alongside the online version of this paper.

\begin{table}
\caption{RV data for TOI-3664}             
\label{rv_table_TOI-3664}      
\centering                          
\begin{tabular}{c c c c}        
\hline                
Time & RV        & RV Error  & Instrument \\
BJD  & [ms$^{-1}$] & [ms$^{-1}$] &  - \\    
\hline                        
2460312.357 &	-16917.48849 &	42.15306 &	CARMENES \\
2460317.391	& -16961.87207 &	78.05758 &	CARMENES \\
2460353.294	& -17008.20173 &    105.88694 &	CARMENES \\
2460353.31	& -16970.19956 &	104.07584 &	CARMENES \\
2460354.294	& -16893.94325 &	49.56977 &	CARMENES \\
2460354.31	& -16904.92121 &	49.06538 &	CARMENES \\
2460355.292	& -16969.35916 &	42.11380 &	CARMENES\\
2460355.308	&-17022.11931  &	54.28219 &	CARMENES\\
2460358.296	&-16996.66665	&60.66222	&CARMENES\\
2460358.313	&-16916.99678	&68.83403	&CARMENES\\
2460359.409	&-17006.97492	&38.70351	&CARMENES\\
2460360.338	&-16958.79632	&30.45402	&CARMENES\\
2460360.356	&-16995.74236	&35.08230	&CARMENES\\
2460361.322	&-16955.73644	&49.14355	&CARMENES\\
2460361.34	&-16893.47251	&49.14944	&CARMENES\\
2460371.324	&-16944.68772	&40.73604	&CARMENES\\
2460371.342	&-16947.42867	&46.67448	&CARMENES\\
2460375.321	&-17019.4548	&36.70504	&CARMENES\\
2460375.339	&-17031.02445	&34.50057	&CARMENES\\
\hline                                   
\\
\end{tabular}
\end{table}

\begin{table}
\caption{RV data for TOI-4034}             
\label{rv_table_TOI-4034}      
\centering                          
\begin{tabular}{c c c c}        
\hline                
Time & RV        & RV Error  & Instrument \\
BJD  & [ms$^{-1}$] & [ms$^{-1}$] &  - \\    
\hline                        
2460311.323543&	-6648.02671	&78.63116	&CARMENES \\
2460317.32424&	-6818.88258	&120.78882	&CARMENES\\
2460317.36384&	-6635.13833	&149.92357	&CARMENES\\
2460442.6425	&-6520.82385&	43.18636&	CARMENES\\
2460442.65876	&-6460.91182&	42.83671&	CARMENES\\
2460443.62291	&-6704.88316&	53.18989&	CARMENES\\
2460443.63993	&-6720.41098&	44.56097&	CARMENES\\
2460447.61369	&-6593.71121&	86.24849&	CARMENES\\
2460447.63136	&-6494.60292&	136.48119&	CARMENES\\
2460449.60282	&-6407.10461&	90.82587	&CARMENES\\
2460449.61965	&-6293.06404&	69.70656	&CARMENES\\
2460455.64061	&-6510.34728&	51.14255	&CARMENES\\
2460455.65621	&-6485.58294&	48.23511	&CARMENES\\
2460458.62617	&-6507.06771&	37.72920	&CARMENES\\
2460458.64236	&-6507.2148	& 37.60009	&CARMENES\\
2460460.61675	&-6411.06859&	38.62132	&CARMENES\\
2460484.61761	&-6661.2072	& 48.96500	&CARMENES\\
2460484.63632	&-6649.54783&	53.69958	&CARMENES\\
2460485.6254	&-6568.54907&	39.20842	&CARMENES\\
2460485.64819	&-6507.46114&	40.74366	&CARMENES\\
2460490.60015	&-6581.23589&	275.60896	&CARMENES\\
2460490.61761	&-6514.34127&	267.9552	&CARMENES\\
\hline                                   
\\
\end{tabular}
\end{table}

\begin{table}
\caption{RV data for TOI-6564}             
\label{rv_table_TOI-6564}      
\centering                          
\begin{tabular}{c c c c}        
\hline                
Time & RV        & RV Error  & Instrument \\
BJD  & [ms$^{-1}$] & [ms$^{-1}$] &  - \\    
\hline                        
2460167.888321      	 &  -8525.299 &	18.765	 &  MINERVA\\
2460167.920541	         &  -8549.063 &	19.279	 &  MINERVA\\
2460175.919779	         &-8552.413	  &  16.954	 &  MINERVA\\
2460177.899948	         &-8379.442	  &  19.882	 &  MINERVA\\
2460178.919678	         &-8477.514	  &  16.528	 &  MINERVA\\
2460179.920636	         &-8521.412	  &  17.732	 &  MINERVA\\
2460181.902220	         &-8393.795	  &  15.221	 &  MINERVA\\
2460425.636623240076	&-7557.43355  &	7.752599 &	CORALIE14\\
2460440.692218369804 &	-7529.524807  &	11.376738 & CORALIE14\\
2460466.56458572997  &	-7678.742939  &	15.927452 &	CORALIE14\\
2460492.58155064983  &	-7534.667595  &	6.336502  &	CORALIE14\\
2460498.64336684998  &	-7685.137505  &	6.492417  &	CORALIE14\\
2460505.56996307988   & -7584.443779  &	8.201031 &	CORALIE14\\
2460507.51850314997   &	-7645.053542  &	6.952943 &	CORALIE14\\
2460510.538426990155  &	-7677.498125  &	7.186331 &	CORALIE14\\
2460514.55492101982	  &  -7691.279090 &	17.339981 &	CORALIE14\\
2460515.49542242987	  &  -7627.871626 &	10.112589 &	CORALIE14\\
2460516.520176849794  &	-7534.520062  &	8.443043  &	CORALIE14\\
2460519.50077578006	  &  -7642.663497 &	7.441667  &	CORALIE14\\
2460542.52890574001	  &  -7695.595509 &	16.942226 &	CORALIE14\\
2460569.50403237995	  &  -7600.939399 &	8.343578  &	CORALIE14\\
2460667.84078198997	  &  -7557.476035 &	6.366845  &	CORALIE24\\
2460696.80468448019	  &  -7589.693804 &	7.265243  &	CORALIE24\\
2460699.82876972994	  &  -7554.467131 &	6.213132 &	CORALIE24\\
2460707.75943840016	  &  -7548.908761 &	6.451489	&CORALIE24\\
2460711.8147889399	  & -7546.974675  &	5.724207	&CORALIE24\\
2460723.84198427992	  & -7547.294835  &	6.556601	&CORALIE24\\
2460736.83497559978	  &  -7600.098830 &	6.373488	&CORALIE24\\
2460737.86745380005	  &  -7705.868519 &	5.763901	&CORALIE24\\
\hline                                   
\\
\end{tabular}
\end{table}

\section{Instrumental fitted parameters}

System and instrumental fitted parameters for all systems are presented in Table \ref{instrumental_table}.

\begin{table*}
\caption{Instrumental parameters for each fit from juliet: median and 68\% confidence interval}             
\label{instrumental_table}      
\centering                          
\begin{tabular}{l l l c}        
\hline               
Parameter  & & Prior distribution$^*$ & Value \\    
\hline                        
$m_{\mathrm{dilution,TESS,all}}$ \dotfill & Dilution parameter \dotfill & \textit{fixed} \dotfill & 1.0 \\

\underline{TOI-3664}\\
$q_{1,\,TESS}$ \dotfill & Quadratic limb-darkening parametrization  \dotfill & $N$(0.360,0.069) \dotfill & 0.368$^{+0.057}_{-0.062}$\\
$q_{2,\,TESS}$ \dotfill & Quadratic limb-darkening parametrization \dotfill & $N$(0.29,0.13) \dotfill & 0.25$^{+0.11}_{-0.10}$\\
$m_{\mathrm{flux,TESS}}$ \dotfill & Photometric offset (relative flux) \dotfill & $N$(0,0.1) \dotfill & -0.00046$^{+0.00023}_{-0.00023}$\\
$\sigma_{\mathrm{TESS}}$ \dotfill & Photometric jitter in data (ppm) \dotfill & log$U$(1e$^{-5}$,1e$^3$) \dotfill & 0.028$^{+5.6}_{-0.028}$ \\
$\sigma_{GP,\mathrm{TESS}}$ \dotfill & Photometric GP amplitude (relative flux) \dotfill & log$U$(1e$^{-6}$,1e$^6$) \dotfill & 0.00213$^{+0.00015}_{-0.00014}$ \\
$\rho_{GP,\mathrm{TESS}}$ \dotfill & Photometric GP time-scale (days) \dotfill & log$U$(1e$^{-3}$,1e$^3$) \dotfill & 0.362$^{+0.037}_{-0.032}$ \\
$\mu_{\mathrm{CARMENES}}$  \dotfill & Systematic RV offset (km/s) \dotfill & $U$(-20,-10) \dotfill & -16.977$^{+0.012}_{-0.011}$  \\
$\sigma_{\mathrm{CARMENES}}$ \dotfill & RV jitter (m/s) \dotfill & $U$(0.001,100) \dotfill & 11.1$^{+11.5}_{-7.7}$ \\

\underline{TOI-4034}\\
$q_{1,\,TESS}$ \dotfill & Quadratic limb-darkening parametrization  \dotfill & $N$(0.337,0.057) \dotfill & 0.289$^{+0.048}_{-0.047}$\\
$q_{2,\,TESS}$ \dotfill & Quadratic limb-darkening parametrization \dotfill & $N$(0.251,0.135) \dotfill & 0.180$^{+0.087}_{-0.100}$\\
$m_{\mathrm{flux,TESS}}$ \dotfill & Photometric offset (relative flux) \dotfill & $N$(0,0.1) \dotfill & -0.000381$^{+0.000036}_{-0.000044}$\\
$\sigma_{\mathrm{TESS}}$ \dotfill & Photometric jitter (ppm) \dotfill & log$U$(1e$^{-5}$,1e$^3$) \dotfill & 0.013$^{+3.352}_{-0.013}$ \\
$\sigma_{GP,\mathrm{TESS}}$ \dotfill & Photometric GP amplitude (relative flux) \dotfill & log$U$(1e$^{-6}$,1e$^6$) \dotfill & 0.000586$^{+0.000034}_{-0.000029}$\\
$\rho_{GP,\mathrm{TESS}}$ \dotfill & Photometric GP time-scale (days) \dotfill & log$U$(1e$^{-3}$,1e$^3$) \dotfill & 0.473$^{+0.062}_{-0.044}$ \\
$\mu_{\mathrm{CARMENES}}$  \dotfill & Systematic RV offset (km/s) \dotfill & $U$(-10,-5) \dotfill & -6.572$^{+0.017}_{-0.017}$  \\
$\sigma_{\mathrm{CARMENES}}$ \dotfill & RV jitter (m/s) \dotfill & $U$(0.001,100) \dotfill & 25$^{+22}_{-17}$ \\

\underline{TOI-6564}\\
$m_{\mathrm{dilution,Bfield}}$ \dotfill & Dilution parameter for Brierfield data \dotfill & $U$(0,1) \dotfill &  0.961$^{+0.027}_{-0.038}$\\
$m_{\mathrm{dilution,PEST}}$ \dotfill & Dilution parameter for PEST data \dotfill & $U$(0,1) \dotfill &  0.931$^{+0.050}_{-0.098}$\\
$q_{1,\,TESS}$ \dotfill & Quadratic limb-darkening parametrization  \dotfill & $N$(0.335,0.057) \dotfill & 0.302$^{+0.041}_{-0.044}$\\
$q_{2,\,TESS}$ \dotfill & Quadratic limb-darkening parametrization \dotfill & $N$(0.250,0.147) \dotfill & 0.213$^{+0.064}_{-0.060}$\\
$q_{1,\,Bfield}$ \dotfill & Quadratic limb-darkening parametrization  \dotfill & $N$(0.666,0.025) \dotfill & 0.664$^{+0.023}_{-0.024}$\\
$q_{2,\,Bfield}$ \dotfill & Quadratic limb-darkening parametrization \dotfill & $N$(0.423,0.073) \dotfill & 0.375$^{+0.061}_{-0.061}$\\
$q_{1,\,PEST}$ \dotfill & Quadratic limb-darkening parametrization  \dotfill & $U$(0,1) \dotfill & 0.20$^{+0.28}_{-0.14}$\\
$q_{2,\,PEST}$ \dotfill & Quadratic limb-darkening parametrization \dotfill & $U$(0,1) \dotfill & 0.42$^{+0.33}_{-0.30}$\\
$m_{\mathrm{flux,TESS}}$ \dotfill & Photometric offset for TESS data (relative flux) \dotfill & $N$(0,0.1) \dotfill & -0.000007$^{+0.000076}_{-0.000067}$\\
$m_{\mathrm{flux,Bfield}}$ \dotfill & Photometric offset for Brierfield data (relative flux) \dotfill & $N$(0,0.1) \dotfill & -0.0007$^{+0.0093}_{-0.0225}$\\
$m_{\mathrm{flux,PEST}}$ \dotfill & Photometric offset for PEST data (relative flux) \dotfill & $N$(0,0.1) \dotfill & -0.000081$^{+0.00068}_{-0.00061}$\\
$\sigma_{\mathrm{TESS}}$ \dotfill & Photometric jitter for TESS data (ppm) \dotfill & log$U$(1e$^{-5}$,1e$^3$) \dotfill & 247$^{+17}_{-18}$ \\
$\sigma_{\mathrm{Bfield}}$ \dotfill & Photometric jitter for Brierfield data (ppm) \dotfill & log$U$(1e$^{-5}$,1e$^3$) \dotfill & 0.48$^{+100}_{-0.48}$ \\
$\sigma_{\mathrm{PEST}}$ \dotfill & Photometric jitter for PEST data (ppm) \dotfill & log$U$(0.1,1e$^5$) \dotfill & 3570$^{+150}_{-140}$ \\
$\sigma_{GP,\mathrm{TESS}}$ \dotfill & Photometric GP amplitude for TESS data (relative flux) \dotfill & log$U$(1e$^{-6}$,1e$^6$) \dotfill & 0.000344$^{+0.000047}_{-0.000038}$\\
$\sigma_{GP,\mathrm{Bfield}}$ \dotfill & Photometric GP amplitude for Brierfield data (relative flux) \dotfill & log$U$(1e$^{-6}$,1e$^6$) \dotfill & 0.013$^{+0.065}_{-0.012}$\\
$\sigma_{GP,\mathrm{PEST}}$ \dotfill & Photometric GP amplitude for PEST data (relative flux) \dotfill & log$U$(1e$^{-6}$,1e$^6$) \dotfill & 0.00159$^{+0.00058}_{-0.00038}$\\
$\rho_{GP,\mathrm{TESS}}$ \dotfill & Photometric GP time-scale for TESS data (days) \dotfill & log$U$(1e$^{-3}$,1e$^3$) \dotfill & 0.54$^{+0.10}_{-0.08}$ \\
$\rho_{GP,\mathrm{Bfield}}$ \dotfill & Photometric GP time-scale for Brierfield data (days) \dotfill & log$U$(1e$^{-3}$,1e$^3$) \dotfill & 2.60$^{+13.78}_{-2.24}$ \\
$\rho_{GP,\mathrm{PEST}}$ \dotfill & Photometric GP time-scale for PEST data (days) \dotfill & log$U$(1e$^{-3}$,1e$^3$) \dotfill & 0.0143$^{+0.0087}_{-0.0049}$ \\
$\mu_{\mathrm{MINERVA}}$  \dotfill & Systematic RV offset for MINERVA-Australis data (km/s) \dotfill & $U$(-10,-5) \dotfill & -8461.3$^{+7.1}_{-7.6}$  \\
$\mu_{\mathrm{CORALIE14}}$  \dotfill & Systematic RV offset for CORALIE14 data (km/s) \dotfill & $U$(-10,-5) \dotfill & -7608.6$^{+2.2}_{-2.2}$  \\
$\mu_{\mathrm{CORALIE24}}$  \dotfill & Systematic RV offset for CORALIE24 data (km/s) \dotfill & $U$(-10,-5) \dotfill & -7624.7$^{+2.4}_{-2.5}$  \\
$\sigma_{\mathrm{MINERVA}}$ \dotfill & RV jitter (m/s) for MINERVA-Australis data \dotfill & $U$(0.001,100) \dotfill & 0.16$^{+3.17}_{-0.15}$ \\
$\sigma_{\mathrm{CORALIE14}}$ \dotfill & RV jitter (m/s) for CORALIE14 data \dotfill & $U$(0.001,100) \dotfill & 0.6$^{+1.8}_{-0.5}$ \\
$\sigma_{\mathrm{CORALIE24}}$ \dotfill & RV jitter (m/s) for CORALIE24 data \dotfill & $U$(0.001,100) \dotfill & 0.09$^{+0.93}_{-0.09}$ \\

\hline                                   
\\
\end{tabular} \\
*n.b. similar to Table \ref{planetary_table}, $U$(a,b) denotes a uniform prior between a and b, while $N$(a,b) denotes a normal distribution with mean a and standard deviation n. In addition, log$U$(a,b) denotes a log-uniform prior between a and b.
\end{table*}


%


\bsp	
\label{lastpage}
\end{document}